\documentclass[a4paper,fleqn,usenatbib]{mnras}

\usepackage{newtxtext,newtxmath}

\usepackage[T1]{fontenc}
\usepackage{ae,aecompl}

\usepackage{graphicx}	

\usepackage{amsmath}	
\usepackage{amsfonts}
\usepackage{amsbsy}
\usepackage{amssymb}
\usepackage{makecell}
\usepackage{gensymb}
\usepackage{float}
\usepackage{enumerate}
\usepackage{lastpage}
\usepackage{natbib}
\usepackage{threeparttable}
\usepackage{array}
\usepackage{booktabs}
\usepackage{multirow}
\usepackage{longtable}
\usepackage{lscape}
\usepackage{tablefootnote}
\usepackage{url}

\makeatletter

\newcommand{\Rmnum}[1]{\expandafter\@slowromancap\romannumeral #1@}
\makeatother

\title[NH$_2$D in Massive star forming regions]{Deuterated Ammonia  in Galactic massive star forming regions}

\author[Yuqiang Li et al.]{Yuqiang Li$^{1,2}$, 
Junzhi Wang$^{1,3}$\thanks{E-mail: jzwang@shao.ac.cn},  Juan Li $^{1,3}$\thanks{E-mail: lijuan@shao.ac.cn}, Shu Liu$^{4}$, Qiuyi Luo$^{1,2}$\\
$^{1}$Shanghai Astronomical Observatory, Chinese Academy of Sciences,80 Nandan Road, Shanghai, 200030, China\\
$^{2}$University of Chinese Academy of Sciences, 19A Yuquanlu, Beijing 100049, China \\
$^{3}$Key Laboratory of Radio Astronomy, Chinese Academy of Sciences,  10 Yuanhua Road, Nanjing, JiangSu 210033, China\\
$^{4}$CAS Key Laboratory of FAST, National Astronomical Observatories, Chinese Academy of Sciences, Beijing 100012, China\\
}

\date{Accepted XXX. Received YYY; in original form ZZZ}

\pubyear{20XX}

\begin{document}
	\label{firstpage}
	\pagerange{\pageref{firstpage}--\pageref{lastpage}}
	\maketitle
	
	\begin{abstract}
We present sensitive observations of NH$_2$D $1_{11}^a-1_{01}^s$ at 110.153599 GHz toward 50 Galactic massive star-forming regions with IRAM 30-m telescope. The NH$_2$D $1_{11}^a-1_{01}^s$ transition is detected toward 36 objects, yielding a detection rate of 72\%. Column densities of NH$_2$D, HC$_3$N and C$^{18}$O  for each source are derived by assuming local thermal equilibrium conditions with a fixed excitation temperature. The deuterium ratio of NH$_3$, defined as the abundance ratio of NH$_2$D to NH$_3$, for 19 sources is also obtained with the information of NH$_3$ from the literature.   The range of deuterium fractionation bends to be large in the late-stage star-forming regions in this work, with the value from 0.043 to 0.0006. The highest deuterium ratio of NH$_3$ is  0.043 in G081.75+00.78 (DR21). We also find that the deuterium ratio of NH$_3$  increases with the Galactocentric distances and decreases with the line width.
	\end{abstract}
	
	

\section{Introduction}  
\label{sect:intro}
Deuterated molecules can be used as tools to diagnose the evolution phase in star-forming regions   \citep{2019ApJ...883..202F}.  Due to low zero-point energy, the deuterium molecules form more easily than other species \citep{1982A&A...111...76H}, which leads to higher deuterium fractionation ($D_{\rm frac}$) in molecules than the D/H ratio as  $\approx1.5 \times 10^{-5}$   \citep{1995ApJ...451..335L,2003ApJ...587..235O}.

 There are three mechanisms to explain the reason of deuterated molecular enhancement in star-forming regions being several magnitudes higher than the D/H ratio:  (\Rmnum{1}) Deuterium molecules accelerate to form in the dust gains surface and volatilize by heating   \citep{1990ApJ...362L..29T,2003ApJ...593L..51C,2008A&A...492..703C}; (\Rmnum{2}) Shocks associated with energetic outflows may affect the deuterium fractionation   \citep{2002ApJ...569..322L,2008A&A...492..703C}; (\Rmnum{3}) The gas-phase chemistry could also enhance the enormous abundance of deuterated molecules  \citep{2000A&A...361..388R,2000A&A...364..780R}.

There are several deuterated molecules detected in molecular clouds, including DCN, DCO$^+$, DNC, and N$_2$D$^+$ \citep{2002ApJ...565..331C,2005ApJ...619..379C,2015A&A...579A..80G}.   
Deuterium  ammonia (NH$_2$D) is also an  important deuterated molecule in molecular clouds  \citep{2007A&A...470..221C,2010A&A...517L...6B}.  
NH$_2$D was extensively detected in star forming regions with several degree of deuterium fractionation: $D_{\rm frac}$(NH$_3$) $\sim 1-0.1$ in dense cores  \citep{2003A&A...403L..25H,2007A&A...470..221C,2017A&A...600A..61H} and $D_{\rm frac}$(NH$_3$) $\sim 0.1-0.001$ in late stage massive star forming regions  \citep{2010A&A...517L...6B,2011A&A...530A.118P,2015A&A...581A..48G}.

 According to chemical equilibrium conditions, the $D_{\rm frac}$ of NH$_3$  should decrease with the increasing of temperature \citep{1982A&A...111...76H,2005A&A...438..585R}. However, several studies  suggested  that the $D_{\rm frac}$ of NH$_3$ did  not follow such relation in late stage massive star forming regions. For example,  \cite{2010A&A...517L...6B} have investigated $D_{\rm frac}$(NH$_3$) $\simeq 0.1$ toward pre-protostellar cores and young stellar objects around the UCH\Rmnum{2} region, which is higher than expected as $D_{\rm frac}$(NH$_3$) $\sim 0.1-0.001$ \citep{1982A&A...111...76H}.   Moreover, since such studies for  NH$_2$D performed only toward limited samples, e.g. \cite{1990A&A...228..447J} observed NH$_2$D $1_{11}^a-1_{01}^s$ at 110 GHz toward 15 hot cores and \cite{2007A&A...467..207P} observed NH$_2$D  $1_{11}^s-1_{01}^a$ at 85.9 GHz toward 32 high mass pre/protocluster clumps, or  large sample with limited sensitivity \citep{2021A&A...649A..21W}, large sample sensitive survey  toward massive star forming regions are needed for understand astro-chemical process of NH$_2$D . 

This work presents results of NH$_2$D observations toward a relatively large sample of 50  Galactic late-stage massive star-forming regions with 6.7 GHz CH$_3$OH masers with the Institut de Radioastronomie Millim ´etrique (IRAM) 30-m telescope. The  observations are described in Section \ref{sect:Obs}, while the main results are reported  in Section \ref{sect:Result},  discussions is presented  in Section \ref{sect:discussion},  and a brief summary  is given in Section \ref{sect:Conclusions}.

\section{Observations and data reduction}
\label{sect:Obs}

The sample includes 50 late-stage massive star-forming regions with 6.7 GHz CH$_3$OH masers selected from  \cite{2014ApJ...783..130R} with HC$_3$N 12-11 higher than 1 K ($T_{\rm mb}$) based on IRAM 30-m observation. The trigonometric parallaxes of these sources have been measured with class \Rmnum{2} methanol or water masers.  
The observations were performed with  the IRAM 30-m millimeter telescope (IRAM 30-m) at
Pico Veleta, Spain on June and October  2016, and August 2017. We used the 3 mm (E0) and 2mm (E1) band of The Eight Mixer Receiver (EMIR) simultaneously and the  Fourier Transform Spectrometers (FTS) backend to cover 8 GHz bandwidth and 195 kHz spectral resolution for each band with dual polarization. The standard position switching mode with azimuth off of -600 arcsec was used.  The 3mm band (E0) covered about from 105.8 to 113.6 GHz, while the 2mm band (E1) covered from 169.8 to 173.8 GHz.
  
The beam sizes of IRAM 30-m  range from FWHM $\sim$ 22.4 arcsec at 110 GHz to $\sim$ 14.3 arcsec at 172 GHz. The typical system temperatures were about 150 K in the 3-mm band and about
400 K in the 2-mm band. Pointing was checked every 2 hours with nearby strong quasi-stellar objects. Focus was checked and corrected at the beginning of each run and during sunsets/sunrises. The antenna temperature ($T_{\rm A}^{\ast}$) was converted to the main beam brightness temperature ($T_{\rm mb}$), using $T_{\rm mb}$ = T$_{\rm A}^{\ast}\cdot$ $F_{\rm eff}/$$B_{\rm eff}$,
where the forward efficiency $F_{\rm eff}$ is 0.95 and beam efficiency $B_{\rm eff}$ is 0.81 for 3 mm band, while  $F_{\rm eff}$ is 0.93 and   $B_{\rm eff}$ is 0.73 for 2 mm band.  Each scan consists of 2 minutes with an on-source integration of 1 minute.  

 The used lines include NH$_2$D $1_{11}^a-1_{01}^s$ at 110.153599 GHz, HC$_3$N 12-11 at 109.173638 GHz, C$^{18}$O 1-0 at 109.782176 GHz, and HC$_3$N 19-18  at 172.849300 GHz. The integration time for each source was about 12  to 234 minutes. The detailed observation parameters are listed in Table \ref{obs1}. Data reduction was conducted with the CLASS  package, which is a part of the GILDAS\footnote{\url{http://www.iram.fr/IRAMFR/GILDAS}} software. The first-order baseline was used for all lines. 

\section{Result and analysis}
\label{sect:Result}

\subsection{The detected lines}

 NH$_2$D $1_{11}^a-1_{01}^s$ was detected in 36 (Figure\ref{spectrum1}-\ref{spectrum5}) among 50 sources with a detection rate of 72 per cent. The other three lines, C$^{18}$O 1-0, HC$_3$N 12-11, and HC$_3$N 19-18  were detected in 50, 50 and 49 targets, with 100, 100 and 98 percent detection rate, respectively. Fluxes of each line in these sources were derived with the GILDAS/CLASS package. There are five hyperfine structures for NH$_2$D $1_{11}^a-1_{01}^s$ \citep{2000ApJ...535..227S}, which are blended due to line broadening in most of the sources. Thus, the fluxes of NH$_2$D $1_{11}^a-1_{01}^s$ were derived with integrated intensities of the emitting channels of the 5 hyperfine lines instead of Gaussian fitting.  As an example, NH$_2$D $1_{11}^a-1_{01}^s$ in G081.75+00.59 showed that three components  (F=2-1, F=2-2 and F=1-2) are mixed  (see Fig. \ref{HF}). The fluxes of C$^{18}$O 1-0, HC$_3$N 12-11, and HC$_3$N 19-18 were also obtained with integrated intensities of the emitting channels. 
 
For sources with the non-detection of  NH$_2$D and HC$_3$N 19-18,   3$\sigma$  upper limits for $\int$$T_{\rm mb}\rm dv$  were calculated:
\begin{equation}
	\int  T_{\rm mb}dv=3 rms \sqrt{\delta v \cdot \Delta v} (\rm K km\; s^{-1}),
\end{equation}
where $\delta v$ is channel separation in velocity, while $\Delta v$ is the line width in km s$^{-1}$.

 NH$_2$D at 110153.599 MHz is close to the rest frequency of  $^{13}$CO 1-0 at 110201.354 MHz, which is about 130 km s$^{-1}$ offset. One  $^{13}$CO 1-0 emission with velocity of 130 km s$^{-1}$  can be considered as  NH$_2$D at 110153.599 MHz with velocity of 0 km s$^{-1}$. Since $^{13}$CO  molecules can be easily excited to $J$=1 energy level and all of the observed sources are within the Galactic plane,  the contamination of $^{13}$CO 1-0  with +130 km s$^{-1}$  to NH$_2$D line should be checked. For emission feature that can be identified as NH$_2$D emission,  the emission of HC$_3$N 12-11 at the same velocity is also required in our criterion. Furthermore, we also require that there is no C$^{18}$O 1-0 emission at the velocity if the emission feature is identified as  $^{13}$CO 1-0.  In the latter case,  emission feature in G012.88+00.48 is considered as $^{13}$CO 1-0 instead of  NH$_2$D (see Fig. \ref{pile}). On the other hand, almost the other 36 sources with emission features above 3$\sigma$ levels meet both criteria. As examples, 
 NH$_2$D $1_{11}^a-1_{01}^s$, C$^{18}$O 1-0, HC$_3$N 12-11 and  HC$_3$N 19-18  lines in three sources are presented in Figure \ref{overlay}.

\subsection{Column densities of the molecules}

Under local thermal equilibrium   (LTE) condition, the partition function ($Q$($T_{\rm ex}$)) is related to the excitation temperature. From $T_{\rm ex}$ = 9.375 K to $T_{\rm ex}$ = 37.5 K, the partition function increase about 7 times, 3.5 times and 4 times based on the information from CDMS\footnote{\url{https://cdms.astro.uni-koeln.de/classic/predictions/catalog/partition_function.html}} \citep{2001A&A...370L..49M,2005JMoSt.742..215M}  
  for NH$_2$D, HC$_3$N and C$^{18}$O, respectively. The column densities of NH$_2$D were derived in the
  constant excitation temperature approximation \citep{2002ApJ...565..344C}. The excitation temperature is relatively higher in the hot core. Thus, excitation temperature $T_{\rm ex}$ of 18.75 K is used for calculations in this work, with $Q(T$$_{\rm ex}$)s from CDMS, which are  64.9118, 86.2186 and 7.4607 for NH$_2$D, HC$_3$N and C$^{18}$O, respectively. For $T_{\rm ex}$ = 9.375 K, the $Q(T_{\rm ex})$ is 2.5 times lower than that of $T_{\rm ex}$ = 18.75 K, while for $T_{\rm ex}$ = 37.5 K, the $Q(T_{\rm ex})$ is three times higher than that of  $T_{\rm ex}$ = 18.75 K.

Assuming the lines are optically thin, the column densities of the molecules are given by
\begin{equation}
	N_{\rm tot}=\frac{8\pi k\nu^2}{hc^3A_{ul}}\frac{Q(T_{\rm ex})}{\rm g_u}e^{E_u/kT_{\rm ex}}\int T_{\rm mb}\rm dv (\rm cm^{-2})
\end{equation} 
 where $k=1.38\times10^{-16}$ erg\;K$^{-1}$ is the Boltzmann constant, $\nu$ is the frequency of the transition, $h=6.624\times10^{-27} $erg\;s is the Planck constant, $c=2.998\times10^{10}$ cm $\rm s^{-1}$ is the speed of light, $A_{\rm ul}$ is the Einstein emission coefficient, $\rm g_u$ is the upper level degeneracy, and $E_u$ is the energy of the upper level above ground state. The frequency, $\rm g_u$, $A_{\rm ul}$, and  $E_u$  for four emission lines are taken from the Splatalogue database \citep{2007AAS...21113211R}\footnote{\url{https://splatalogue.online//index.php}}. Note that  HC$_3$N 12-11 line is used for calculations of  HC$_3$N molecule.

The column density of H$_2$ can be derived by  C$^{18}$O  \citep{1982ApJ...262..590F} with 
\begin{equation}
	N({\rm H_2})=\frac{N(\rm C^{18}O)}{2.18\times 10^{15}}\times 10^{21} (\rm cm^{-2}).
\end{equation} 

The relative abundances of molecules to H$_2$ shown in Table \ref{Colunm density} are based on the H$_2$ column density estimated with C$^{18}$O.

\subsection{Deuterium fractionation of NH$_3$}
Among the 50 sources, 19 were reported with  NH$_3$ observations, which provided column densities of NH$_3$  \citep{2016AJ....152...92L}. Thus,  the deuterium fractionations $D_{\rm frac}$ (NH$_3$), like the abundance ratio of NH$_2$D to NH$_3$,    can be calculated for these 19 sources.

The deuterium fractionation values $D_{\rm frac}$ (NH$_3$) are given in Table \ref{Colunm density}. One of the important results for this study is that the  range of deuterium fractionation, with the value from 0.043 to 0.0006, is large in the late-stage star-forming regions. The highest deuterium fractionation is 0.043 in G081.75+00.78 (DR21), which is a giant star-forming complex  \citep{2009ApJ...694.1056K}  with massive molecular  outflows \citep{2010A&A...520A..49S}, while the lowest $D_{\rm frac}$ (NH$_3$) with NH$_2$D detection is  in G031.28+00.06, with $D_{\rm frac}$ (NH$_3$) of 0.0006.

\section{Discussion}
\label{sect:discussion}

\subsection{NH$_2$D abundance in massive star forming regions}

NH$_2$D has been detected  toward nearby low mass dense cores  \citep{2000ApJ...535..227S, 2001ApJ...554..933S}. With a detection rate of  50 per cent  for NH$_2$D ($1_{11}^a-1_{01}^s$) line toward 16 dark molecular clouds, \cite{2000ApJ...535..227S}  reported the lowest value of $D_{\rm frac}$(NH$_3$) as 0.025 in L1489N and the highest as 0.178 in $\rho$ Op E.  With observations of NH$_2$D ($1_{11}^s-1_{01}^a$) at 85.9 GHz toward 32 protostellar and prestellar sources, \cite{2001ApJ...554..933S} reported a nearly 70 per cent detection rate with significant emission  above 5$\sigma$  for $D_{\rm frac}$(NH$_3$) from 0.0028 to 0.09. 

 NH$_2$D has also been detected  in  hot cores \citep{1990A&A...228..447J, 2007A&A...467..207P, 2021A&A...649A..21W}. 
With observations of NH$_2$D $1_{11}^a-1_{01}^s$ at 110 GHz toward 15 late stage massive star forming regions, \cite{1990A&A...228..447J} reported a detection rate of 53 per cent, with $D_{\rm frac}$(NH$_3$) from 0.004 to 0.03. 
With the detection of NH$_2$D  $1_{11}^s-1_{01}^a$ at 85.9 GHz toward 22 out of 32  high mass pre/protocluster clumps, \cite{2007A&A...467..207P} reported the lowest value of $D_{\rm frac}$(NH$_3$) of 0.004  in G10.61-0.33 and the highest value of  0.665 in G18.17-0.30.  In a large sample survey of multiple NH$_2$D lines toward 992 APEX Telescope Large Area Survey (ATLASGAL) sources, 
\cite{2021A&A...649A..21W}   reported a detection rate of $\sim$ 39 per cent for NH$_2$D  $1_{11}^s-1_{01}^a$ at 85.9 GHz, while  the detection rate is 17 per cent (65 out of 373  observed sources) for  NH$_2$D $1_{11}^a-1_{01}^s$ at 110 GHz. The detection rate in different evolutionary stages suggested that  NH$_2$D detection rate  is higher
 during the earlier evolutionary phases \citep{2021A&A...649A..21W}.

The formation of deuterated molecules is an active process in cold molecular gas \citep{2012A&ARv..20...56C}. Abundant neutral species are frozen-out on to dust grains in cold molecular gas and some deuterium processes happen on the surface of dust grains \citep{2012A&ARv..20...56C}.  NH$_2$D form via H$_2$D$^+$ \citep{2009A&A...493...89E}, with the abundance ratio of H$_3^+$ to H$_2$D$^+$ showing a sharp drop above $\simeq 15$ K \citep{2008A&A...492..703C}. This might mean that $D_{\rm frac}$(NH$_3$) decreases at relatively high temperature. Also, through simulation, \cite{2005A&A...438..585R} showed in a model that the deuterium fractionation of NH$_3$ decreases with temperature increasing when the temperature exceeds 22 K.

 Deuterium fractionation ($D_{\rm frac}$) of N$_2$H$^+$ was also suggested to decrease with evolution stage in  high-mass star-forming regions \citep{2010ApJ...713L..50C,2011ApJ...743..196C,2015A&A...575A..87F}. Thus,  massive starforming
regions in the late stage are expected to have low NH$_2$D abundance. However,  with a detection rate of 72 percent NH$_2$D in late-stage massive star-forming regions and $D_{\rm frac}$ of NH$_3$ up to higher than 0.1 through our observations, it seems that the $D_{\rm frac}$ of  NH$_3$ does not decrease with the evolutionary stage in high-mass star-forming regions,  as had been suggested for N$_2$H$^+$  \citep{2010ApJ...713L..50C}.  \cite{2010A&A...517L...6B} suggested that $D_{\rm frac}$(NH$_3$) can be used as a tool to distinguish between pre-protostellar and protostellar cores. However,  \cite{2015A&A...575A..87F} suggested that $D_{\rm frac}$(NH$_3$) does not change obviously with evolution, which is supported by the results presented here.

$D_{\rm frac}$ (NH$_3$) in the sample of late-stage massive star-forming regions in our observation range from 0.0006 to 0.043.  The lowest $D_{\rm frac}$ (NH$_3$) (0.0006)  is still about five times higher than the D/H ratio in the interstellar medium (ISM: $\approx1.5 \times 10^{-5}$). Based on the high detection rate, we suggest that NH$_{2}$D can commonly exist in late-stage massive star-forming regions.

\subsection{NH$_2$D abundance and  dense gas properties}
The density and temperature of molecular clouds might affect the chemical enhancement of  NH$_2$D.  Information from simultaneously obtained  HC$_3$N lines was used to probe dense gas properties in these sources. As a molecule with large dipole moment, HC$_3$N can only be excited to high $J$ level in dense molecular gas. Thus, millimetre HC$_3$N lines, which are normally optically thin in molecular clouds,  are good dense gas tracers \citep{1983ApJ...271..161V}.   No correlation was found between the relative abundance of NH$_2$D to molecular hydrogen and that of HC$_3$N (see Fig. \ref{HC3N} left).  Deuterium fractionations ($D_{\rm frac}$) of NH$_3$ show no correlation with relative abundance of HC$_3$N  (see Fig. \ref{HC3N} right).  Note that the relative abundance of HC$_3$N  derived with emission lines of  HC$_3$N   mainly reflects the dense gas fraction instead of the abundance of HC$_3$N.  It seems that the gas density is not the main reason for the enhancement of NH$_2$D.

The line ratio of two transitions of HC$_3$N, 19-18 and 12-11, can be used to infer the excitation conditions of the molecular gas.  No correlation with the line ratio of HC$_3$N 19-18/12-11 was found for both relative abundance of NH$_2$D and  $D_{\rm frac}$ of NH$_3$ (Fig. \ref{HC3Nratio}). Thus, we suggest that the gas temperature in the dense part cannot be the main contribution to the enhancement of NH$_2$D.

\subsection{NH$_2$D enhancement along $D_{\rm GC}$}

The relations between relative abundance of NH$_2$D or  $D_{\rm frac}$ of NH$_3$  and Galactocentric distance ($D_{\rm GC}$) of each source are also tested (see Fig. \ref{DGC}). No correlation was found for   relative abundance of NH$_2$D  and $D_{\rm GC}$. However, a clear trend was found that $D_{\rm frac}$ of NH$_3$ increases with the increasing $D_{\rm GC}$.
Two methods of  fitting the relationship between  $D_{\rm frac}$ of NH$_3$   and Galactocentric distance were used and similar results were obtained.  The detailed results are: 
\begin{equation}
	\log(D_{\rm frac}({\rm NH_3}))=(0.168\pm 0.072)D_{\rm GC}-(3.452\pm 0.487).
	\label{lq_eq}
\end{equation}
with least-squares method,  and  
\begin{equation}
	\log(D_{\rm frac}({\rm NH_3}))=0.156^{+0.372}_{-0.376}D_{\rm GC}-3.604^{+2.723}_{-2.713}.
	\label{MCMC_eq}
\end{equation}
with Markov Chain Monte Carlo(MCMC) method. With the Pearson correlation coefficient, the linear relationships between $D_{\rm frac}$ of NH$_3$ and $D_{\rm GC}$ were tested. We obtained the correlation coefficient $\rho$ = 0.49 and {\textit{p}-value p = 0.06. Since $D_{\rm frac}$ of NH$_3$ is an astro-chemical process rather than the chemical evolution caused during the star-formation history.

\subsection{NH$_2$D enhancement  and turbulence?}

Through the observations toward a sample including infrared dark cloud, high-mass protostellar object, hot molecular cores and UCH\Rmnum{2} region,   $D_{\rm frac}$ of HNC, HCO$^+$ and N$_2$H$^+$ was found to decrease with increasing line width (FWHM) of HN$^{13}$C, H$^{13}$CO$^+$ and N$_2$H$^+$, respectively, which was explained by the possible decrease of deuterium ratio with increasing microturbulent velocity field \citep{2015A&A...579A..80G}.  The relation between $D_{\rm frac}$ of NH$_3$ and FWHM of HC$_3$N 12-11 is presented in Fig. \ref{FWHM}, which shows a good anti-correlation, except for G049.48-00.38, which is W51M. 
A least-squarest to fit these 14 targets yields:
\begin{equation}
	\log(D_{\rm frac}({\rm NH_3}))=-(0.406\pm 0.115)FWHM+(0.883\pm 0.435).
	\label{FWHM_eq}
\end{equation}
with a correlation coefficient of 0.66 and {\textit{p}-value of 0.01 which were tested using the Pearson correlation coefficient. We exclude W51 M when fitted Equation (\ref{FWHM_eq}).
  {The process of the NH$_2$D enhancement may be similar to} DNC, DCO$^+$ and N$_2$D$^+$ seen in \citep{2015A&A...579A..80G}, while the active turbulence may decrease such enhancement.

\section{Summary and conclusion remarks}
\label{sect:Conclusions}

We performed observations of  NH$_2$D $1_{11}^a-1_{01}^s$ transition toward 50 Galactic late-stage massive star forming regions with the IRAM 30-m telescope. NH$_2$D $1_{11}^a-1_{01}^s$ emission was detected toward 36 targets, with a detection rate of 72 per cent. C$^{18}$O 1-0, HC$_3$N 12-11, and HC$_3$N 19-18 were also detected in 50 sources, except  for G168.06+00.82,  in which HC$_3$N 19-18 was not detected. Our main results include:
\begin{enumerate}[1.]
	\item Column densities of NH$_2$D,  C$^{18}$O,  and  HC$_3$N in each source, were derived based on the detected lines,  with the assumption of  LTE condition with a constant excitation temperature. 
	\item Combining with the NH$_3$ data of \cite{2016AJ....152...92L}, we calculated the $D_{\rm frac}$(NH$_3$) for 19 targets. We detected a  large range of deuterium fractionation in the late-stage star-forming regions. The value of the range is from 0.043 to 0.0006. The highest $D_{\rm frac}$(NH$_3$) is 0.043 in G081.75+00.78 (DR21).
	\item A trend of increasing $D_{\rm frac}$(NH$_3$) with increasing D$_{GC}$ was found.
	\item A clear trend of increasing $D_{\rm frac}$(NH$_3$) with decreasing line width was found.
	\item Because of the high detection rate of NH$_2$D, we suggest that NH$_2$D can commonly exist in late-stage massive star-forming regions.
\end{enumerate}

\section*{Acknowledgements}

This work is supported by the National Key R$\&$D Program of China (No. 2017YFA0402604)  and the National Natural Science Foundation
of China grant U1731237.  This study is based on observations
carried out under project numbers 012-16 and 023-17  with the IRAM 30-m telescope. IRAM is
supported by INSU/CNRS (France), MPG (Germany) and IGN (Spain).  

\section*{Data availability}
The original data observed with IRAM 30-m can be accessed by IRAM archive system at \url{https://www.iram-institute.org/EN/content-page-386-7-386-0-0-0.html}. If anyone is interested in the  reduced data presented in this paper, please contact  Junzhi Wang at jzwang@shao.ac.cn.










\clearpage
\onecolumn 
\centering

\begin{longtable}[c]{lccccc}
	\caption{Information for source}
	\label{obs1}\\
	\hline
	\hline
	\multicolumn{1}{c}{\multirow{2}{*}{Source}} & \multicolumn{1}{c}{\multirow{2}{*}{Alias}}            & R.A.         & Decl.       & $D_{\rm GC}$$^1$ & $V_{\rm LSR}$$^2$  \\
	\multicolumn{1}{c}{}                        &                 & (hh:mm:ss)   & (dd:mm:ss)  & (kpc)    & (km s$^{-1}$)   \\ \hline
	\endfirsthead
	\multicolumn{6}{c}%
	{{\bfseries Table \thetable\ continued from previous page}} \\
	\hline
	\hline
	\multicolumn{1}{c}{\multirow{2}{*}{Source}} & \multicolumn{1}{c}{\multirow{2}{*}{Alias}}           & R.A.         & Decl.       & $D_{\rm GC}$$^1$ & $V_{\rm LSR}$$^2$  \\
	\multicolumn{1}{c}{}                        &                 & (hh:mm:ss)   & (dd:mm:ss)  & (kpc)    & (km s$^{-1}$)   \\ \hline
	\endhead
	\hline
	\hline
	\endfoot
	\endlastfoot
	G000.67-00.03           & Sgr B2          & 17:47:20.00  & -28:22:40.0 & 0.2      & $62\pm5$   \\
	G005.88-00.39           &                 & 18:00:30.31  & -24:04:04.5 & 5.3      & $9\pm3$    \\
	G009.62+00.19           &                 & 18:06:14.66  & -20:31:31.7 & 3.3      & $2\pm3$    \\
	G010.47+00.02           &                 & 18:08:38.23  & -19:51:50.3 & 1.6      & $69\pm5$   \\
	G010.62-00.38           & W 31            & 18:10:28.55  & -19:55:48.6 & 3.8      & $-3\pm5$   \\
	G011.49-01.48           &                 & 18:16:22.13  & -19:41:27.2 & 7.1      & $11\pm3$   \\
	G011.91-00.61           &                 & 18:13:58.12  & -18:54:20.3 & 5.1      & $37\pm5$   \\
	G012.80-00.20           &                 & 18:14:14.23  & -17:55:40.5 & 5.5      & $34\pm5$   \\
	G012.88+00.48           & IRAS 18089-1732 & 18:11:51.42  & -17:31:29.0 & 5.9      & $31\pm7$   \\
	G012.90-00.24           &                 & 18:14:34.42  & -17:51:51.9 & 5.9      & $36\pm10$  \\
	G012.90-00.26           &                 & 18:14:39.57  & -17:52:00.4 & 5.9      & $39\pm10$  \\
	G014.33-00.64           &                 & 18:18:54.67  & -16:47:50.3 & 7.2      & $22\pm5$   \\
	G015.03-00.67           & M 17            & 18:20:24.81  & -16:11:35.3 & 6.4      & $22\pm3$   \\
	G016.58-00.05           &                 & 18:21:09.08  & -14:31:48.8 & 5.0      & $60\pm5$   \\
	G023.00-00.41           &                 & 18:34:40.20  & -09:00:37.0 & 4.5      & $80\pm3$   \\
	G023.44-00.18           &                 & 18:34:39.19  & -08:31:25.4 & 3.7      & $97\pm3$   \\
	G027.36-00.16           &                 & 18:41:51.06  & -05:01:43.4 & 3.9      & $92\pm3$   \\
	G028.86+00.06           &                 & 18:43:46.22  & -03:35:29.6 & 4.0      & $100\pm10$ \\
	G029.95-00.01           & W 43S           & 18:46:03.74  & -02:39:22.3 & 4.6      & $98\pm3$   \\
	G031.28+00.06           &                 & 18:48:12.39  & -01:26:30.7 & 5.2      & $109\pm3$  \\
	G031.58+00.07           & W 43Main        & 18:48:41.68  & -01:09:59.0 & 4.9      & $96\pm5$   \\
	G032.04+00.05           &                 & 18:49:36.58  & -00:45:46.9 & 4.8      & $97\pm5$   \\
	G034.39+00.22           &                 & 18:53:18.77  & +01:24:08.8 & 7.1      & $57\pm5$   \\
	G035.02+00.34           &                 & 18:54:00.67  & +02:01:19.2 & 6.5      & $52\pm5$   \\
	G035.19-00.74           &                 & 18:58:13.05  & +01:40:35.7 & 6.6      & $30\pm7$   \\
	G035.20-01.73           &                 & 19:01:45.54  & +01:13:32.5 & 5.9      & $42\pm3$   \\
	G037.43+01.51           &                 & 18:54:14.35  & +04:41:41.7 & 6.9      & $41\pm3$   \\
	G043.16+00.01           & W 49N           & 19:10:13.41  & +09:06:12.8 & 7.6      & $10\pm5$   \\
	G043.79-00.12           & OH 43.8-0.1     & 19:11:51.99  & +09:35:50.3 & 5.7      & $44\pm10$  \\
	G049.48-00.36           & W 51 IRS2       & 19:23:39.82  & +14:31:05.0 & 6.3      & $56\pm3$   \\
	G049.48-00.38           & W 51M           & 19:23:43.87  & +14:30:29.5 & 6.3      & $58\pm4$   \\
	G059.78+00.06           &                 & 19:43:11.25  & +23:44:03.3 & 7.5      & $25\pm3$   \\
	G069.54-00.97           & ON 1            & 20:10:09.07  & +31:31:36.0 & 7.8      & $12\pm5$   \\
	G075.76+00.33           &                 & 20:21:41.09  & +37:25:29.3 & 8.2      & $-9\pm9$   \\
	G078.12+03.63           & IRAS 20126+4104 & 20:14:26.07  & +41:13:32.7 & 8.1      & $-4\pm5$   \\
	G081.75+00.59           & DR 21           & 20:39:01.99  & +42:24:59.3 & 8.2      & $-3\pm3$   \\
	G081.87+00.78           & W 75N           & 20:38:36.41  & +42:37:34.8 & 8.2      & $7\pm3$    \\
	G092.67+03.07           &                 & 21:09:21.73  & +52:22:37.1 & 8.5      & $-5\pm10$  \\
	G109.87+02.11           & Cep A           & 22:56:18.10  & +62:01:49.5 & 8.6      & $-7\pm5$   \\
	G111.54+00.77            & NGC 7538        & 23:13:45.36  & +61:28:10.6 & 9.6      & $-57\pm5$  \\
	G121.29+00.65           & L 1287          & 00:36:47.35  & +63:29:02.2 & 8.8      & $-23\pm5$  \\
	G123.06-06.30           & NGC 281         & 00:52:24.70  & +56:33:50.5 & 10.1     & $-30\pm5$  \\
	G133.94+01.06           & W 3OH           & 02:27:03.82  & +61:52:25.2 & 9.8      & $-47\pm3$  \\
	G168.06+00.82           & IRAS 05137+3919 & 05:17:13.74  & +39:22:19.9 & 15.9     & $-27\pm5$  \\
	G176.51+00.20           &                 & 05:37:52.14  & +32:00:03.9 & 9.3      & $-17\pm5$  \\
	G183.72-03.66           &                 & 05:40:24.23  & +23:50:54.7 & 10.0     & $3\pm5$    \\
	G188.94+00.88           & S 252           & 06:08:53.35  & +21:38:28.7 & 10.4     & $8\pm5$    \\
	G192.60-00.04           & S 255           & 06:12:54.02  & +17:59:23.3 & 9.9      & $6\pm5$    \\
	G209.00-19.38           & Orion Nebula    & 05:35:15.80  & -05:23:14.1 & 8.6      & $3\pm5$    \\
	G232.62+00.99           &                 & 07:32:09.78  & -16:58:12.8 & 9.4      & $21\pm3$   \\ \hline
\end{longtable}
\vspace{-1.5em}
\begin{minipage}{\linewidth}
	\renewcommand{\footnoterule}{}
	\footnotetext{$^{1,2}$ Values are refered from \cite{2014ApJ...783..130R}}
\end{minipage}

\tiny
\begin{longtable}[c]{ccccccccccc}
	\caption{The data of NH$_2$D$1_{11}^a-1_{01}^s$,HC$_3$N(12-11),C$^{18}$O(1-0) and HC$_3$N(19-18)}
	\label{data}\\
	\hline
	\hline
	source        & \multicolumn{7}{c}{2mm}                                                         & \multicolumn{3}{c}{3mm}       \\ \cmidrule(lr){2-8} \cmidrule(lr){9-11}\\
	\multirow{2}{*}{} & rms & \multicolumn{2}{c}{NH$_2$D($1_{11}^a-1_{01}^s$)} & \multicolumn{2}{c}{C$^{18}$O(1-0)} & \multicolumn{2}{c}{HC$_3$N(12-11)} & rms & \multicolumn{2}{c}{HC$_3$N(19-18)} \\ \cmidrule(lr){3-4} \cmidrule(lr){5-6} \cmidrule(lr){7-8} \cmidrule(lr){10-11} 
	&    (mK) &  $\int T_{\rm mb}\rm  dv$    & \makecell[c]{Velocity \\ range} &   $\int T_{\rm mb}\rm  dv$    & \makecell[c]{Velocity \\ range}  &  $\int T_{\rm mb}\rm  dv$   & \makecell[c]{Velocity \\ range}    &    (mK)  &  $\int T_{\rm mb}\rm  dv$    & \makecell[c]{Velocity \\ range}     \\
	& & (K$\cdot$km $\rm s^{-1}$) & (km $\rm s^{-1}$) & (K$\cdot$km $\rm s^{-1}$) & (km $\rm s^{-1}$) & (K$\cdot$km $\rm s^{-1}$) & (km $\rm s^{-1}$) &  &(K$\cdot$km $\rm s^{-1}$) & (km $s^{-1}$) \\ \hline
	\endfirsthead
	\multicolumn{11}{c}%
	{{\bfseries Table \thetable\ continued from previous page}} \\
	\hline
	\hline
	source        & \multicolumn{7}{c}{2mm}                                                         & \multicolumn{3}{c}{3mm}       \\ \cmidrule(lr){2-8} \cmidrule(lr){9-11}\\
	\multirow{2}{*}{} & rms & \multicolumn{2}{c}{NH$_2$D($1_{11}^a-1_{01}^s$)} & \multicolumn{2}{c}{C$^{18}$O(1-0)} & \multicolumn{2}{c}{HC$_3$N(12-11)} & rms & \multicolumn{2}{c}{HC$_3$N(19-18)} \\ \cmidrule(lr){3-4} \cmidrule(lr){5-6} \cmidrule(lr){7-8} \cmidrule(lr){10-11} 
	&    (mK) &  $\int T_{\rm mb}\rm  dv$    & \makecell[c]{Velocity \\ range} &   $\int T_{\rm mb}\rm  dv$    & \makecell[c]{Velocity \\ range}  &  $\int T_{\rm mb}\rm  dv$   & \makecell[c]{Velocity \\ range}    &   (mK)  &  $\int T_{\rm mb}\rm  dv$    & \makecell[c]{Velocity \\ range}     \\
	& & (K$\cdot$km $\rm s^{-1}$) & (km $\rm s^{-1}$) & (K$\cdot$km $\rm s^{-1}$) & (km $\rm s^{-1}$) & (K$\cdot$km $\rm s^{-1}$) & (km $\rm s^{-1}$) &  &(K$\cdot$km $\rm s^{-1}$) & (km $\rm s^{-1}$) \\ \hline
	\endhead
	\hline
	\endfoot
	\endlastfoot
	G000.67-00.03           & 29 & 1.99$\pm$0.08 & 57.5$-$65.6  & 88.6$\pm$0.2   & -37.7$-$106.6  & 138.8$\pm$0.1  & 30.0$-$104.3  & 178 & 51.0$\pm$1.3  & 39.9$-$82.9  \\
	G005.88-00.39           & 26 & 0.37$\pm$0.07   & 5.2$-$12.6  & 24.93$\pm$0.08   & 4.6$-$13.6   & 77.2$\pm$0.1   & -1.9$-$22.0   & 136 & 98.4$\pm$0.8  & -3.4$-$24.7   \\
	G009.62+00.19           & 19 & $\le$0.072         &       & 37.16$\pm$0.07 & -4.1$-$11.0   & 21.0$\pm$0.1   & -14.9$-$20.0  & 112 & 20.7$\pm$0.5  & -4.7$-$13.7   \\
	G010.47+00.02           & 19 & 1.82$\pm$0.08   & 57.4$-$76.4  & 30.95$\pm$0.07   & 58.8$-$74.8  & 34.0$\pm$0.1   & 48.2$-$83.5  & 122 & 49.4$\pm$0.7  & 53.2$-$80.6  \\
	G010.62-00.38           & 23 & $\le$0.100         &       & 76.1$\pm$0.1   & -12.0$-$7.6  & 36.9$\pm$0.1   & -19.9$-$10.2  & 96 & 43.5$\pm$0.5  & -15.3$-$8.8  \\
	G011.49-01.48           & 25 & $\le$0.086         &       & 5.14$\pm$0.09    & 4.7$-$16.7  & 4.75$\pm$0.06  & 6.3$-$13.1  & 172 & 3.8$\pm$0.4   & 8.7$-$12.3  \\
	G011.91-00.61           & 14 & 1.08$\pm$0.05 & 30.0$-$41.5  & 9.98$\pm$0.04  & 34.0$-$40.6  & 12.76$\pm$0.08 & 15.0$-$50.0  & 37 & 6.6$\pm$0.2   & 23.6$-$43.9  \\
	G012.80-00.20           & 16 & $\le$0.066         &       & 59.27$\pm$0.07   & 29.7$-$46.7  & 28.96$\pm$0.07   & 24.5$-$44.9  & 32 & 20.4$\pm$0.1  & 28.9$-$42.6  \\
	G012.88+00.48           & 14 & $\le$1.12$\pm$0.05 & 29.1$-$40.5  & 28.52$\pm$0.04 & 28.8$-$37.5  & 14.28$\pm$0.05 & 25.8$-$41.2  & 32 & 9.9$\pm$0.1   & 25.5$-$40.0  \\
	G012.90-00.24           & 25 & 1.34$\pm$0.08 & 31.0$-$41.9  & 27.40$\pm$0.08 & 31.2$-$40.3  & 1.88$\pm$0.08  & 31.1$-$42.4  & 83 & 2.4$\pm$0.4   & 28.5$-$44.0  \\
	G012.90-00.26           & 18 & 0.91$\pm$0.06 & 33.2$-$43.0  & 32.03$\pm$0.07 & 32.0$-$45.2  & 14.91$\pm$0.08 & 26.0$-$46.4  & 52 & 10.8$\pm$0.2  & 28.7$-$44.3  \\
	G014.33-00.64           & 27 & 2.8$\pm$0.1   & 16.2$-$31.0  & 10.91$\pm$0.06 & 20.0$-$25.3  & 19.9$\pm$0.1   & 9.8$-$30.0  & 148 & 13.4$\pm$0.5  & 16.4$-$27.0  \\
	G015.03-00.67           & 19 & $\le$0.157         &       & 6.30$\pm$0.05  & 15.1$-$22.9  & 11.00$\pm$0.06 & 13.8$-$23.0  & 57 & 8.9$\pm$0.2   & 15.9$-$23.9  \\
	G016.58-00.05           & 16 & 0.56$\pm$0.05 & 54.7$-$64.3  & 21.96$\pm$0.05 & 54.4$-$64.1  & 8.38$\pm$0.06  & 50.7$-$66.9  & 44 & 4.6$\pm$0.2   & 54.7$-$64.5  \\
	G023.00-00.41           & 14 & 0.74$\pm$0.05 & 71.7$-$83.8  & 25.40$\pm$0.05   & 68.1$-$82.8  & 10.94$\pm$0.08   & 61.9$-$94.1  & 40 & 6.2$\pm$0.2   & 68.4$-$88.7  \\
	G023.44-00.18           & 14 & 0.75$\pm$0.05 & 93.8$-$106.2 & 26.87$\pm$0.04 & 98.0$-$105.5 & 13.52$\pm$0.07 & 87.8$-$115.8 & 32 & 3.8$\pm$0.2   & 92.5$-$113.7 \\
	G027.36-00.16           & 13 & 1.89$\pm$0.05 & 84.4$-$99.0  & 18.64$\pm$0.04 & 86.9$-$97.8  & 17.44$\pm$0.08   & 75.2$-$110.2  & 30 & 12.3$\pm$0.2  & 67.6$-$115.9  \\
	G028.86+00.06           & 12 & $\le$0.123         &       & 22.34$\pm$0.04 & 95.5$-$108.2 & 8.34$\pm$0.05  & 91.6$-$113.5 & 27 & 5.5$\pm$0.1   & 87.7$-$113.6 \\
	G029.95-00.01           & 18 & 0.31$\pm$0.05   & 92.5$-$100.0  & 28.28$\pm$0.07   & 91.3$-$105.7  & 16.55$\pm$0.09   & 81.6$-$110.2  & 103 & 18.8$\pm$0.6  & 84.4$-$109.5  \\
	G031.28+00.06           & 17 & 0.18$\pm$0.04 & 106.9$-$113.3 & 20.67$\pm$0.05 & 105.0$-$114.0 & 13.48$\pm$0.07   & 96.6$-$116.1 & 81 & 6.0$\pm$0.3   & 104.5$-$114.6 \\
	G031.58+00.07           & 15 & $\le$0.137         &       & 18.52$\pm$0.05 & 90.9$-$101.1  & 9.26$\pm$0.05  & 90.7$-$101.5  & 41 & 4.5$\pm$0.2   & 90.7$-$102.1  \\
	G032.04+00.05           & 10 & 0.36$\pm$0.03 & 91.8$-$99.3  & 17.68$\pm$0.03 & 89.0$-$100.9  & 11.12$\pm$0.05 & 86.0$-$109.7  & 39 & 7.1$\pm$0.2   & 88.0$-$103.2  \\
	G034.39+00.22           & 8  & 1.73$\pm$0.03 & 51.9$-$64.7  & 13.19$\pm$0.02 & 52.1$-$62.3  & 5.06$\pm$0.03  & 49.3$-$64.7  & 24 & 1.33$\pm$0.08 & 53.3$-$62.8  \\
	G035.02+00.34           & 8  & $\le$0.065         &       & 12.99$\pm$0.02 & 47.6$-$55.6  & 8.72$\pm$0.03  & 44.8$-$58.2  & 23 & 7.5$\pm$0.1   & 43.8$-$61.2  \\
	G035.19-00.74           & 13 & 0.57$\pm$0.04 & 26.9$-$38.9  & 16.13$\pm$0.04 & 27.9$-$40.3  & 17.43$\pm$0.05 & 25.6$-$42.4  & 55 & 12.1$\pm$0.2  & 25.6$-$41.0  \\
	G035.20-01.73           & 10 & 0.05$\pm$0.02 & 41.0$-$44.4  & 8.66$\pm$0.03  & 40.9$-$48.7  & 4.45$\pm$0.04  & 35.3$-$48.9  & 28 & 1.15$\pm$0.06 & 40.4$-$44.9  \\
	G037.43+01.51           & 9  & 0.18$\pm$0.02 & 40.1$-$46.8  & 11.67$\pm$0.03 & 40.0$-$48.9  & 5.40$\pm$0.03  & 39.2$-$47.6  & 40 & 2.6$\pm$0.1   & 40.6$-$47.7  \\
	G043.16+00.01           & 14 & 0.42$\pm$0.06  & 7.1$-$24.6  & 45.82$\pm$0.08   & -9.2$-$22.3   & 22.52$\pm$0.08   & -8.1$-$26.4   & 44 & 26.9$\pm$0.3  & -5.2$-$23.6   \\
	G043.79-00.12           & 10 & $\le$0.129         &       & 17.95$\pm$0.04 & 33.6$-$52.8  & 7.87$\pm$0.04  & 36.9$-$53.5  & 36 & 5.1$\pm$0.2   & 35.5$-$53.0  \\
	G049.48-00.36           & 15 & 0.55$\pm$0.05   & 54.0$-$66.0  & 29.19$\pm$0.06   & 54.1$-$73.1  & 27.47$\pm$0.07   & 46.8$-$72.1  & 44 & 36.4$\pm$0.3  & 45.1$-$72.8  \\
	G049.48-00.38           & 12 & 2.31$\pm$0.06  & 47.5$-$65.0  & 51.09$\pm$0.06   & 49.3$-$64.9  & 60.77$\pm$0.08   & 40.2$-$75.0  & 51 & 75.6$\pm$0.4  & 26.6$-$82.9  \\
	G059.78+00.06           & 11 & 0.75$\pm$0.05 & 14.0$-$28.9  & 5.95$\pm$0.04  & 17.7$-$28.1  & 5.77$\pm$0.03  & 18.3$-$25.8  & 32 & 3.0$\pm$0.1   & 16.4$-$28.0  \\
	G069.54-00.97           & 10 & 0.55$\pm$0.04 & 4.7$-$16.2  & 14.54$\pm$0.03 & 3.1$-$11.4  & 9.90$\pm$0.04  & 4.3$-$16.9  & 25 & 6.5$\pm$0.1   & 2.1$-$17.5  \\
	G075.76+00.33           & 8  & $\le$0.096         &       & 7.31$\pm$0.03  & -9.7$-$5.8  & 9.69$\pm$0.03  & -9.9$-$6.1  & 19 & 4.35$\pm$0.07 & -7.8$-$3.4  \\
	G078.12+03.63           & 16 & 2.16$\pm$0.05 & -9.4$-$1.6  & 5.05$\pm$0.06  & -8.4$-$3.7  & 8.99$\pm$0.06  & -10.4$-$2.9  & 45 & 7.6$\pm$0.2   & -13.0$-$3.5  \\
	G081.75+00.59           & 9  & 4.10$\pm$0.03   & -9.7$-$2.3  & 13.91$\pm$0.02   & -9.0$-$-1.5  & 9.67$\pm$0.03    & -8.1$-$2.3  & 27 & 3.31$\pm$0.08 & -6.6$-$0.4  \\
	G081.87+00.78           & 11 & 0.31$\pm$0.03   & 4.5$-$11.9   & 20.43$\pm$0.04   & 3.2$-$16.1   & 16.82$\pm$0.04   & 2.2$-$13.2   & 48 & 18.4$\pm$0.2  & 0.1$-$18.7   \\
	G092.67+03.07           & 12 & 0.17$\pm$0.03 & -9.1$-$-0.9  & 5.26$\pm$0.03  & -11.1$-$-2.9  & 8.81$\pm$0.03  & -9.0$-$-0.9  & 36 & 6.3$\pm$0.2   & -12.5$-$4.3  \\
	G109.87+02.11           & 10 & 0.08$\pm$0.02   & -12.8$-$-8.0 & 18.81$\pm$0.04 & -17.7$-$-4.5 & 10.75$\pm$0.04 & -17.7$-$-3.3 & 52 & 10.3$\pm$0.2  & -18.1$-$-2.3 \\
	G111.54+00.77           & 6 & 0.20$\pm$0.02 & -64.5$-$-51.1 & 11.93$\pm$0.03 & -67.2$-$-46.5 & 5.59$\pm$0.02  & -64.2$-$-52.6 & 13 & 4.10$\pm$0.07 & -65.2$-$-41.3 \\
	G121.29+00.65           & 6 & 0.91$\pm$0.02 & -23.5$-$-11.6 & 6.92$\pm$0.02  & -24.0$-$-12.5 & 6.66$\pm$0.02  & -21.5$-$-13.7 & 16 & 2.15$\pm$0.05 & -22.4$-$-13.7 \\
	G123.06-06.30           & 8  & 0.70$\pm$0.02 & -34.7$-$-24.0 & 5.92$\pm$0.03  & -37.6$-$-22.8 & 7.68$\pm$0.03  & -37.5$-$-25.6 & 17 & 3.21$\pm$0.06 & -37.1$-$-25.2 \\
	G133.94+01.06           & 13 & 0.22$\pm$0.04   & -52.3$-$-41.6 & 14.93$\pm$0.05   & -53.6$-$-39.7 & 12.54$\pm$0.05   & -56.0$-$-40.4 & 31 & 11.7$\pm$0.1  & -54.9$-$-40.2 \\
	G168.06+00.82           & 17 & $\le$0.135    &               & 2.61$\pm$0.04    & -28.6$-$-21.5 & 0.30$\pm$0.04    & -27.5$-$-22.2 & 112 & $\le$0.85    &       \\
	G176.51+00.20           & 6  & 0.39$\pm$0.02 & -22.7$-$-12.4 & 4.42$\pm$0.02  & -24.1$-$-12.7 & 2.42$\pm$0.01  & -20.6$-$-14.8 & 20 & 0.89$\pm$0.05 & -20.2$-$-16.1 \\
	G183.72-03.66           & 5  & 0.49$\pm$0.02 & -2.9$-$8.8   & 2.07$\pm$0.02  & -4.5$-$8.6   & 2.95$\pm$0.01  & -1.0$-$5.7   & 17 & 0.93$\pm$0.04 & -0.3$-$5.4   \\
	G188.94+00.88           & 9 & 0.14$\pm$0.02 & -1.0$-$6.7   & 4.65$\pm$0.03  & -3.0$-$9.2   & 3.90$\pm$0.03  & -3.5$-$7.2   & 24 & 1.09$\pm$0.07 & 0.3$-$6.1   \\
	G192.60-00.04           & 9  & 0.13$\pm$0.03   & -0.9$-$11.5   & 7.76$\pm$0.03    & 2.4$-$13.6   & 5.09$\pm$0.03    & 3.0$-$11.5   & 27 & 2.17$\pm$0.09 & 2.2$-$11.5   \\
	G209.00-19.38           & 6  & $\le$0.057         &       & 3.25$\pm$0.02  & 3.3$-$13.9   & 2.70$\pm$0.02  & 3.7$-$11.9   & 16 & 2.08$\pm$0.05 & 4.1$-$12.0   \\
	G232.62+00.99           & 8  & $\le$0.089         &       & 5.13$\pm$0.03    & 8.0$-$23.4  & 3.24$\pm$0.02  & 12.6$-$21.0  & 35 & 2.27$\pm$0.10 & 13.4$-$19.4   \\ \hline
\end{longtable}
\normalsize

\begin{longtable}[c]{ccccccc}
	\caption{Colunm density and deuterium fractionation}
	\label{Colunm density}\\
	\hline
	\hline
	\multirow{2}{*}{source} & $N$(NH$_2$D)       & $N$(HC$_3$N) & $N$(C$^{18}$O)    & $N$(NH$_3$) $^*$       & $N$(H$_2$)         & $N$(NH$_2$D)/$N$(NH$_3$)$^**$ \\
	& $10^{12}$cm$^{-2}$ & $10^{12}$cm$^{-2}$  & $10^{15}$cm$^{-2}$ & $10^{14}$cm$^{-2}$ & $10^{21}$cm$^{-2}$ &                          \\ \hline
	\endfirsthead
	\multicolumn{7}{c}%
	{{\bfseries Table \thetable\ continued from previous page}} \\
	\hline
	\hline
	\multirow{2}{*}{source} & $N$(NH$_2$D)       & $N$(HC$_3$N) & $N$(C$^{18}$O)    & $N$(NH$_3$)$^*$        & $N$(H$_2$)         & $N$(NH$_2$D)/$N$(NH$_3$)$^**$ \\
	& $10^{12}$cm$^{-2}$ & $10^{12}$cm$^{-2}$  & $10^{15}$cm$^{-2}$ & $10^{14}$cm$^{-2}$ & $10^{21}$cm$^{-2}$ &                          \\ \hline
	\endhead
	\hline
	\endfoot
	\endlastfoot
	G009.62+00.19 & $\le$6.9  & 101.8$ \pm 0.5$ & 45.75$ \pm 0.09$ & 54.82$ \pm 26.48$  & 200.0$ \pm 0.4$ & $\le$0.126 \\
	G012.88+00.48 & $\le$36.0  & 69.3$ \pm 0.3$ & 35.11$ \pm 0.05$ & 72.48$ \pm 7.57$  & 153.3$ \pm 0.2$ & $\le$0.496 \\
	G012.90-00.24 & 43.0$ \pm 2.6$  & 9.1$ \pm 0.4$ & 33.73$ \pm 0.09$ & 73.48$ \pm 5.66$  & 147.3$ \pm 0.4$ & 0.59$ \pm 0.06$ \\
	G014.33-00.64 & 89.8$ \pm 3.3$  & 96.6$ \pm 0.6$ & 13.43$ \pm 0.08$ & 38.35$ \pm 6.64$  & 58.6$ \pm 0.3$ & 2.3$ \pm 0.4$ \\
	G015.03-00.67 & $\le$5.0  & 53.4$ \pm 0.3$ & 7.76$ \pm 0.06$ & 56.02$ \pm 27.20$  & 33.9$ \pm 0.3$ & $\le$0.090 \\
	G016.58-00.05 & 18.0$ \pm 1.5$  & 40.7$ \pm 0.3$ & 27.04$ \pm 0.06$ & 47.29$ \pm 6.08$  & 118.1$ \pm 0.3$ & 0.38$ \pm 0.06$ \\
	G023.00-00.41 & 23.8$ \pm 1.6$  & 53.1$ \pm 0.4$ & 31.27$ \pm 0.07$ & 93.50$ \pm 5.02$  & 136.5$ \pm 0.3$ & 0.25$ \pm 0.02$ \\
	G023.44-00.18 & 23.9$ \pm 1.5$  & 65.6$ \pm 0.3$ & 33.08$ \pm 0.05$ & 103.65$ \pm 8.85$ & 144.4$ \pm 0.2$ & 0.23$ \pm 0.02$ \\
	G029.95-00.01 & 9.9$ \pm 1.5$  & 80.3$ \pm 0.5$ & 34.81$ \pm 0.08$ & 57.78$ \pm 8.93$  & 152.0$ \pm 0.4$ & 0.17$ \pm 0.04$ \\
	G031.28+00.06 & 5.7$ \pm 1.3$  & 65.4$ \pm 0.4$ & 25.44$ \pm 0.06$ & 89.03$ \pm 10.21$  & 111.1$ \pm 0.3$ & 0.06$ \pm 0.02$ \\
	G035.19-00.74 & 18.3$ \pm 1.4$  & 84.6$ \pm 0.2$ & 19.86$ \pm 0.05$ & 45.05$ \pm 3.37$  & 86.7$ \pm 0.2$ & 0.41$ \pm 0.04$ \\
	G043.79-00.12 & $\le$4.1  & 38.2$ \pm 0.2$ & 22.09$ \pm 0.05$ & 22.19$ \pm 23.00$  & 96.5$ \pm 0.2$ & $\le$0.185 \\
	G049.48-00.36 & 17.7$ \pm 1.6$  & 133.3$ \pm 0.4$ & 35.93$ \pm 0.09$ & 226.65$ \pm 20.31$ & 156.9$ \pm 0.3$ & 0.078$ \pm 0.010$ \\
	G049.48-00.38 & 74.1$ \pm 1.9$  & 294.9$ \pm 0.4$ & 62.89$ \pm 0.07$ & 92.43$ \pm 10.02$  & 274.6$ \pm 0.3$ & 0.80$ \pm 0.09$ \\
	G069.54-00.97 & 17.7$ \pm 1.1$  & 48.0$ \pm 0.2$ & 17.90$ \pm 0.04$ & 44.04$ \pm 20.07$  & 78.2$ \pm 0.2$ & 0.4$ \pm 0.2$ \\
	G078.12+03.63 & 69.3$ \pm 1.7$  & 43.6$ \pm 0.3$ & 6.21$ \pm 0.07$ & 43.15$ \pm 3.87$  & 27.1$ \pm 0.3$ & 1.6$ \pm 0.1$ \\
	G081.75+00.59 & 131.4$ \pm 1.0$ & 46.9$ \pm 0.1$ & 17.13$ \pm 0.03$ & 30.36$ \pm 1.56$  & 74.8$ \pm 0.1$ & 4.3$ \pm 0.2$ \\
	G109.87+02.11 & 2.7$ \pm 0.7$  & 52.1$ \pm 0.2$ & 23.16$ \pm 0.05$ & 18.84$ \pm 6.43$  & 101.1$ \pm 0.2$ & 0.14$ \pm 0.06$ \\
	G111.54+00.77 & 6.5$ \pm 0.7$  & 27.1$ \pm 0.1$ & 14.68$ \pm 0.03$ & 4.32$ \pm 1.39$   & 64.1$ \pm 0.1$ & 1.5$ \pm 0.5$ \\  \hline
	G000.67-00.03 & 63.6$ \pm 2.6$  & 673.4$ \pm 1.2$ & 109.1$ \pm 0.3$ &    & 476.3$ \pm 1.3$ &  \\
	G005.88-00.39 & 11.8$ \pm 2.2$  & 374.5$ \pm 0.6$ & 30.69$ \pm 0.09$ &    & 134.0$ \pm 0.4$ &  \\
	G010.47+00.02 & 58.2$ \pm 2.6$  & 165.2$ \pm 0.5$ & 38.10$ \pm 0.09$ &    & 166.4$ \pm 0.4$ &  \\
	G010.62-00.38 & $\le$9.6  & 179.2$ \pm 0.6$ & 93.7$ \pm 0.1$ &    & 409.1$ \pm 0.5$ &  \\
	G011.49-01.48 & $\le$8.2  & 23.0$ \pm 0.3$ & 6.3$ \pm 0.1$ &   & 27.6$ \pm 0.5$ &  \\
	G011.91-00.61 & 34.4$ \pm 1.5$  & 61.9$ \pm 0.4$ & 12.29$ \pm 0.05$ &    & 54.7$ \pm 0.2$ &  \\
	G012.80-00.20 & $\le$6.3  & 140.5$ \pm 0.4$ & 72.97$ \pm 0.08$ &    & 318.6$ \pm 0.4$ &  \\
	G012.90-00.26 & 29.3$ \pm 1.8$  & 72.4$ \pm 0.4$ & 39.43$ \pm 0.08$ &    & 172.2$ \pm 0.4$ &  \\
	G027.36-00.16 & 60.5$ \pm 1.6$  & 84.6$ \pm 0.4$ & 22.95$ \pm 0.05$ &    & 100.2$ \pm 0.2$ &  \\
	G028.86+00.06 & $\le$4.0  & 40.5$ \pm 0.3$ & 27.50$ \pm 0.05$ &    & 120.1$ \pm 0.2$ &  \\
	G031.58+00.07 & $\le$4.4  & 44.9$ \pm 0.2$ & 22.80$ \pm 0.06$ &    & 100.0$ \pm 0.2$ &  \\
	G032.04+00.05 & 11.5$ \pm 0.9$  & 53.9$ \pm 0.2$ & 21.77$ \pm 0.04$ &    & 95.1$ \pm 0.2$ &  \\
	G034.39+00.22 & 55.5$ \pm 0.7$  & 24.5$ \pm 0.1$ & 16.23$ \pm 0.03$ &    & 70.9$ \pm 0.1$ &  \\
	G035.02+00.34 & $\le$2.1  & 42.3$ \pm 0.1$ & 15.99$ \pm 0.03$ &    & 69.8$ \pm 0.1$ &  \\
	G035.20-01.73 & 1.8$ \pm 0.6$  & 21.6$ \pm 0.2$ & 10.66$ \pm 0.03$ &    & 46.5$ \pm 0.1$ &  \\
	G037.43+01.51 & 5.7$ \pm 0.7$  & 26.2$ \pm 0.1$ & 14.37$ \pm 0.03$ &    & 62.7$ \pm 0.1$ &  \\
	G043.16+00.01 & 13.4$ \pm 1.8$  & 109.3$ \pm 0.4$ & 56.41$ \pm 0.09$ &    & 246.3$ \pm 0.4$ &  \\
	G059.78+00.06 & 23.9$ \pm 1.5$  & 28.0$ \pm 0.2$ & 7.28$ \pm 0.05$ &    & 31.8$ \pm 0.2$ &  \\
	G075.76+00.33 & $\le$3.1  & 47.0$ \pm 0.2$ & 9.00$ \pm 0.04$ &    & 39.3$ \pm 0.2$ &  \\
	G081.87+00.78 & 9.8$ \pm 1.0$  & 81.6$ \pm 0.2$ & 25.15$ \pm 0.05$ &    & 109.8$ \pm 0.2$ &  \\
	G092.67+03.07 & 5.3$ \pm 1.1$  & 42.7$ \pm 0.2$ & 6.47$ \pm 0.04$ &    & 28.3$ \pm 0.2$ &  \\
	G121.29+00.65 & 29.2$ \pm 0.7$  & 32.30$ \pm 0.08$ & 8.52$ \pm 0.03$ &    & 37.2$ \pm 0.1$ &  \\
	G123.06-06.30 & 22.4$ \pm 0.8$  & 37.2$ \pm 0.1$ & 7.29$ \pm 0.04$ &    & 31.8$ \pm 0.2$ &  \\
	G133.94+01.06 & 6.9$ \pm 1.3$  & 60.8$ \pm 0.2$ & 18.37$ \pm 0.06$ &    & 80.2$ \pm 0.3$ &  \\
	G168.06+00.82 & $\le$4.3  & 1.5$ \pm 0.2$ & 3.21$ \pm 0.06$ &    & 14.0$ \pm 0.2$ &  \\
	G176.51+00.20 & 12.4$ \pm 0.6$  & 11.72$ \pm 0.07$ & 5.44$ \pm 0.02$ &    & 23.8$ \pm 0.1$ &  \\
	G183.72-03.66 & 15.6$ \pm 0.6$  & 14.30$ \pm 0.07$ & 2.55$ \pm 0.02$ &    & 11.1$ \pm 0.1$ &  \\
	G188.94+00.88 & 4.6$ \pm 0.8$  & 18.9$ \pm 0.1$ & 5.72$ \pm 0.04$ &    & 25.0$ \pm 0.2$ &  \\
	G192.60-00.04 & 4.1$ \pm 0.9$  & 24.7$ \pm 0.1$ & 9.55$ \pm 0.04$ &    & 41.7$ \pm 0.2$ &  \\
	G209.00-19.38 & $\le$1.8  & 13.11$ \pm 0.08$ & 40.0$ \pm 0.02$ &    & 17.5$ \pm 0.1$ &  \\
	G232.62+00.99 & $\le$2.8  & 15.7$ \pm 0.1$ & 6.32$ \pm 0.04$ &    & 27.6$ \pm 0.2$ &  \\ \hline
\end{longtable}
\vspace{-1.5em}
\begin{minipage}{\linewidth}
	\renewcommand{\footnoterule}{}
	\footnotetext{$^{*}$ NH$_3$ data are taken from \cite{2016AJ....152...92L}}
	\footnotetext{$^{**}$ The $N$(NH$_2$D)/$N$(NH$_3$) ratio in percentage.}
\end{minipage}

\newpage

\begin{figure}
	\centering
	\includegraphics[width=0.5\textwidth]{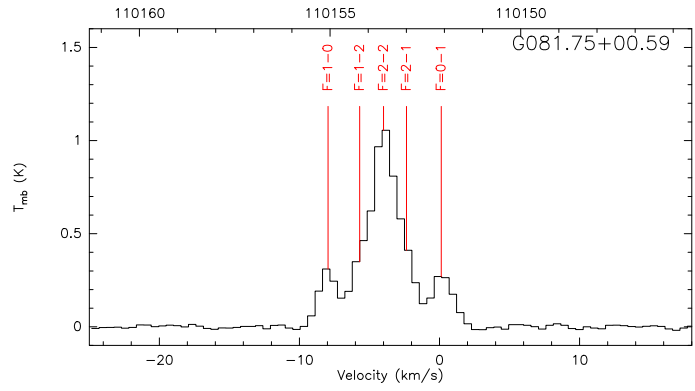}
	\caption{The hyperfine structure of NH$_2$D in G081.75+00.59. It can be seen clearly that F=2-1, F=2-2 and F=1-2 are mingled.}
	\label{HF}
\end{figure}

\begin{figure}
	\centering
	\includegraphics[width=0.5\textwidth]{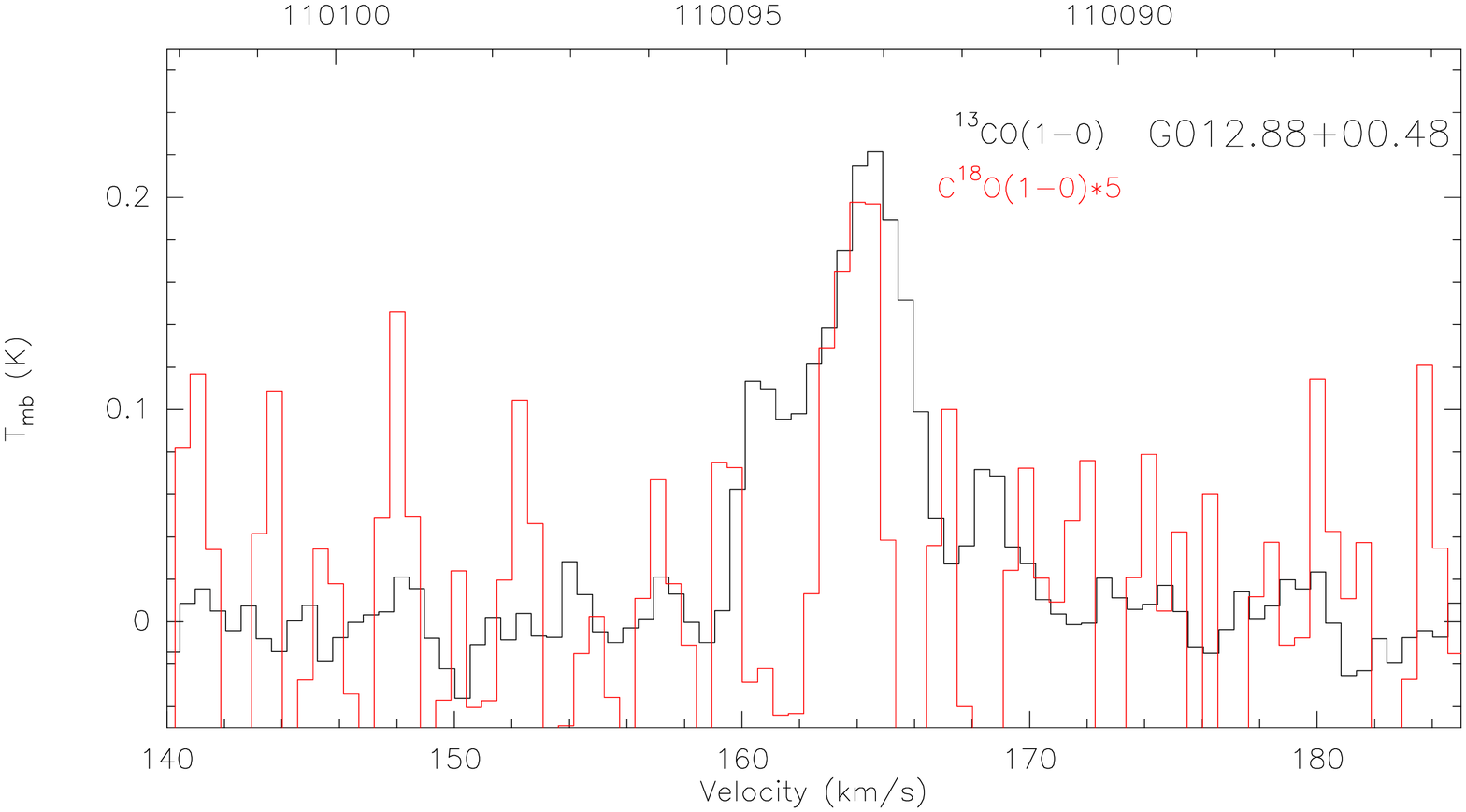}\includegraphics[width=0.5\textwidth]{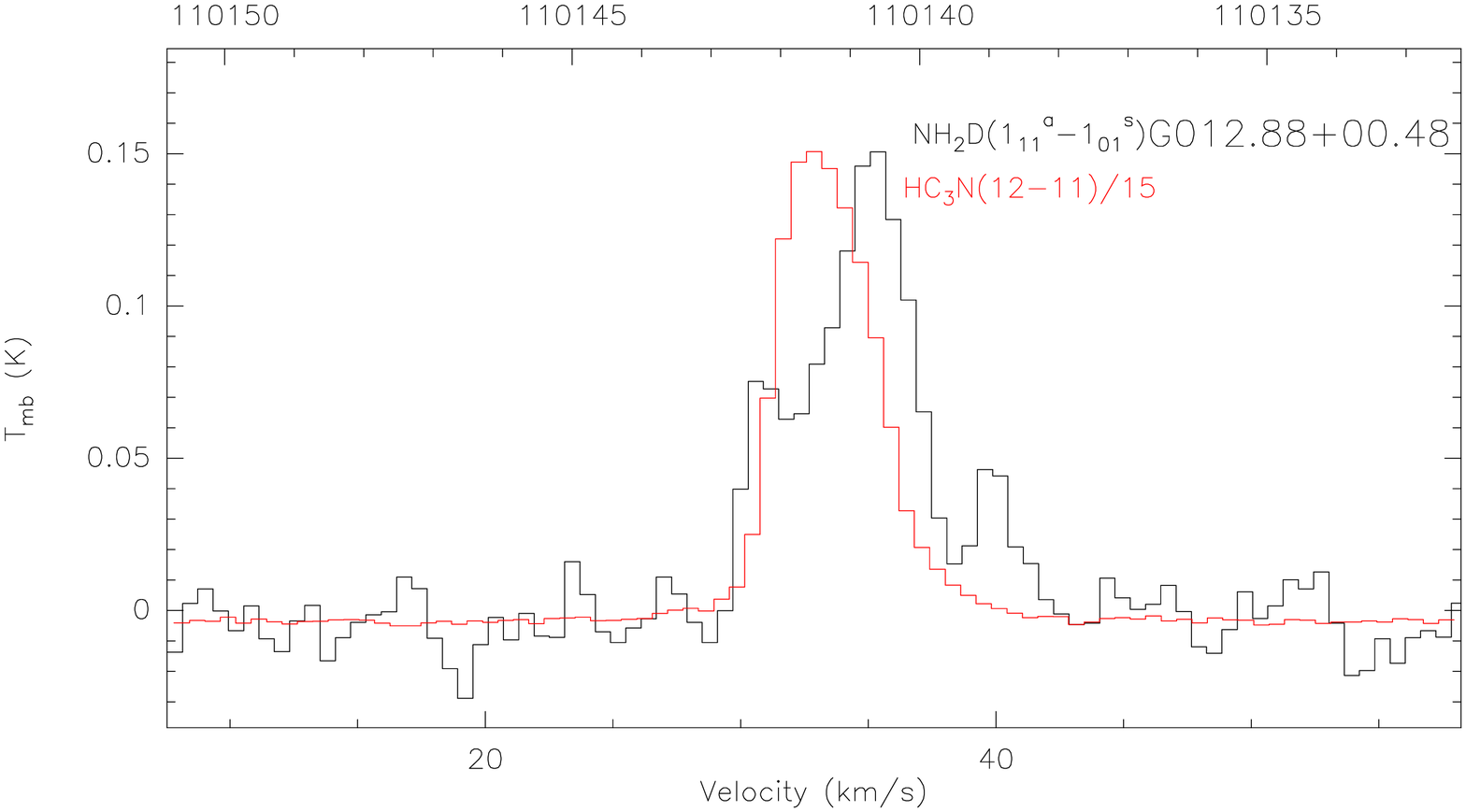}
	\caption{G012.88+00.48 is obviously affected by $^{13}$CO in the direction of sight. The left figure is $^{13}$CO and C$^{18}$O, which NH$_2$D is identified $^{13}$CO. The $^{13}$CO and C$^{18}$O are coincide. Beside, the C$^{18}$O is 5 times weaker than $^{13}$CO. The right figure is NH$_2$D and HC$_3$N(12-11). These two lines are clearly staggered. Thus, it is possible that this NH$_2$D is polluted by $^{13}$CO}.
	\label{pile}
\end{figure}

\begin{figure}
	\centering
	\includegraphics[width=0.5\textwidth]{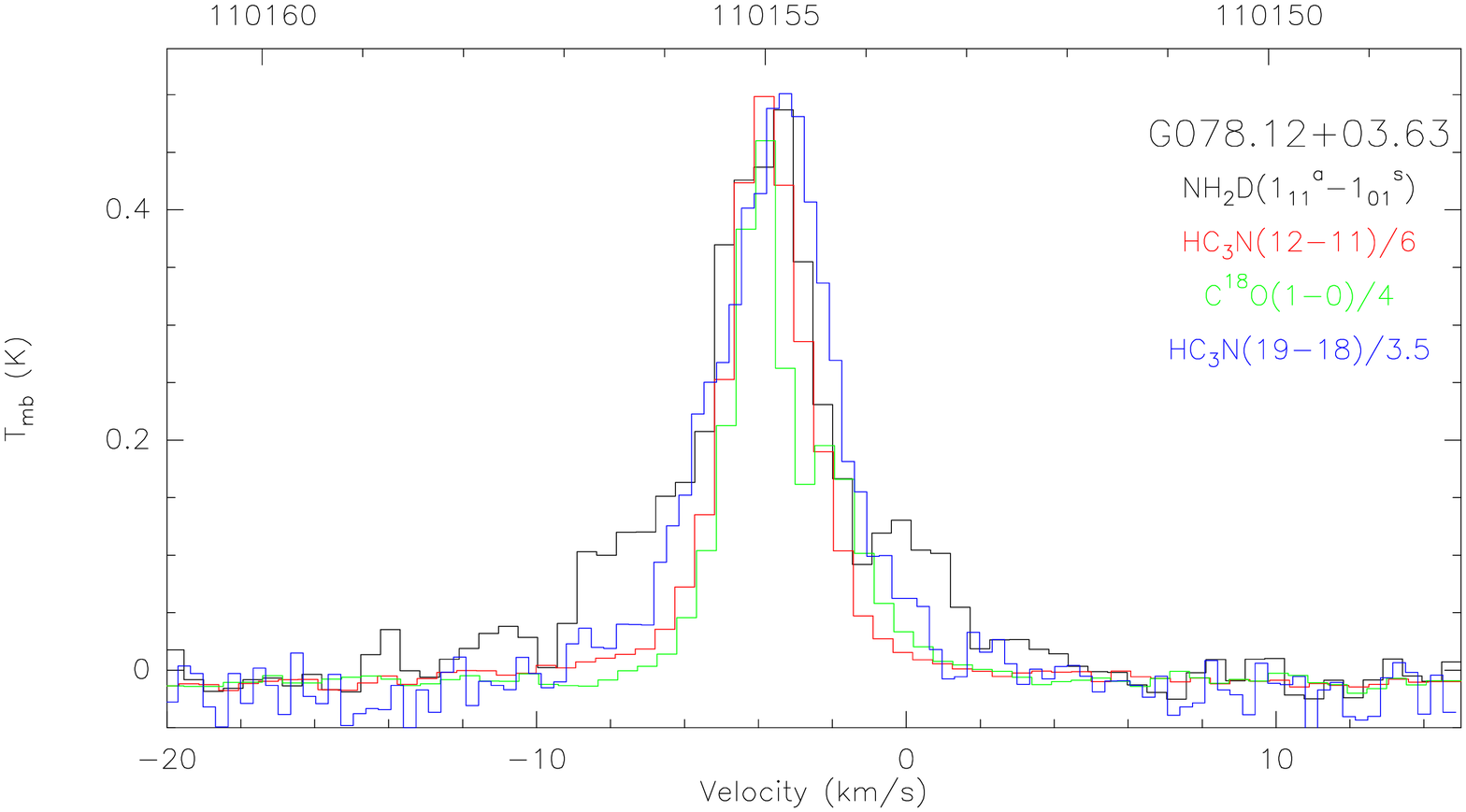}\includegraphics[width=0.5\textwidth]{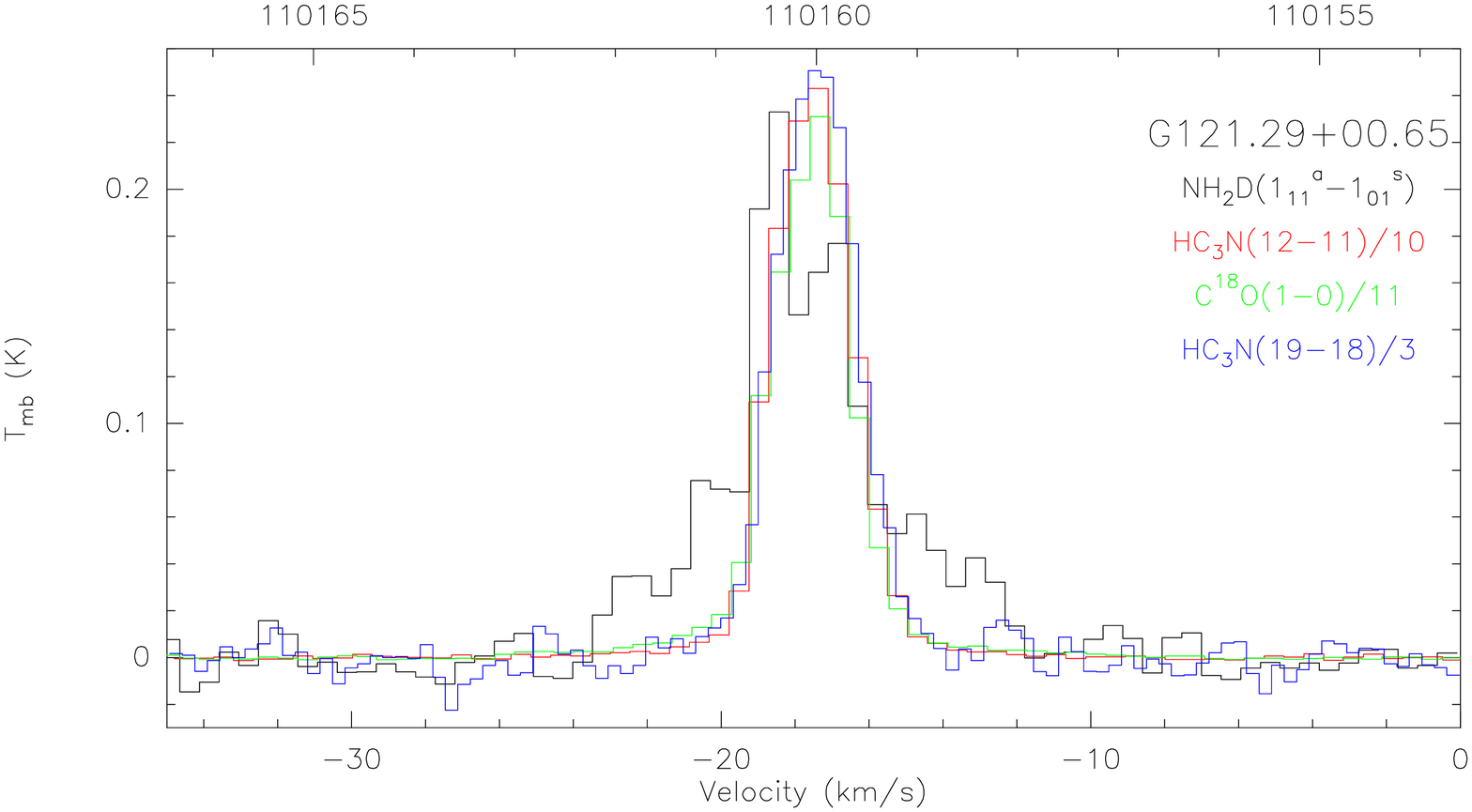}
	\caption{The four lines are the overlay of  G078.12+03.63 and G121.29+00.65. }
	\label{overlay}
\end{figure}

\begin{figure}
	\centering
	\includegraphics[width=0.5\textwidth]{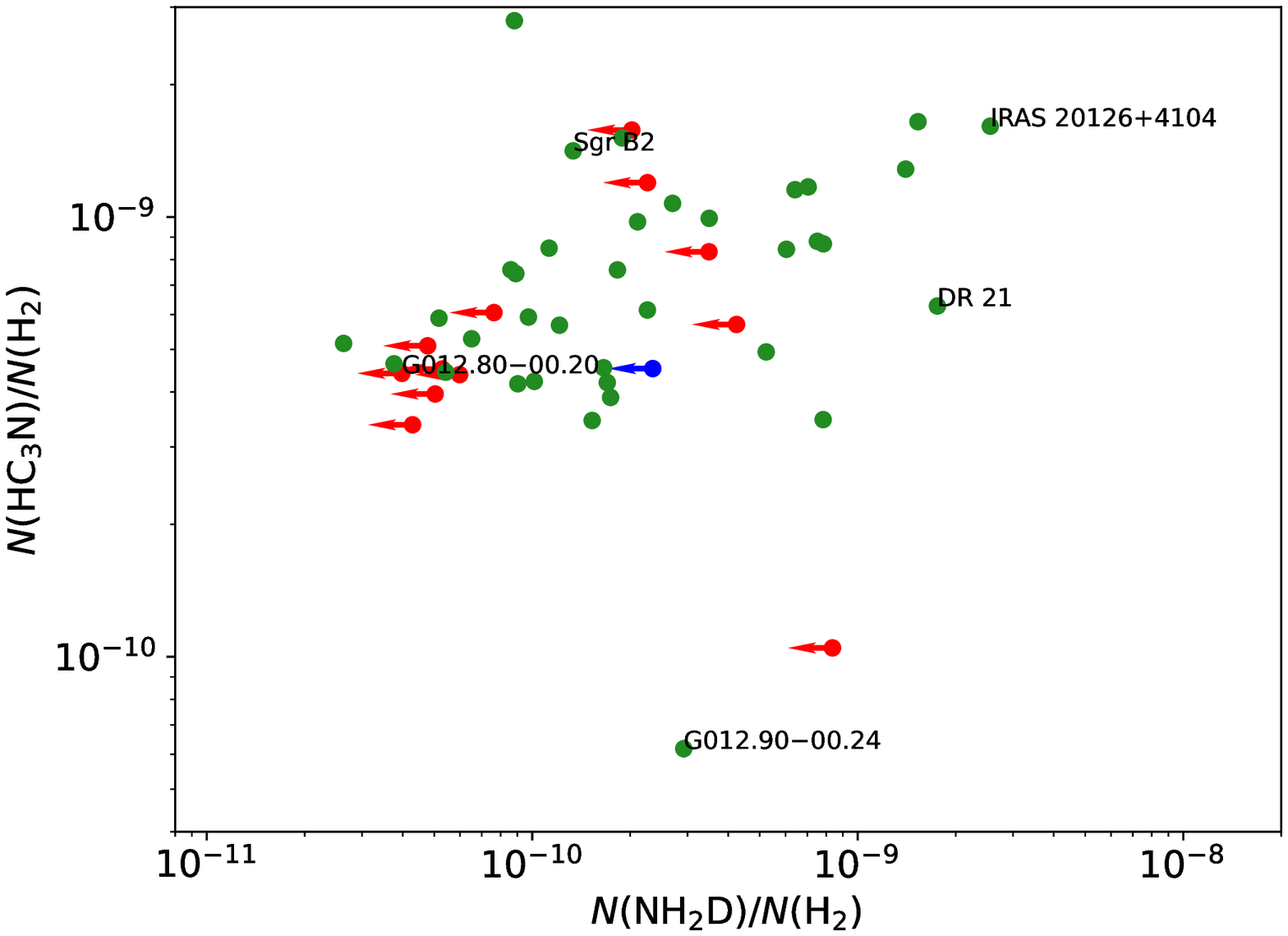}\includegraphics[width=0.5\textwidth]{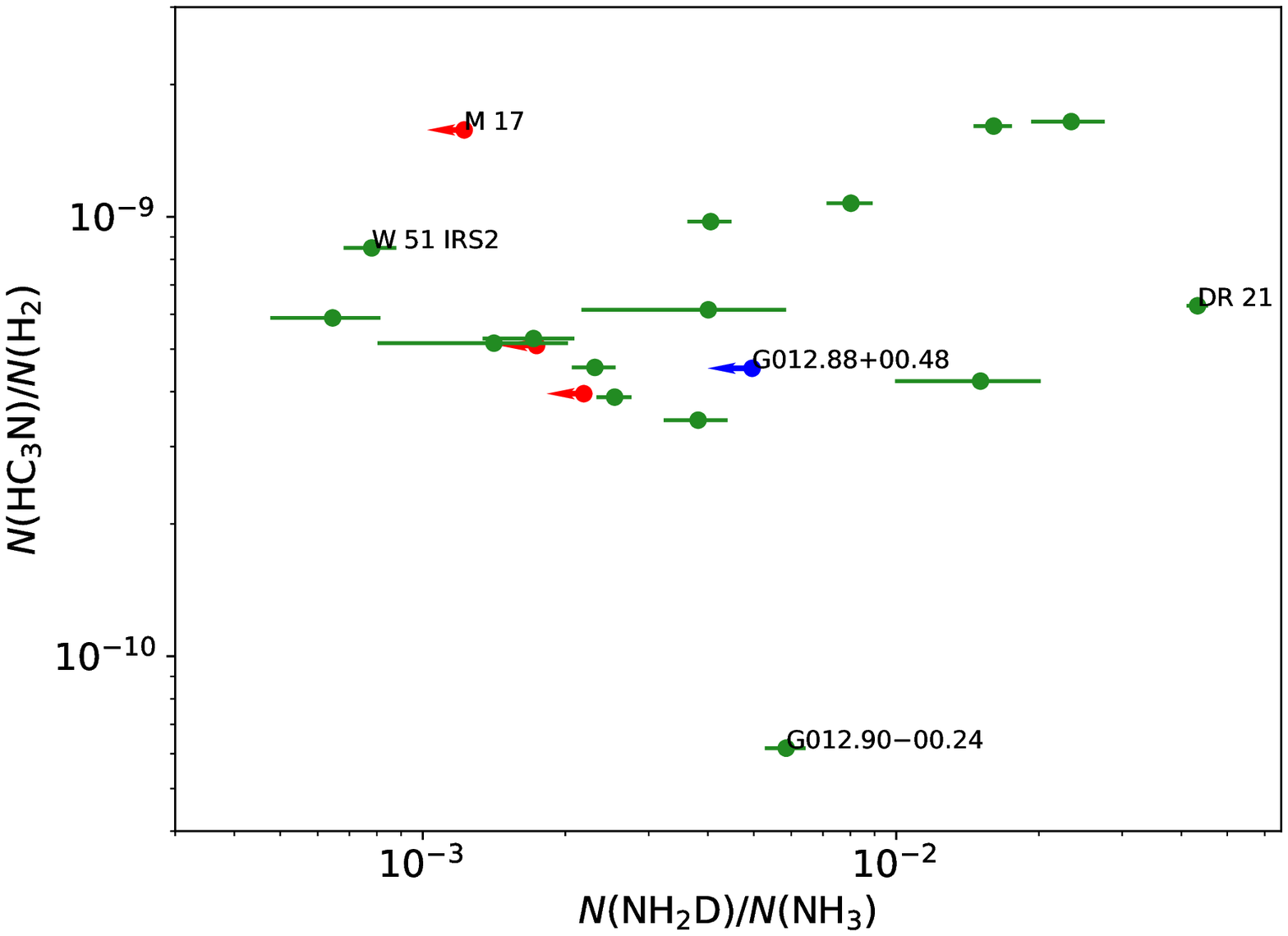}
	\caption{The relationship between deuterated gas and dense gas. The left figure shows the relative abundance of NH$_2$D and HC$_3$N, where $N$(H$_2$) is derived by C$^{18}$O. The right figure is $D_{\rm frac}$(NH$_3$) and the relative abundance of HC$_3$N. The green marks denote the detected targets, red marks denote the targets given by 3$\sigma$ upper limit, the blue mark denotes G012.88+00.48. As it was affected by CO, only the upper limit was given.}
	\label{HC3N}
\end{figure}

\begin{figure}
	\centering
	\includegraphics[width=0.5\textwidth]{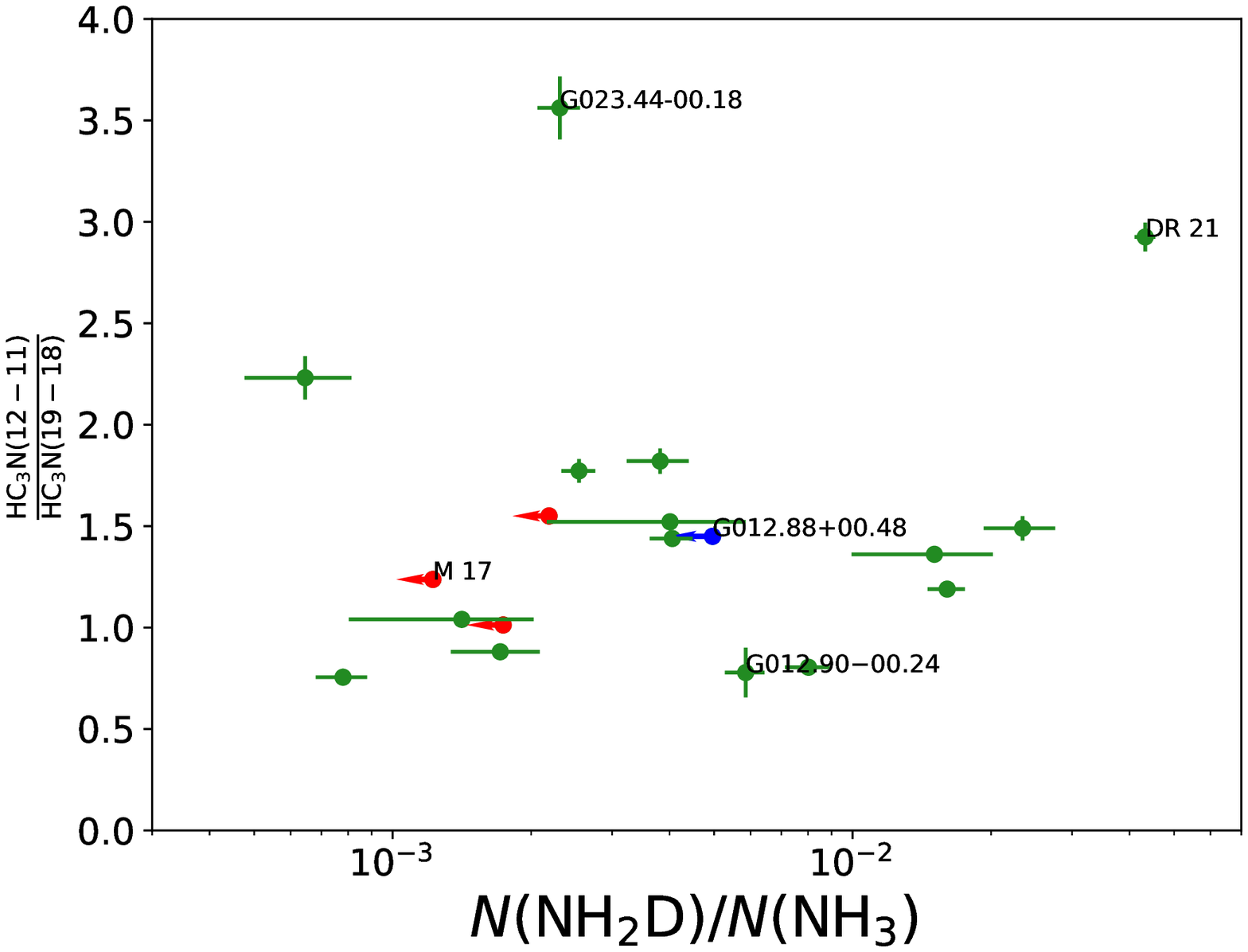}\includegraphics[width=0.5\textwidth]{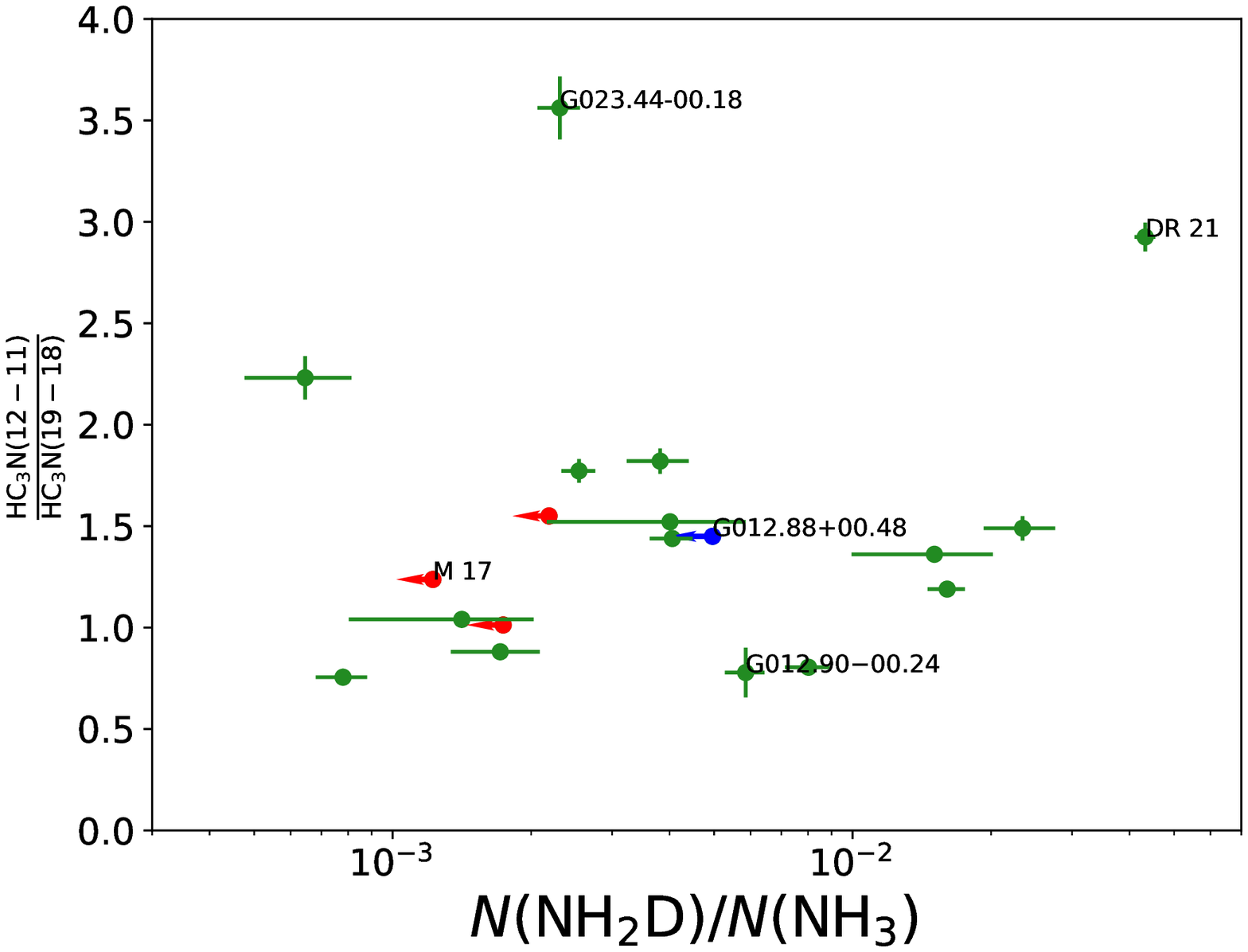}
	\caption{The relationship between deuterated gas and excitation condition. The left figure is the relative abundance of NH$_2$D and excitation condition traced by the ratio of integral flux HC$_3$N(12-11) and HC$_3$N(19-18). The $N$(H$_2$) is derived by C$^{18}$O. The right figure shows $D_{\rm frac}$(NH$_3$) and the excitation condition. The green marks denote the detected targets, red marks denote the targets given by 3$\sigma$ upper limit, and the blue mark denotes G012.88+00.48. As it was affected by CO, only the upper limit was given.}
	\label{HC3Nratio}
\end{figure}

\begin{figure}
	\centering
	\includegraphics[width=0.5\textwidth]{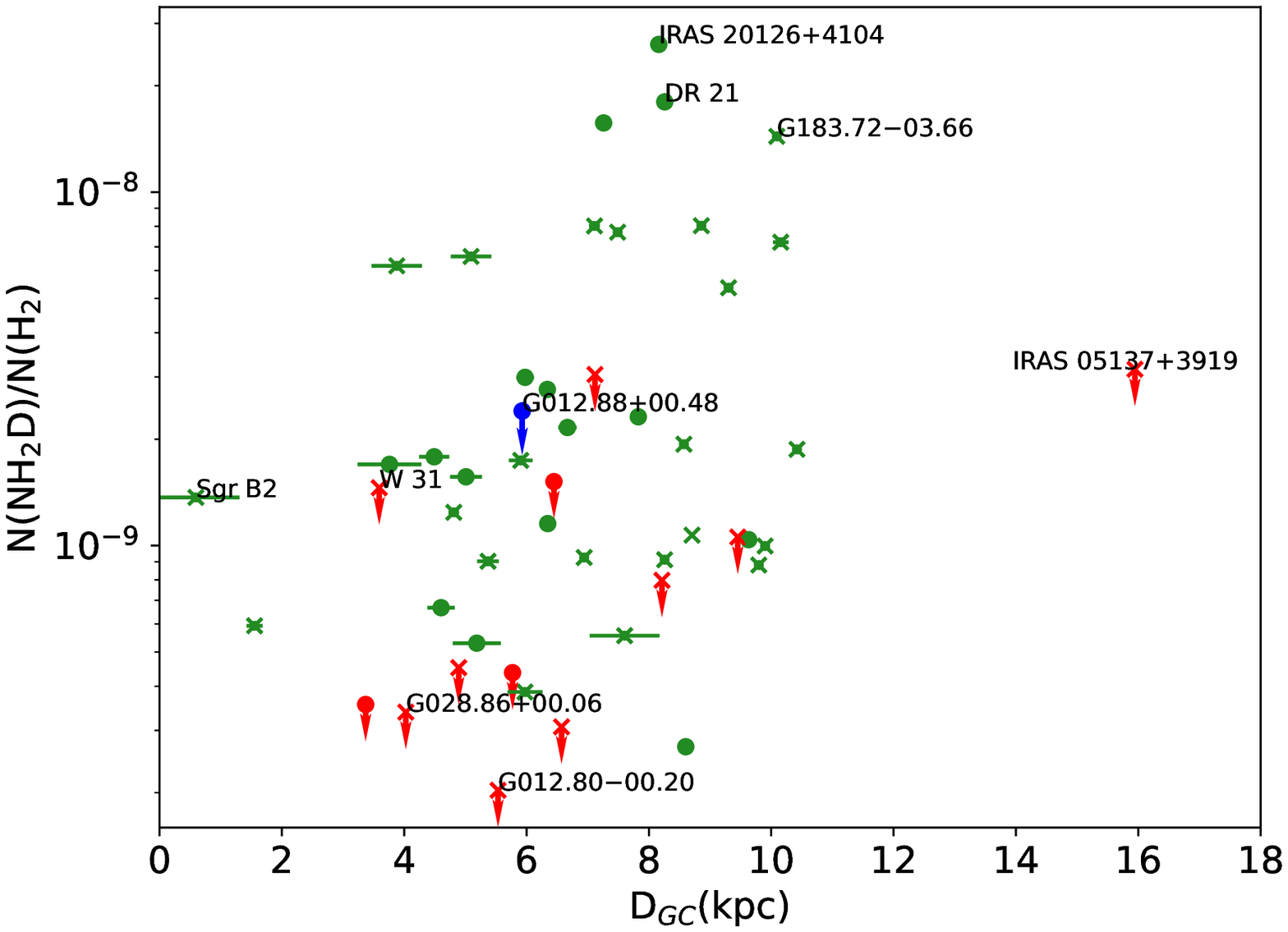}\includegraphics[width=0.5\textwidth]{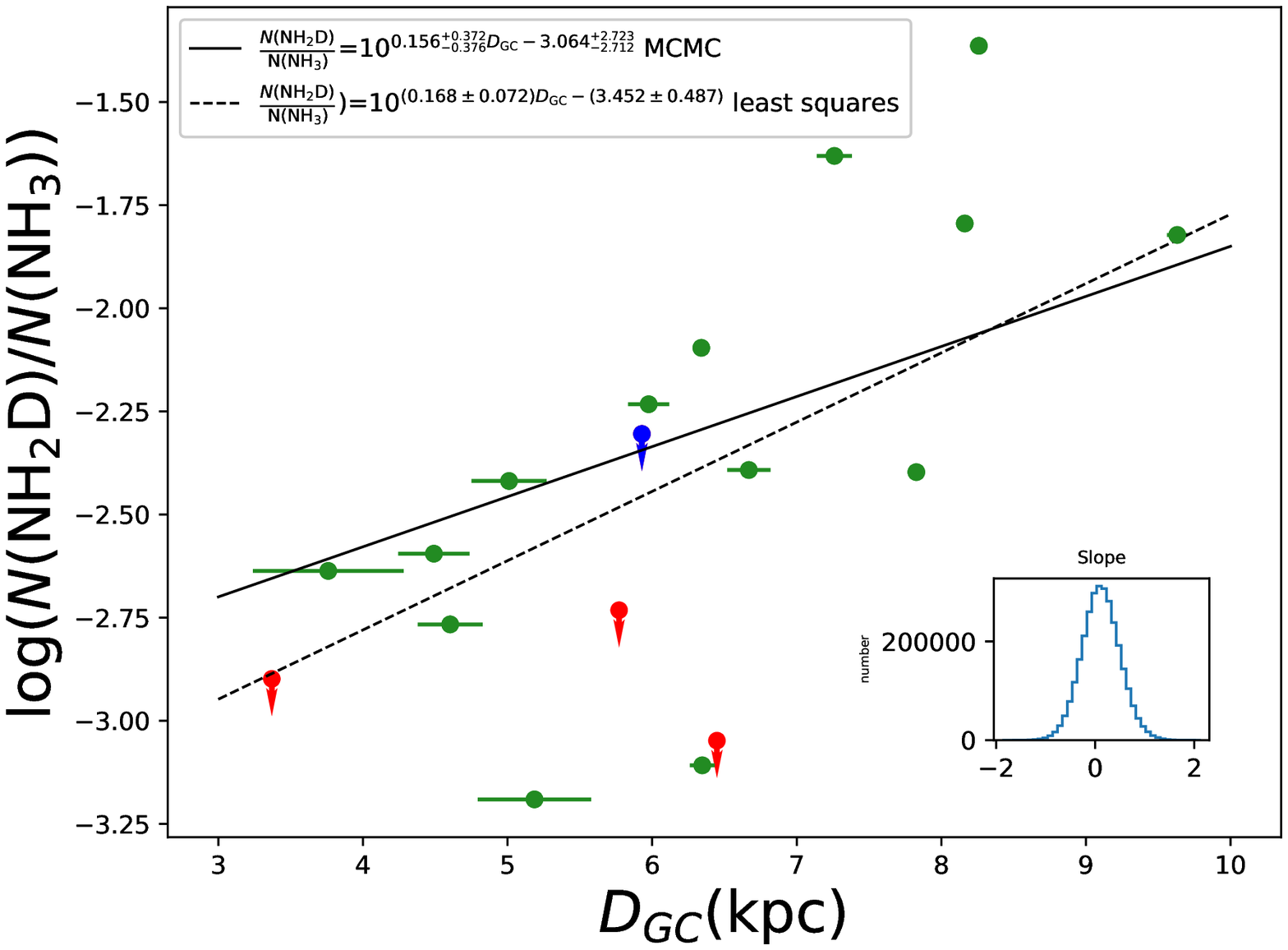}
	\caption{The relationship between deuterated gas and $D_{\rm GC}$. The left panel shows the relative abundance of NH$_2$D and $D_{\rm GC}$. The right figure shows $D_{\rm frac}$(NH$_3$) and the $D_{\rm GC}$. In the left panel, the circles denote sources with reference observation data and the fork marks those that do not have reference observation data. The green marks denote the detected targets, red marks denote the targets given by 3$\sigma$ upper limit,the blue mark denotes G012.88+00.48. Because NH$_2$D is effected by CO in G012.88+00.48 that also given the upper limit.  In the right figure, the solid line is fitted by MCMC method and the dash line is uses the least-squares method.}
	\label{DGC}
\end{figure}

\begin{figure}
	\centering
	\includegraphics[width=0.5\textwidth]{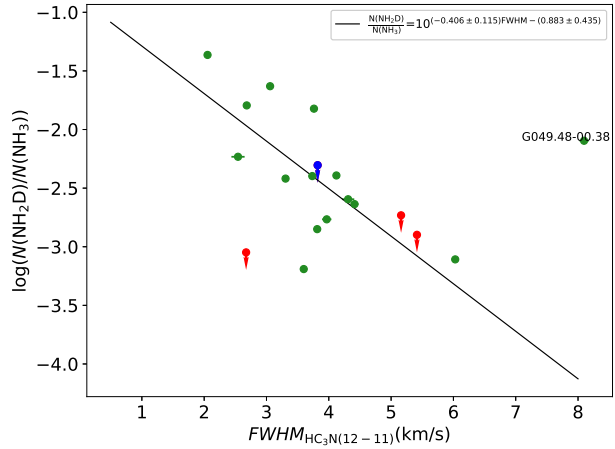}
	\caption{The relationship between deuterated gas and full width at half maxima (FWHM) of HC$_3$N.}
	\label{FWHM}
\end{figure}

\clearpage

\appendix

\section{Detected NH$_2$D lines}

\begin{figure}
\centering
\includegraphics[width=0.5\textwidth]{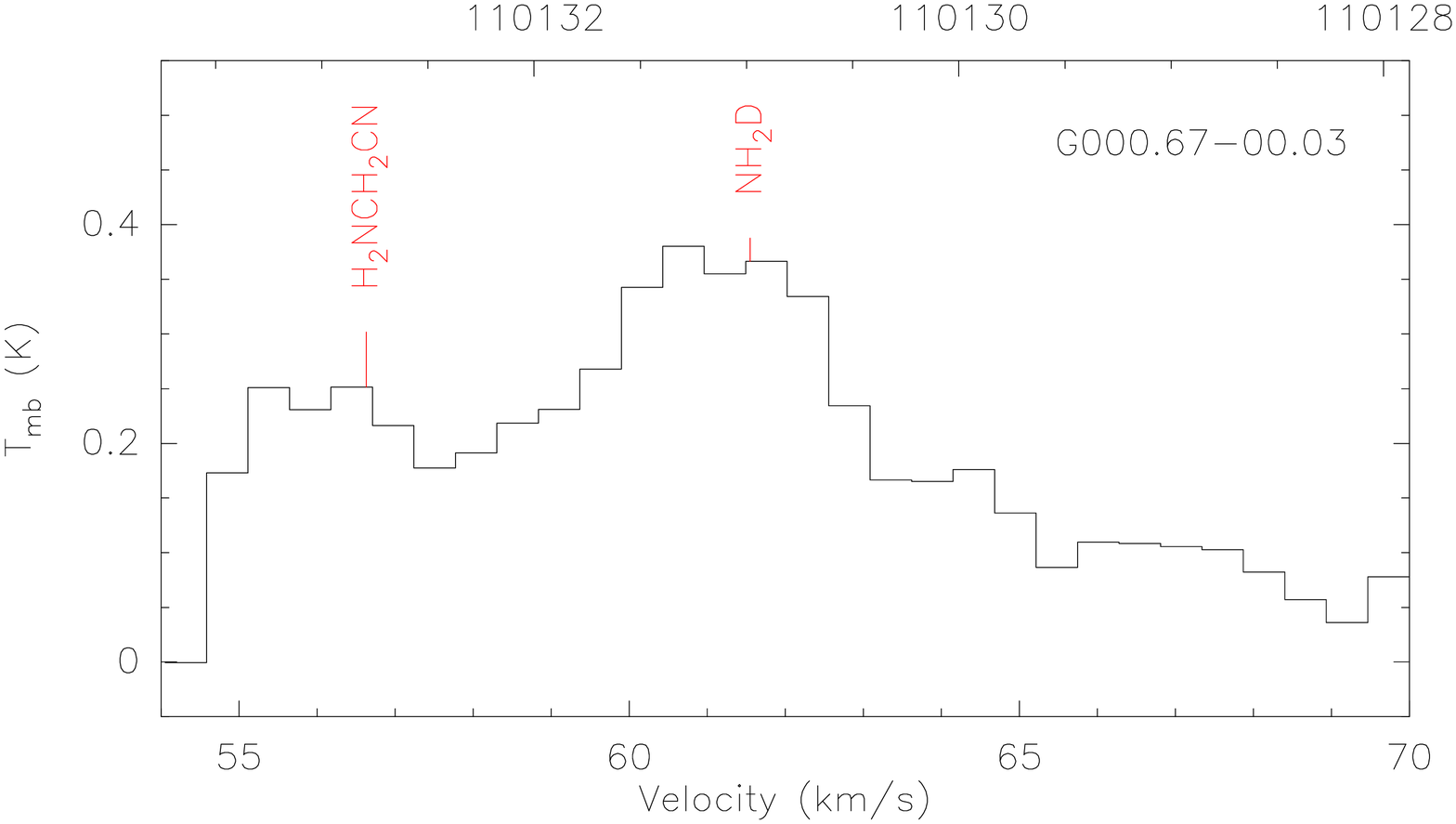}\includegraphics[width=0.5\textwidth]{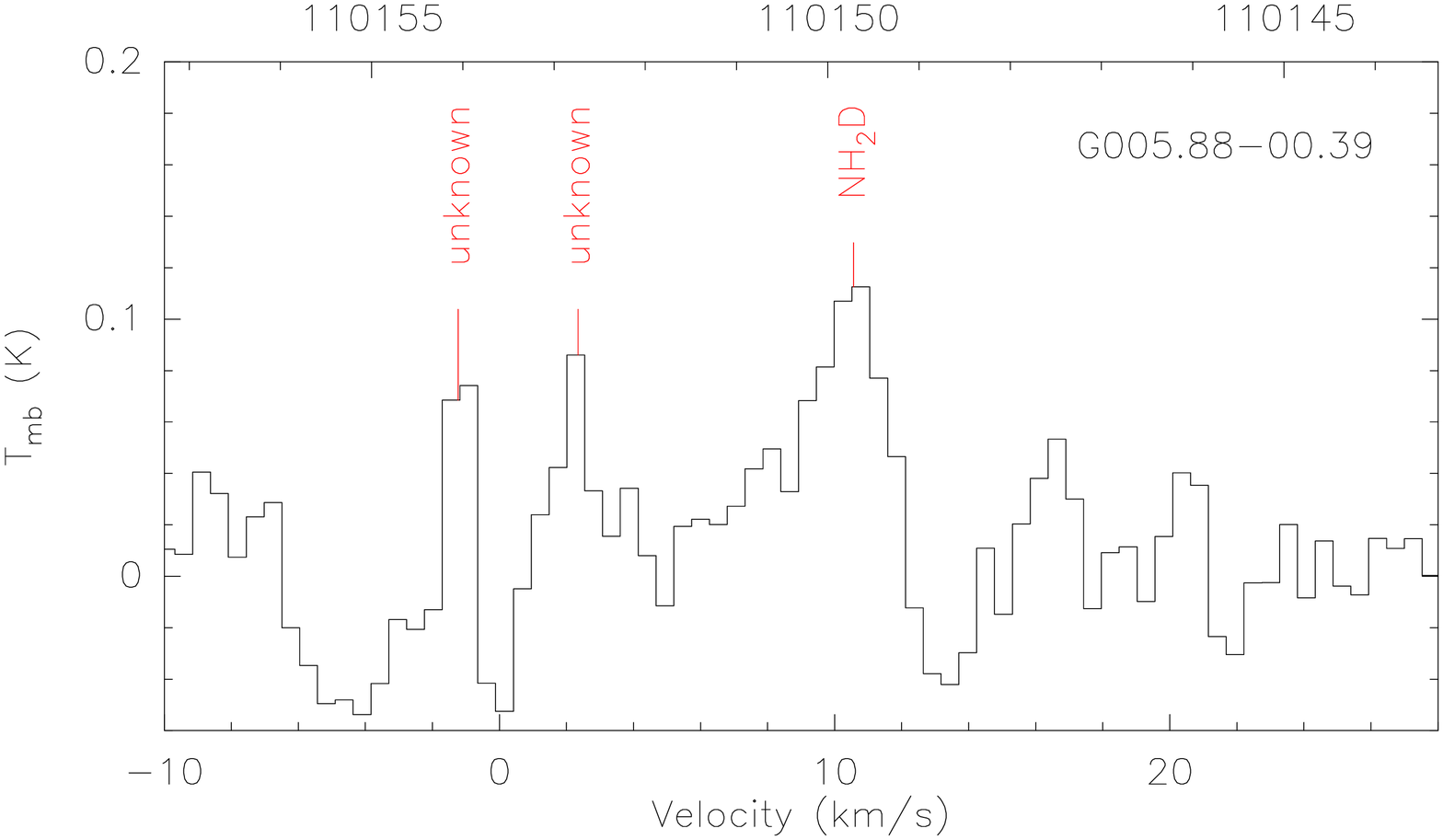}\\\includegraphics[width=0.5\textwidth]{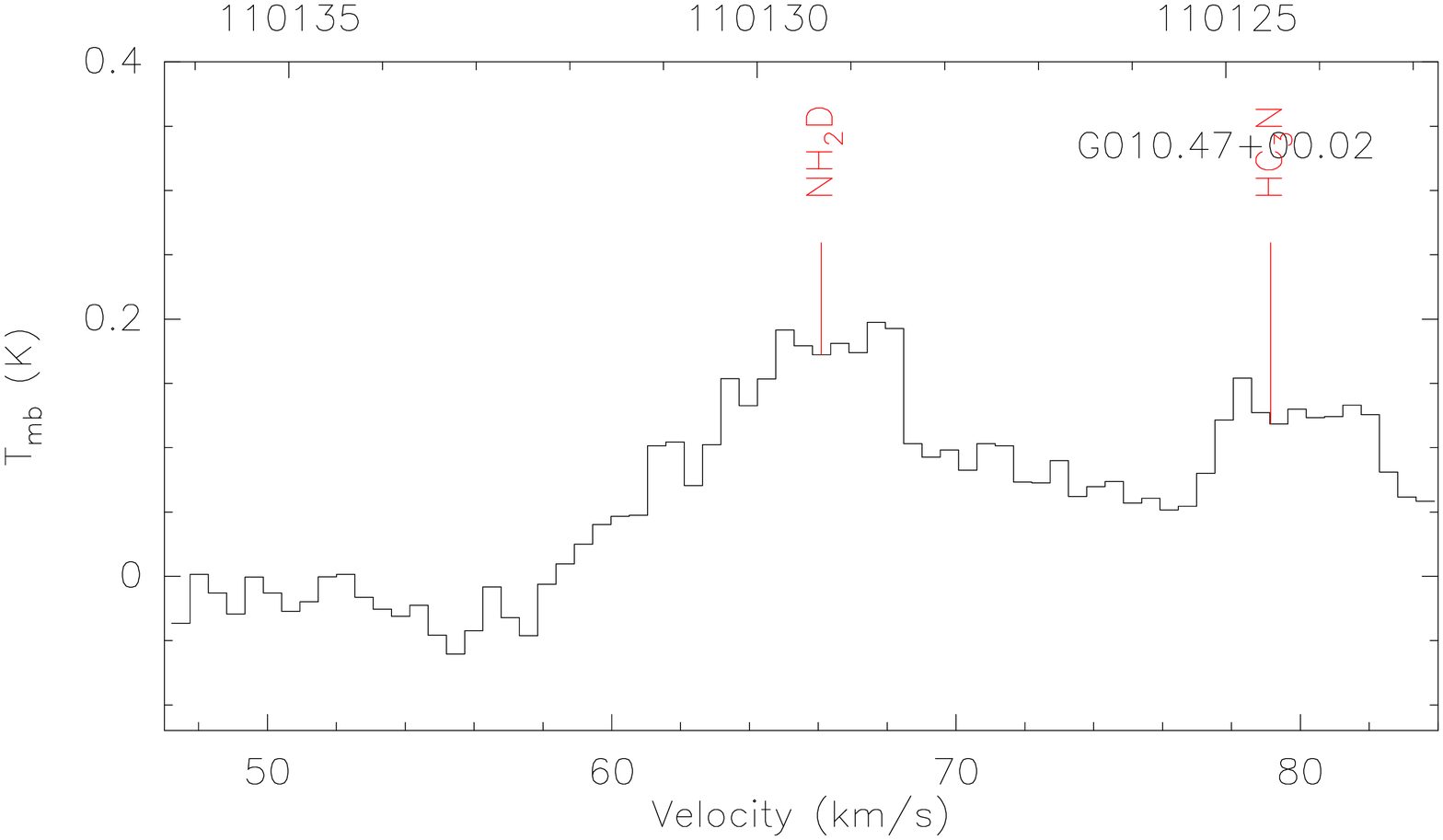}\includegraphics[width=0.5\textwidth]{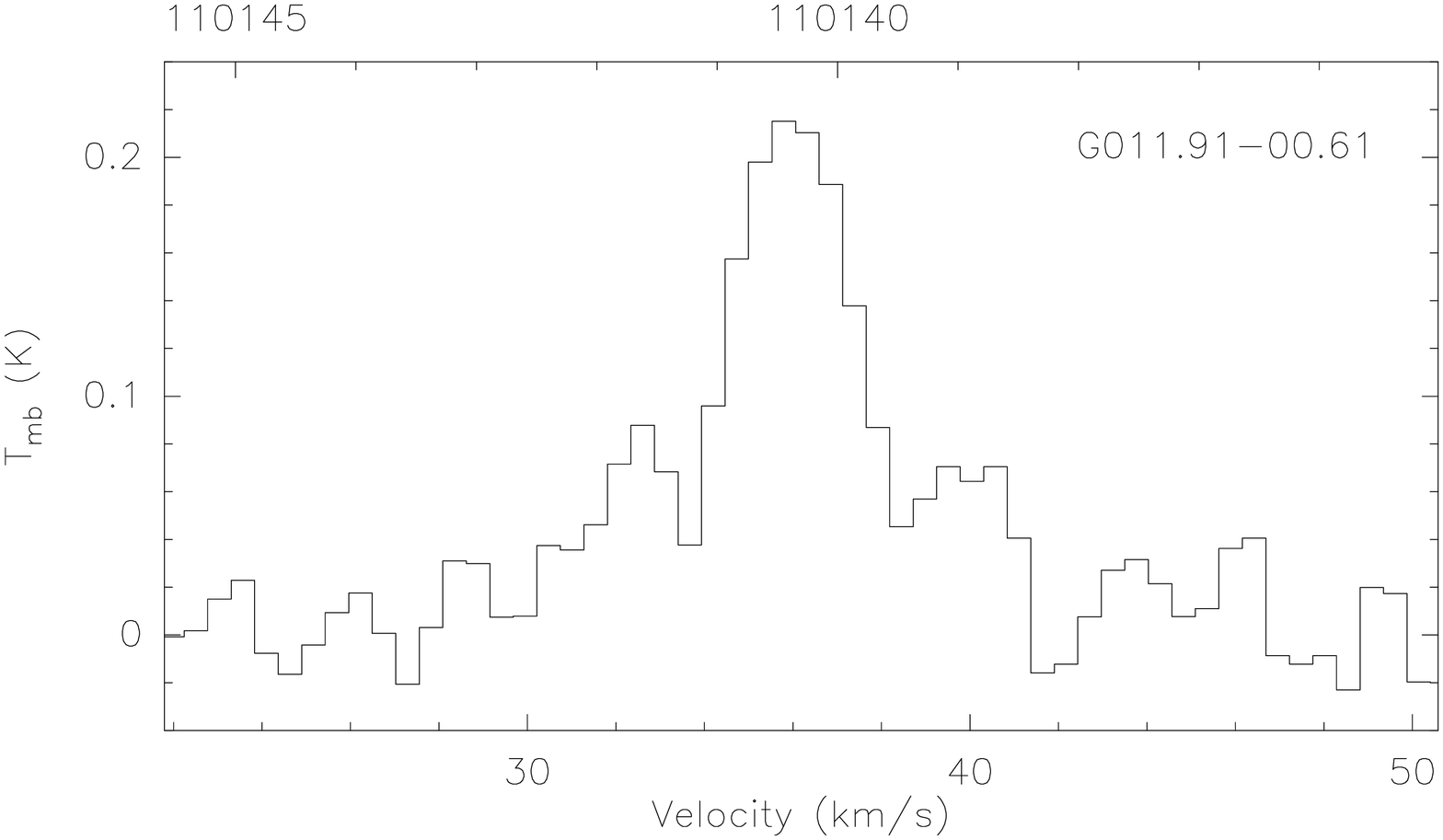}\\\includegraphics[width=0.5\textwidth]{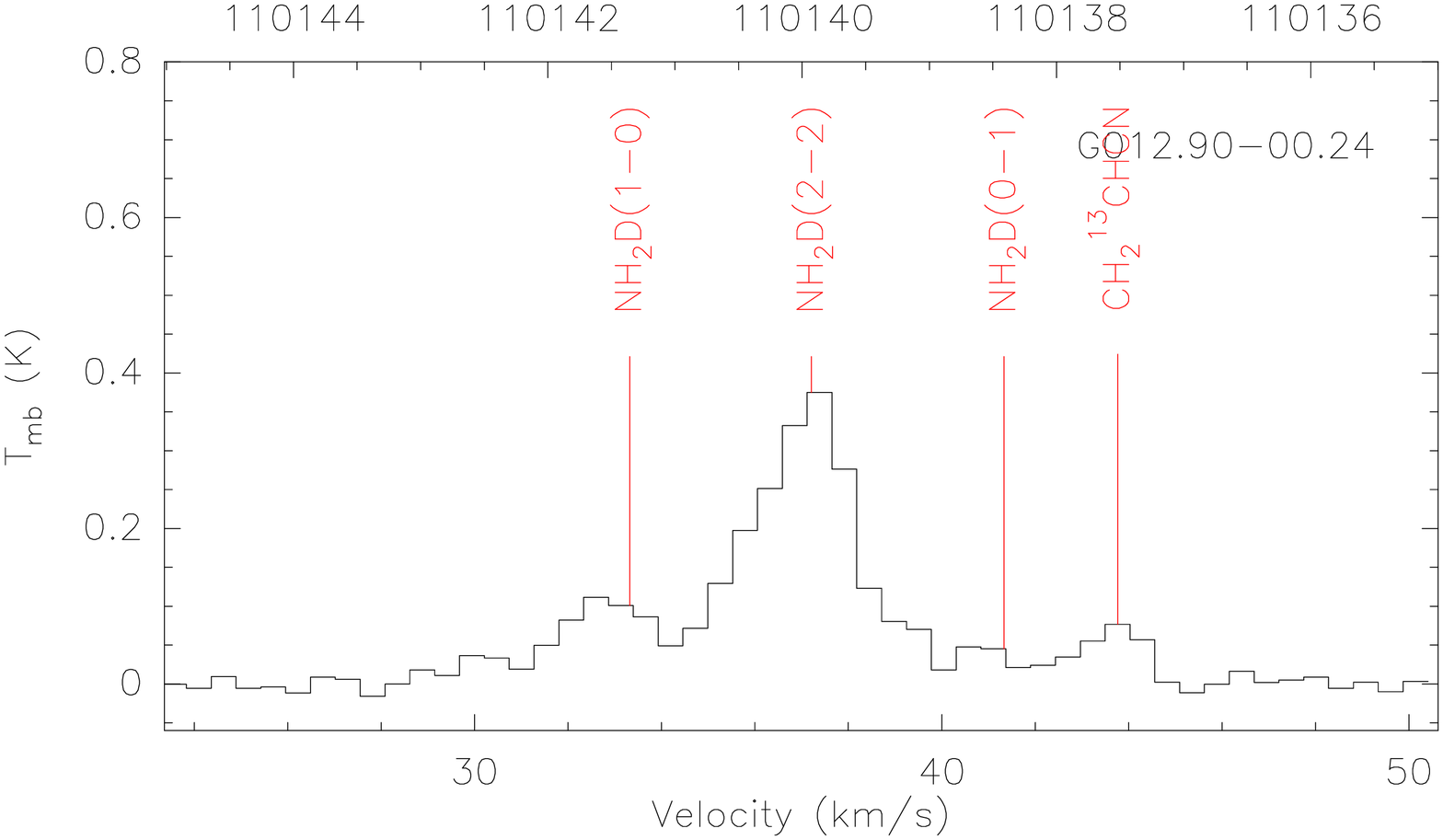}\includegraphics[width=0.5\textwidth]{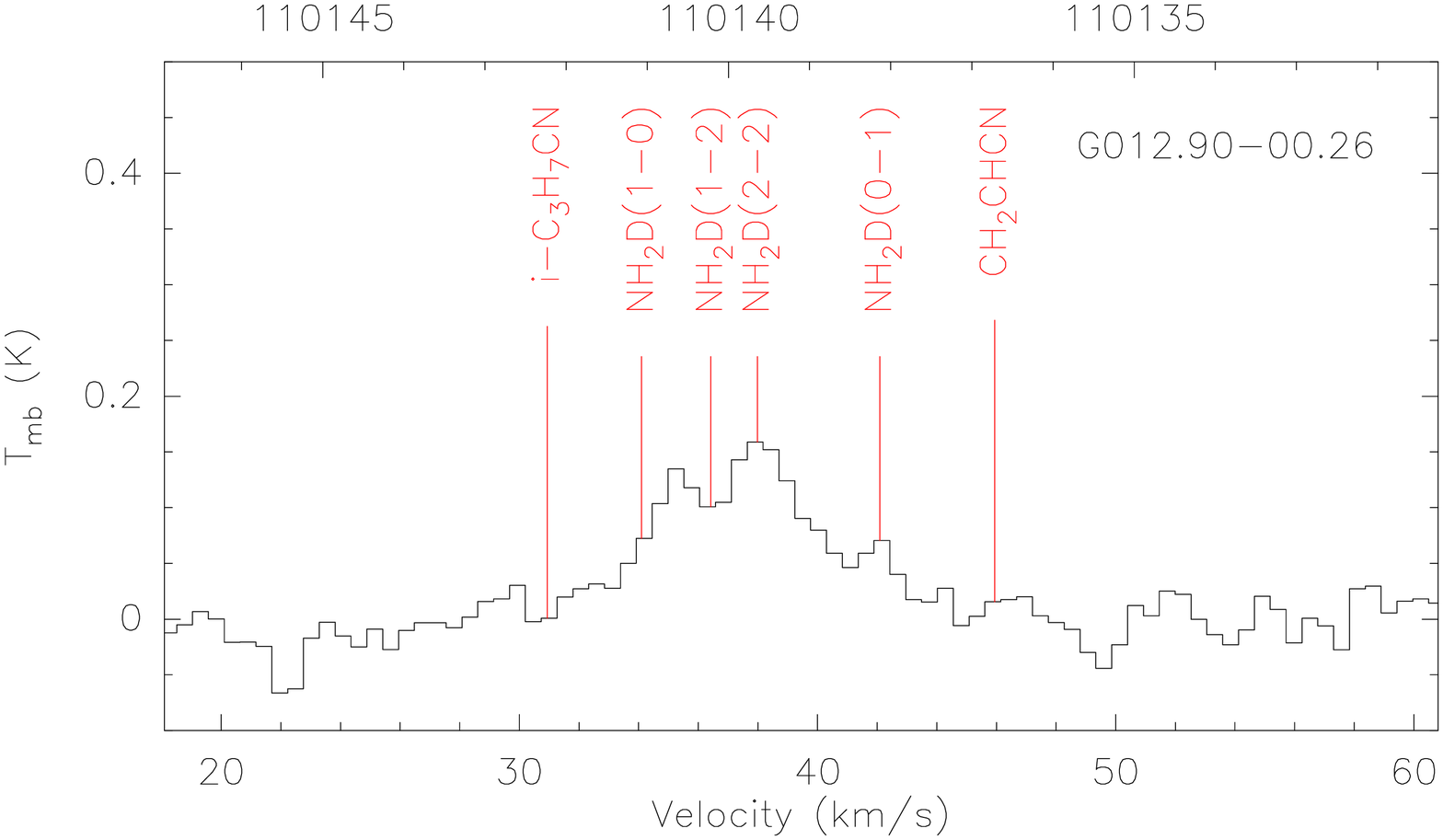}\\
\includegraphics[width=0.5\textwidth]{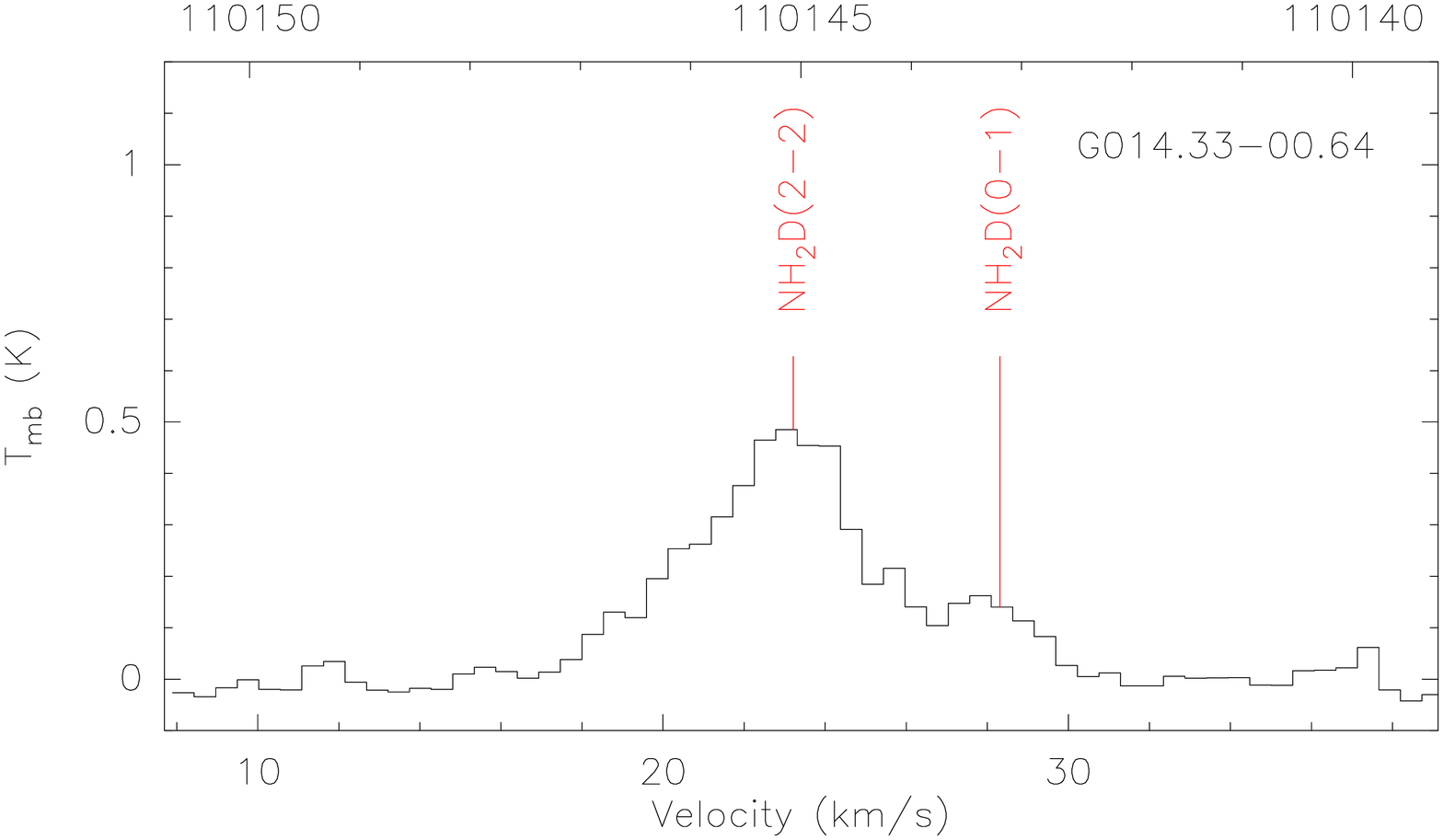}\includegraphics[width=0.5\textwidth]{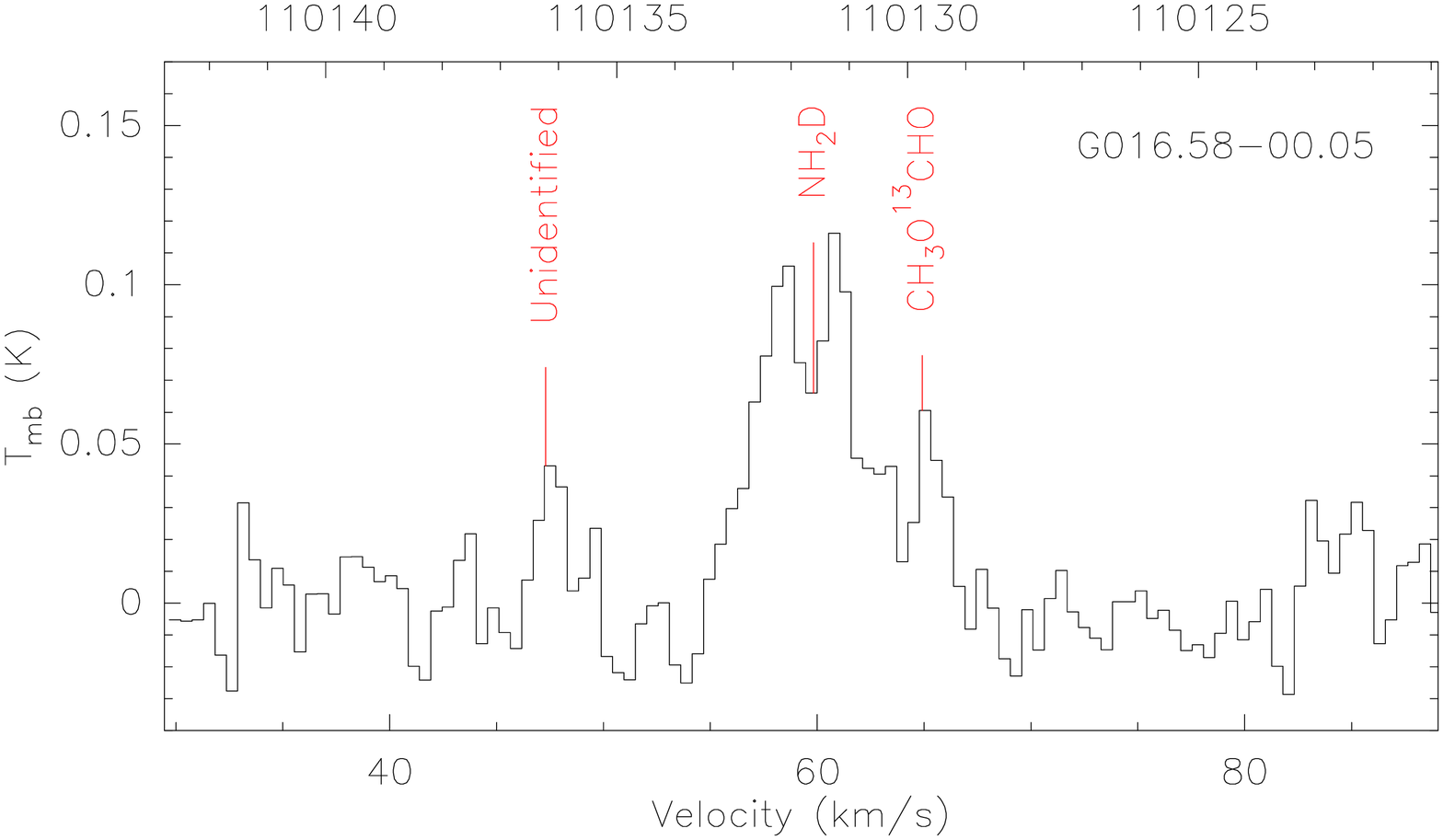}\\
	\caption{The spectra detected for NH$_2$D$1_{11}^a-1_{01}^s$.}
	\label{spectrum1}
\end{figure}

\begin{figure}
\centering
\includegraphics[width=0.5\textwidth]{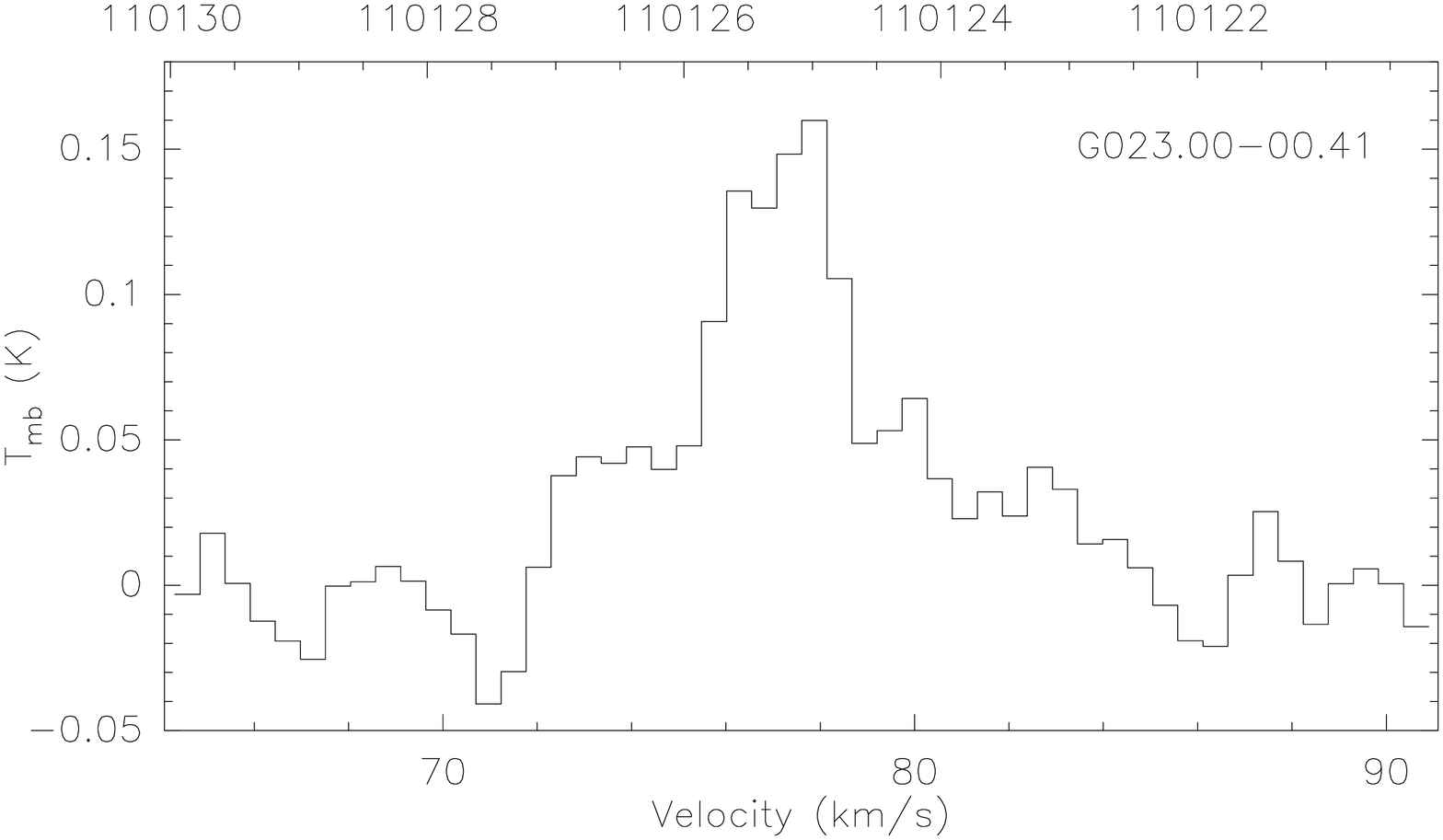}\includegraphics[width=0.5\textwidth]{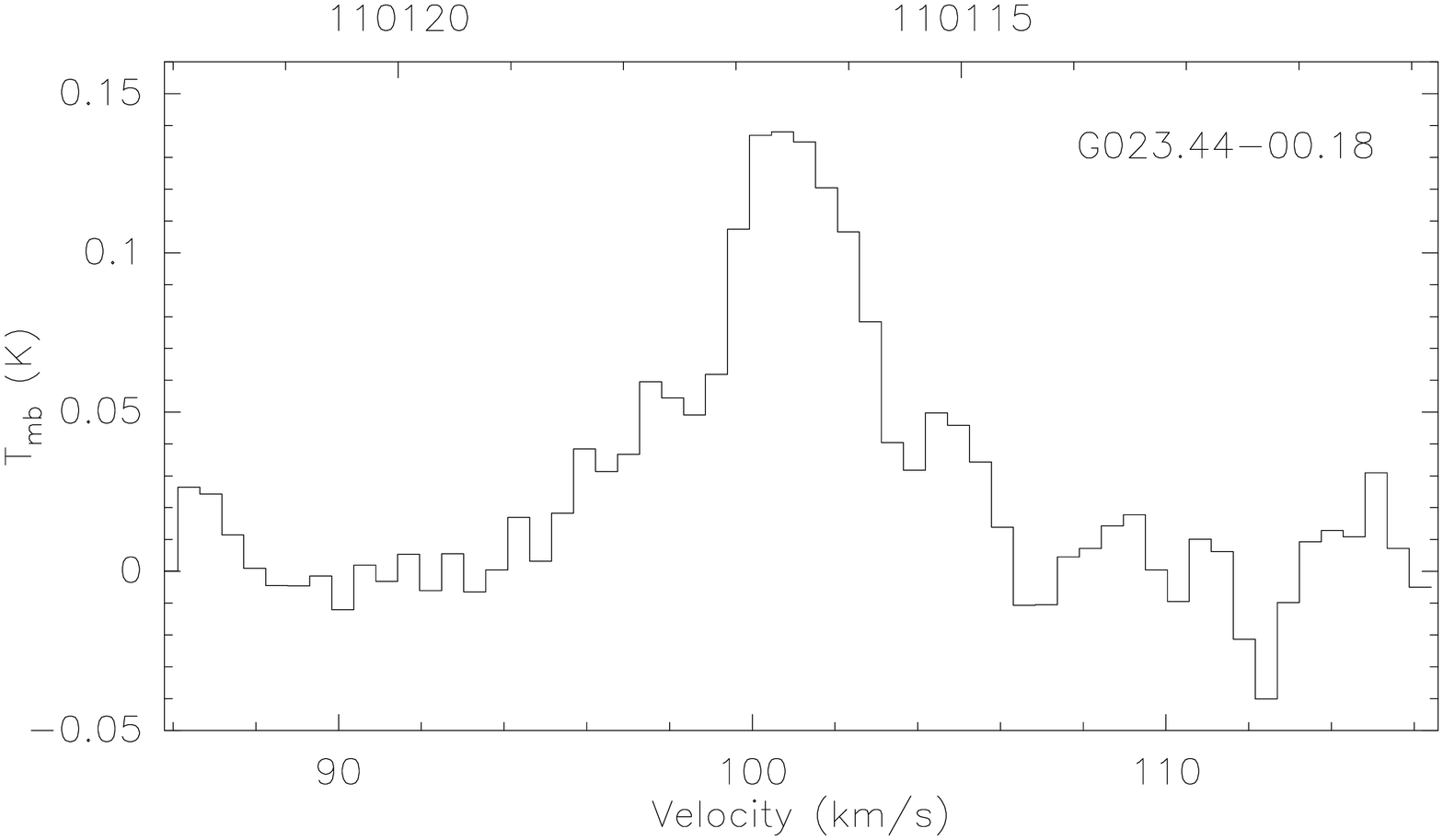}\\
\includegraphics[width=0.5\textwidth]{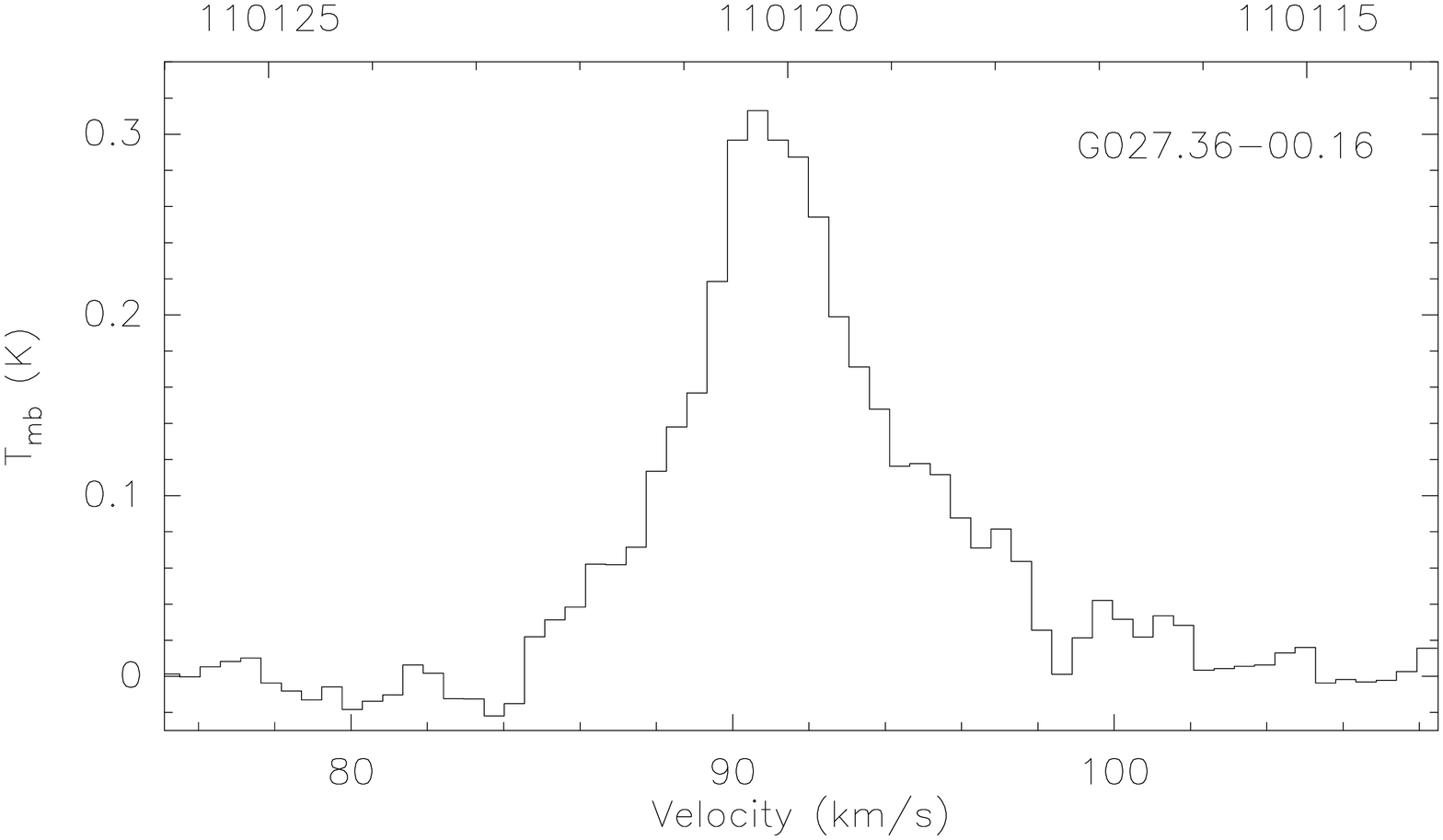}\includegraphics[width=0.5\textwidth]{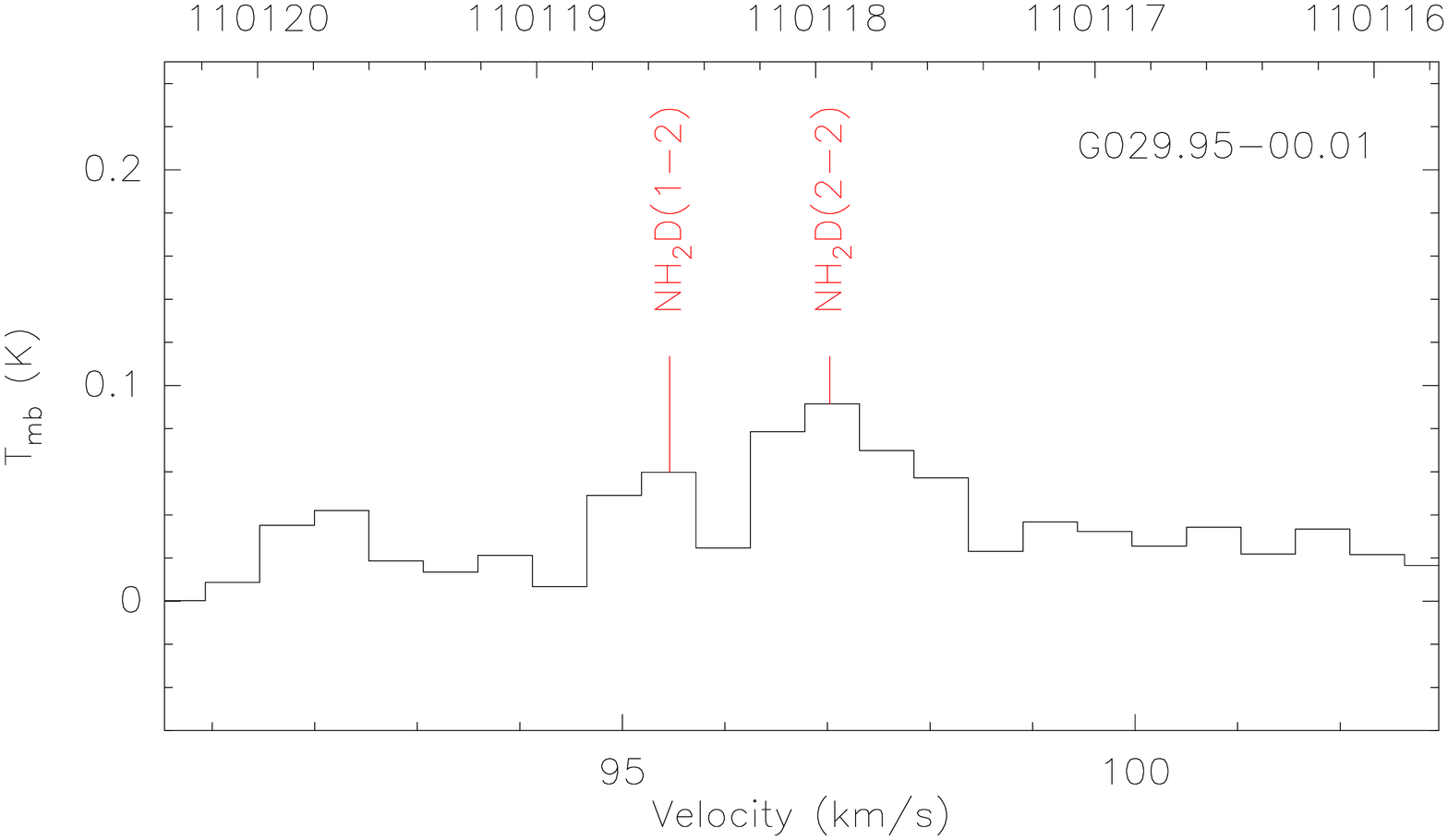}\\
\includegraphics[width=0.5\textwidth]{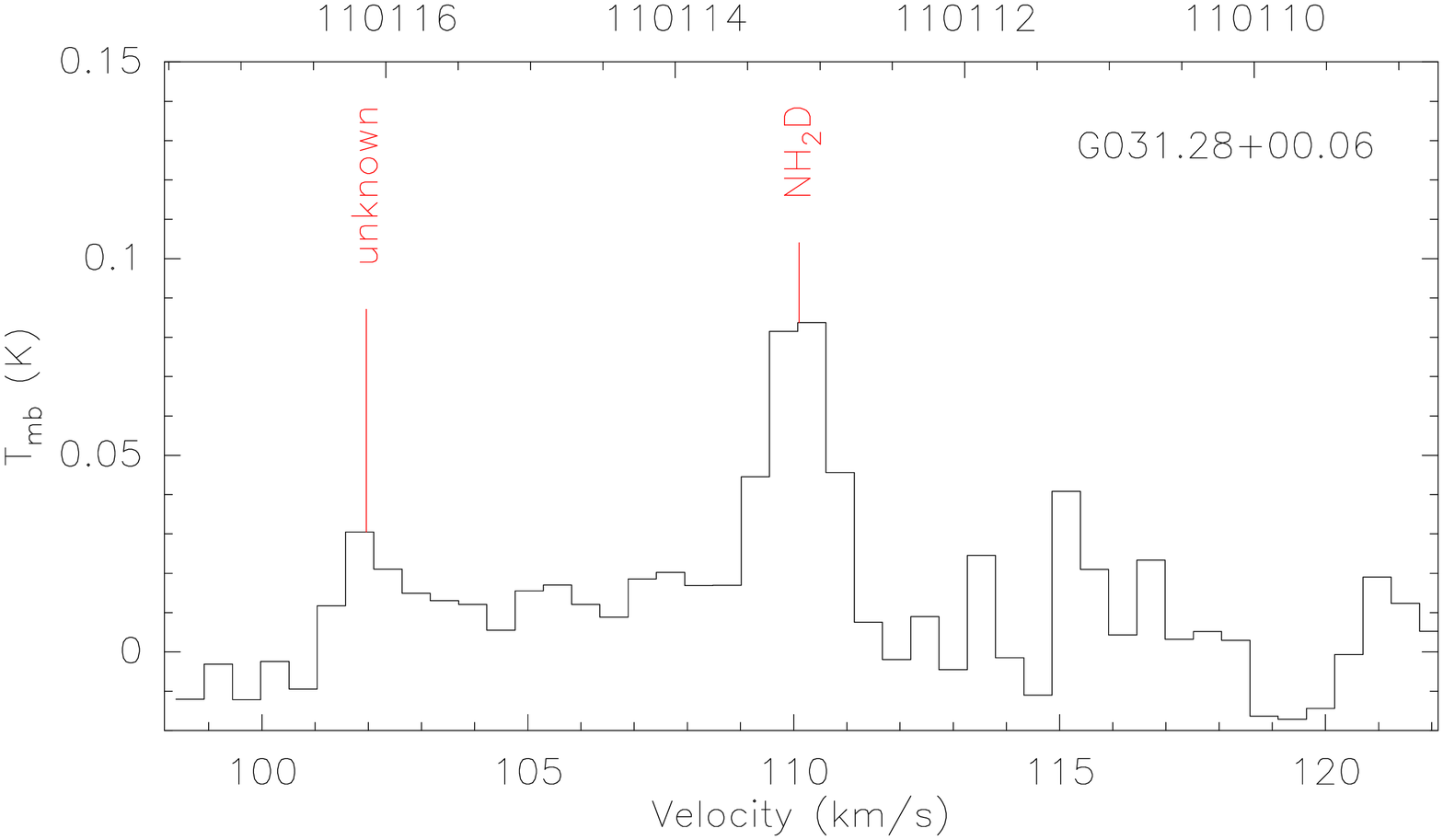}\includegraphics[width=0.5\textwidth]{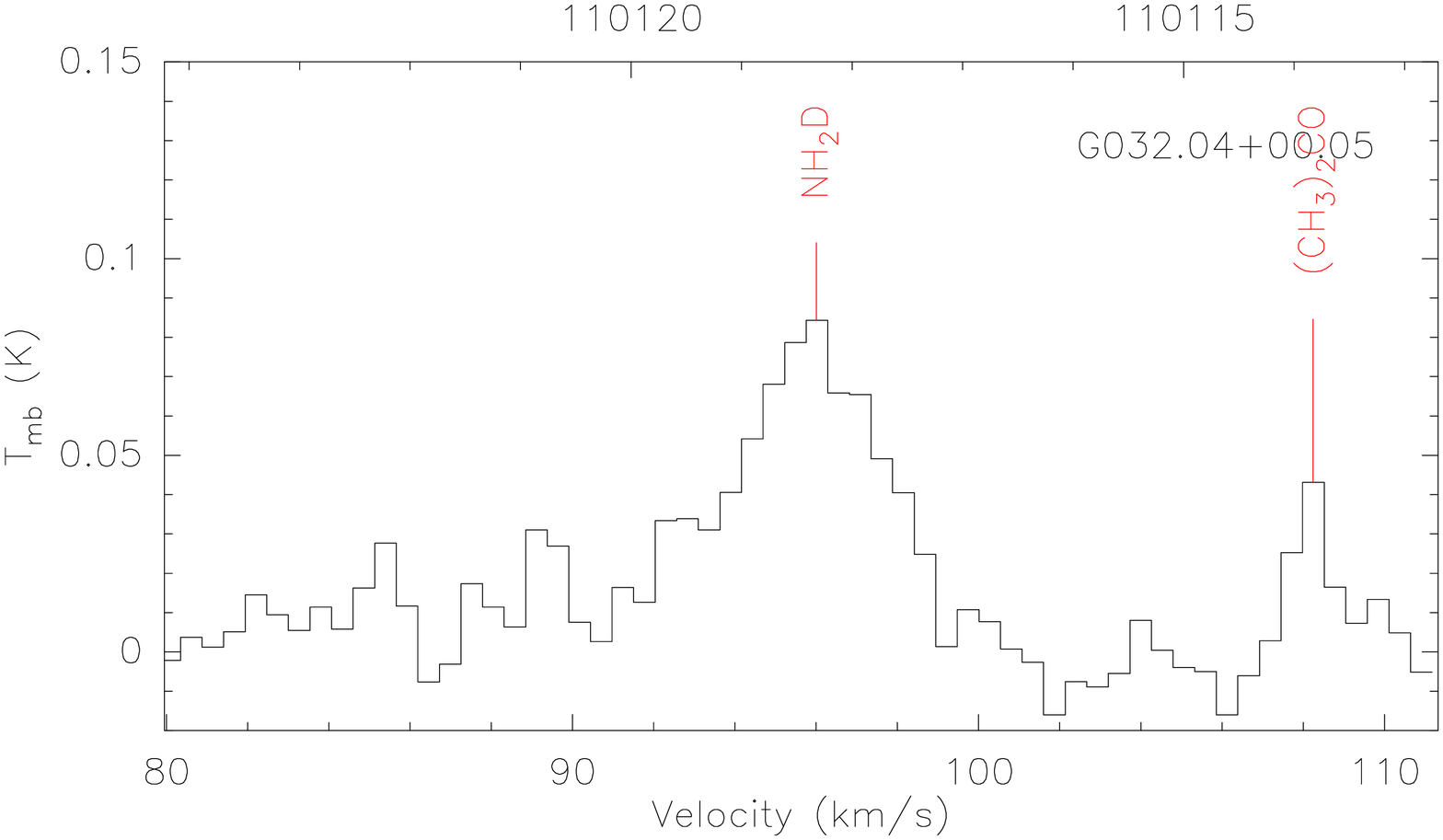}\\
\includegraphics[width=0.5\textwidth]{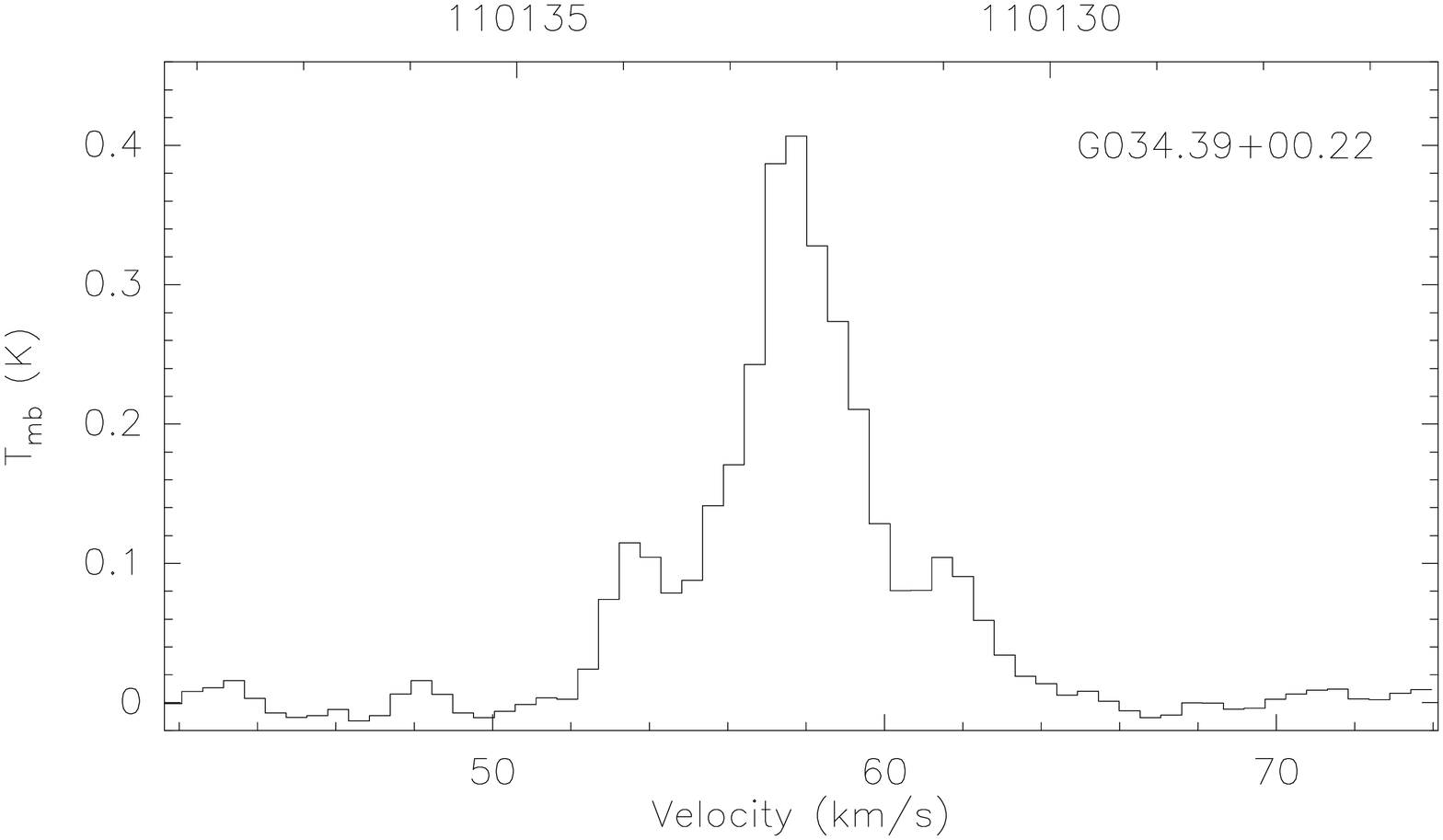}\includegraphics[width=0.5\textwidth]{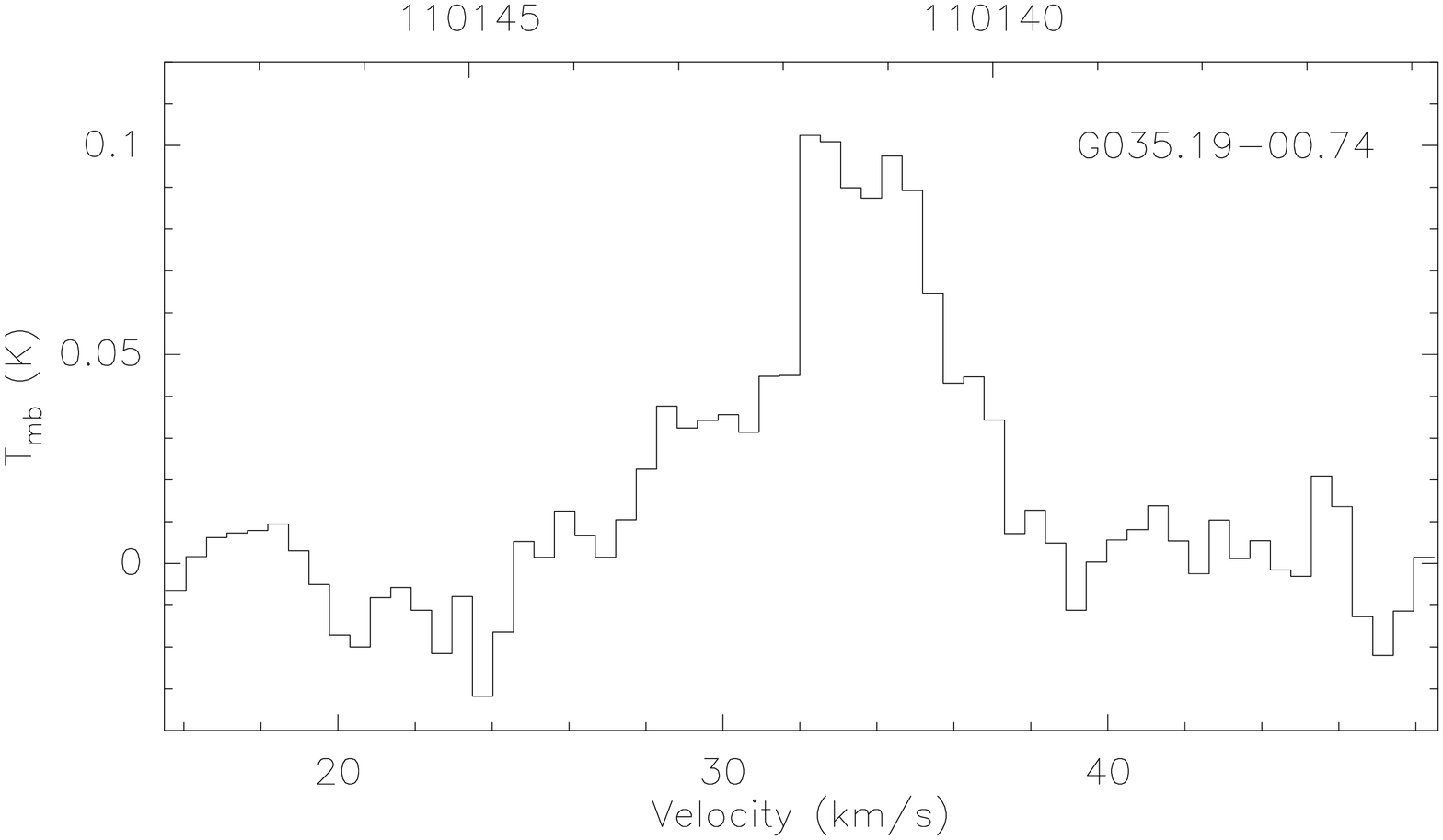}\\
	\caption{The spectra detected for NH$_2$D$1_{11}^a-1_{01}^s$.}
	\label{spectrum2}
\end{figure}

\begin{figure}
\centering
\includegraphics[width=0.5\textwidth]{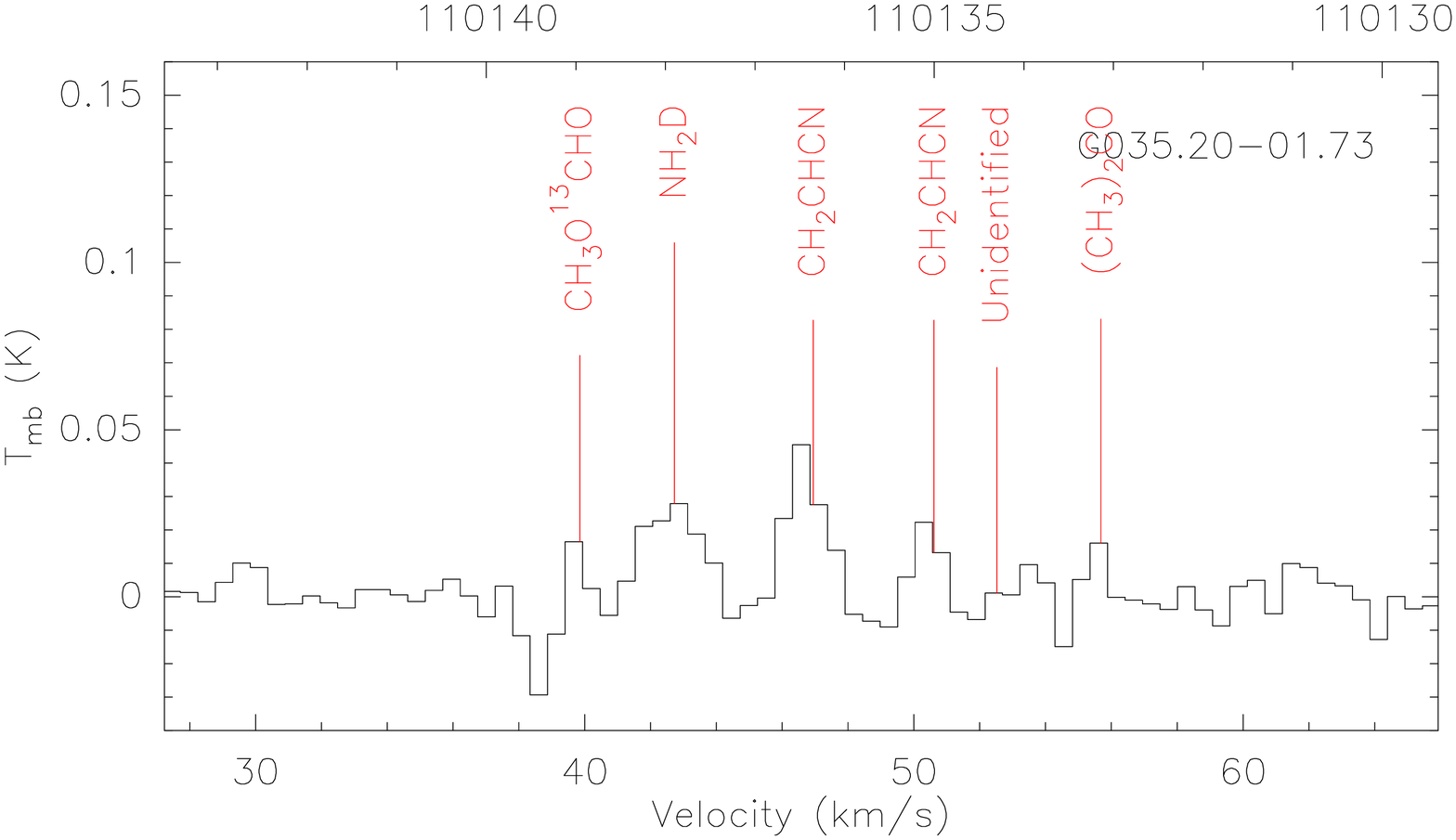}\includegraphics[width=0.5\textwidth]{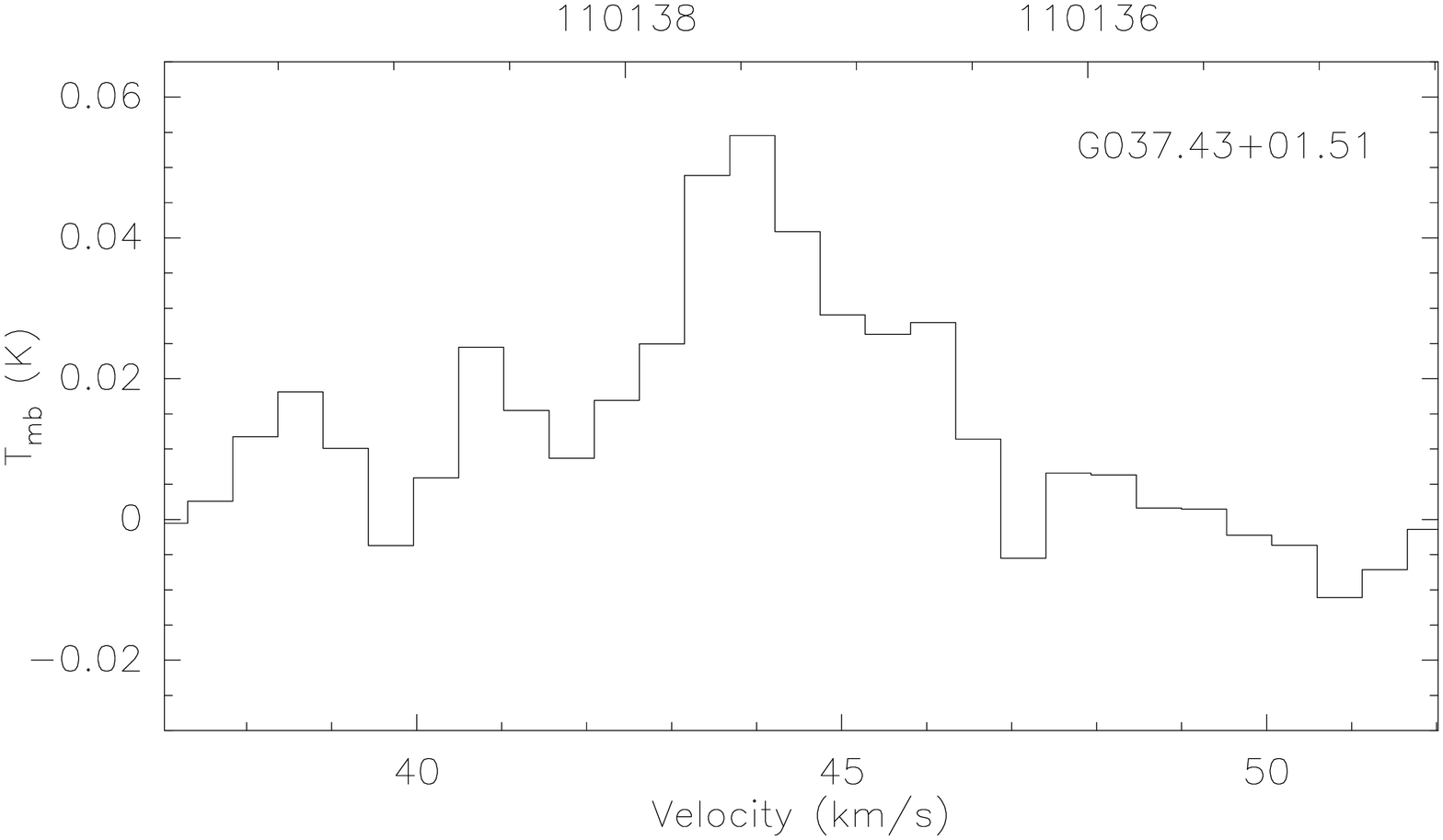}\\
\includegraphics[width=0.5\textwidth]{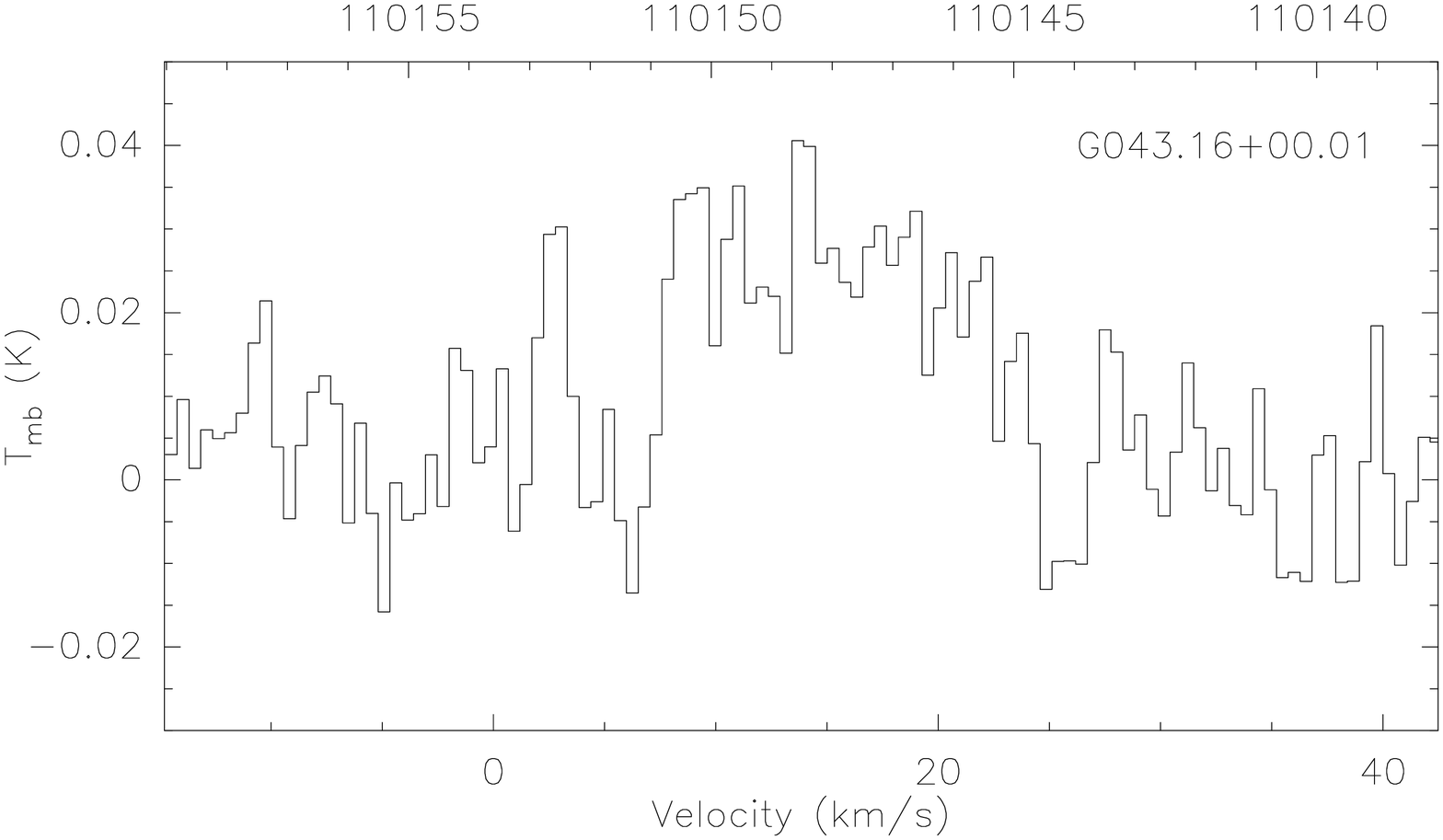}\includegraphics[width=0.5\textwidth]{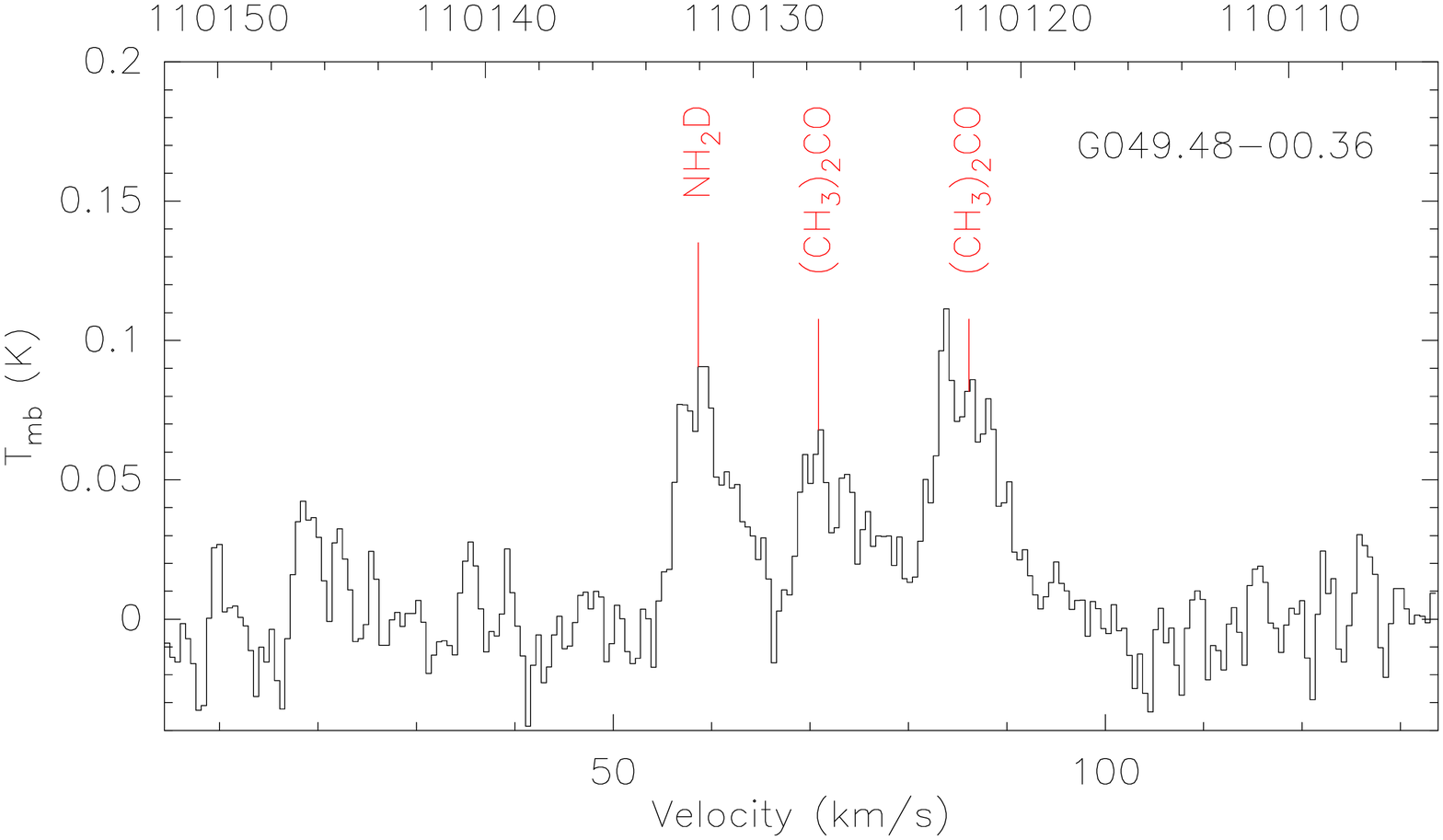}\\
\includegraphics[width=0.5\textwidth]{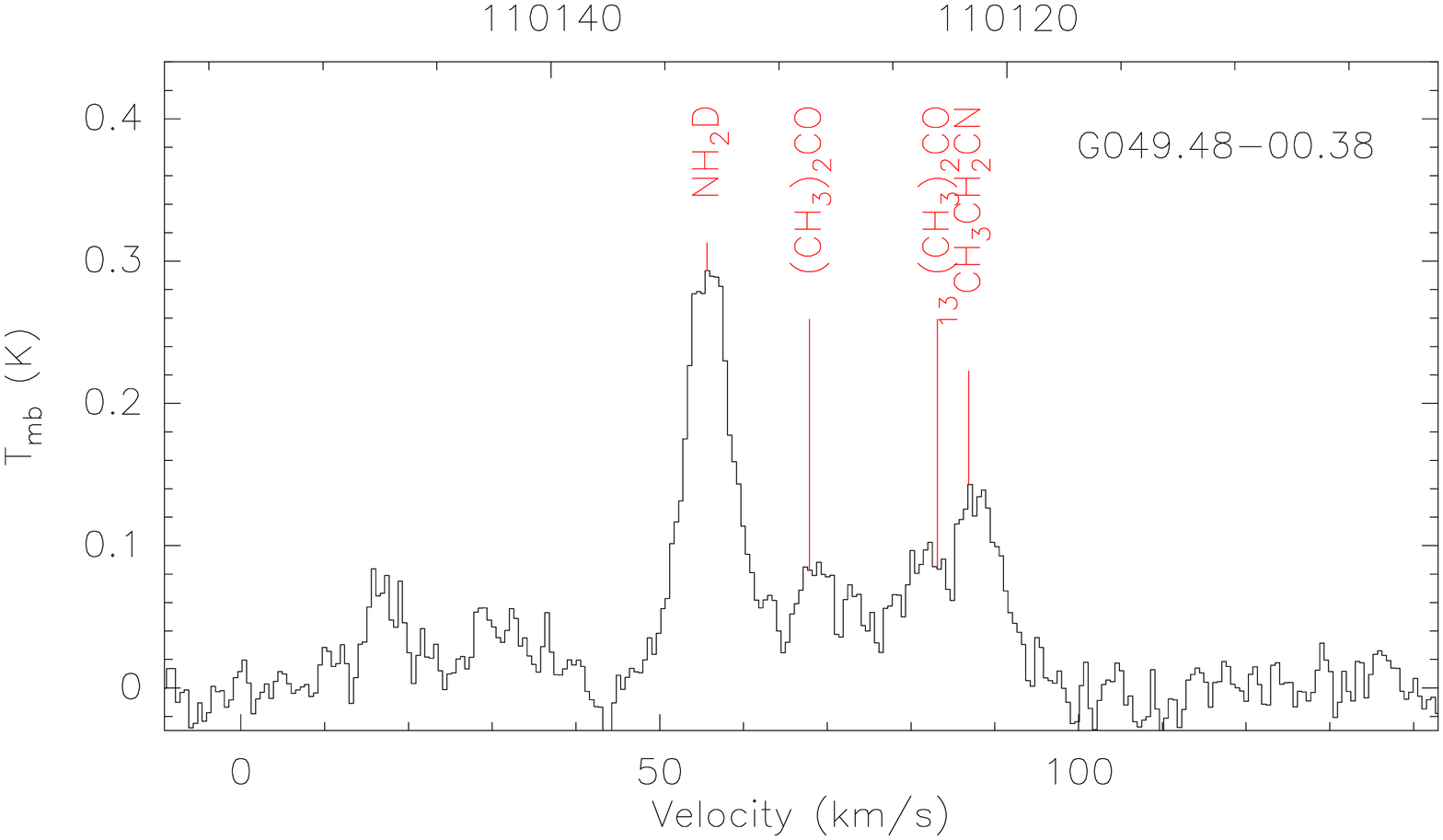}\includegraphics[width=0.5\textwidth]{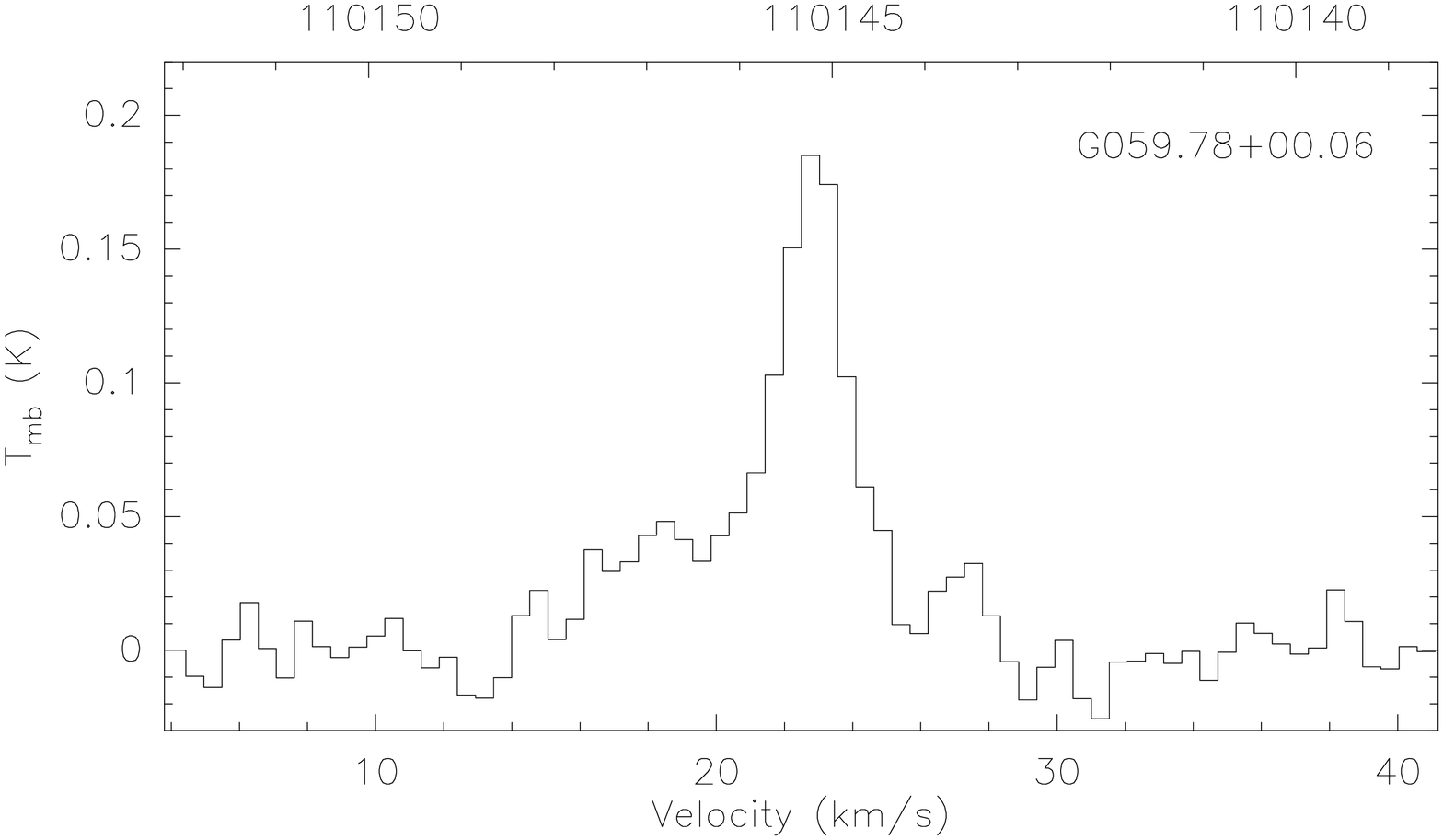}\\
\includegraphics[width=0.5\textwidth]{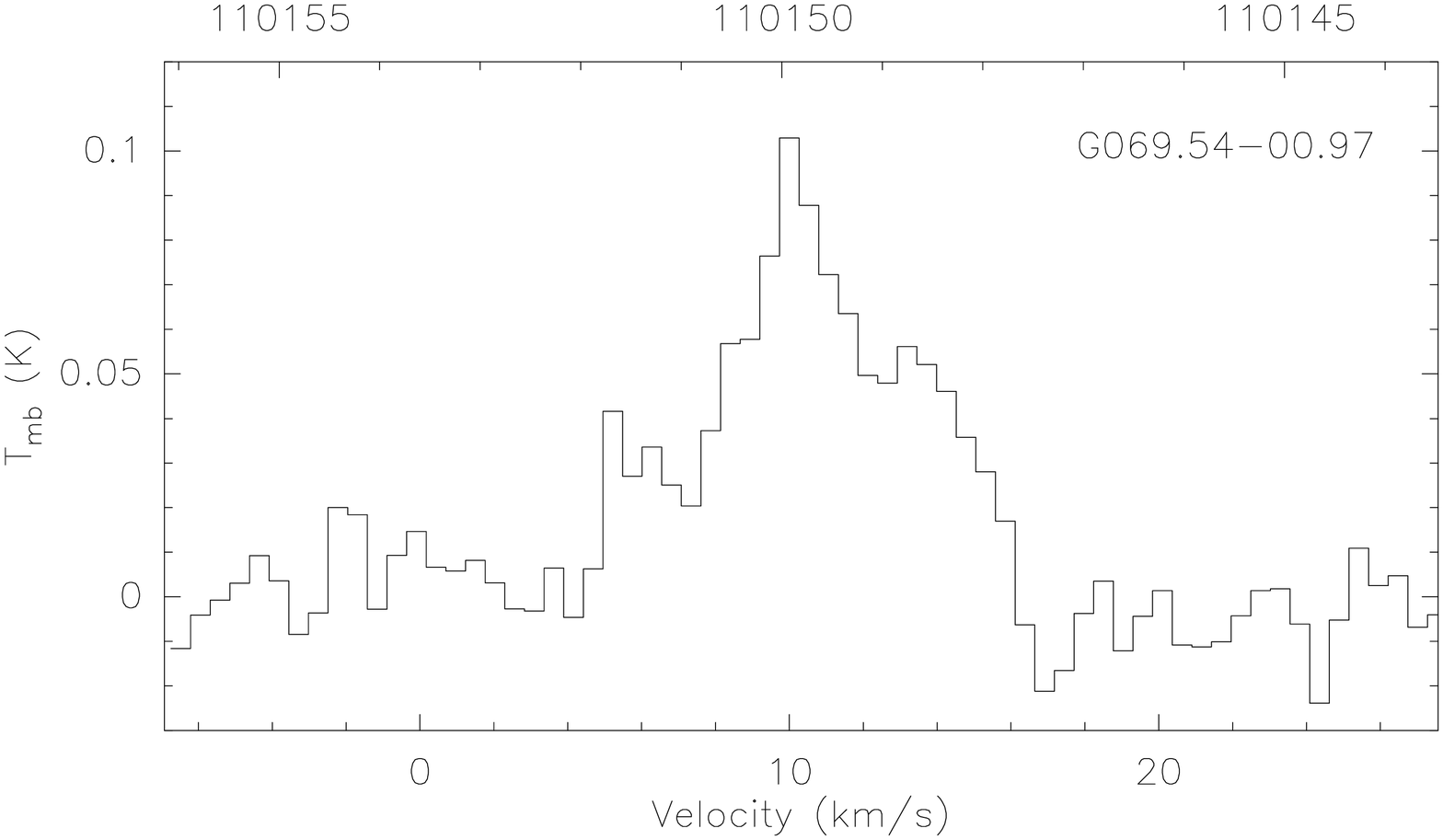}\includegraphics[width=0.5\textwidth]{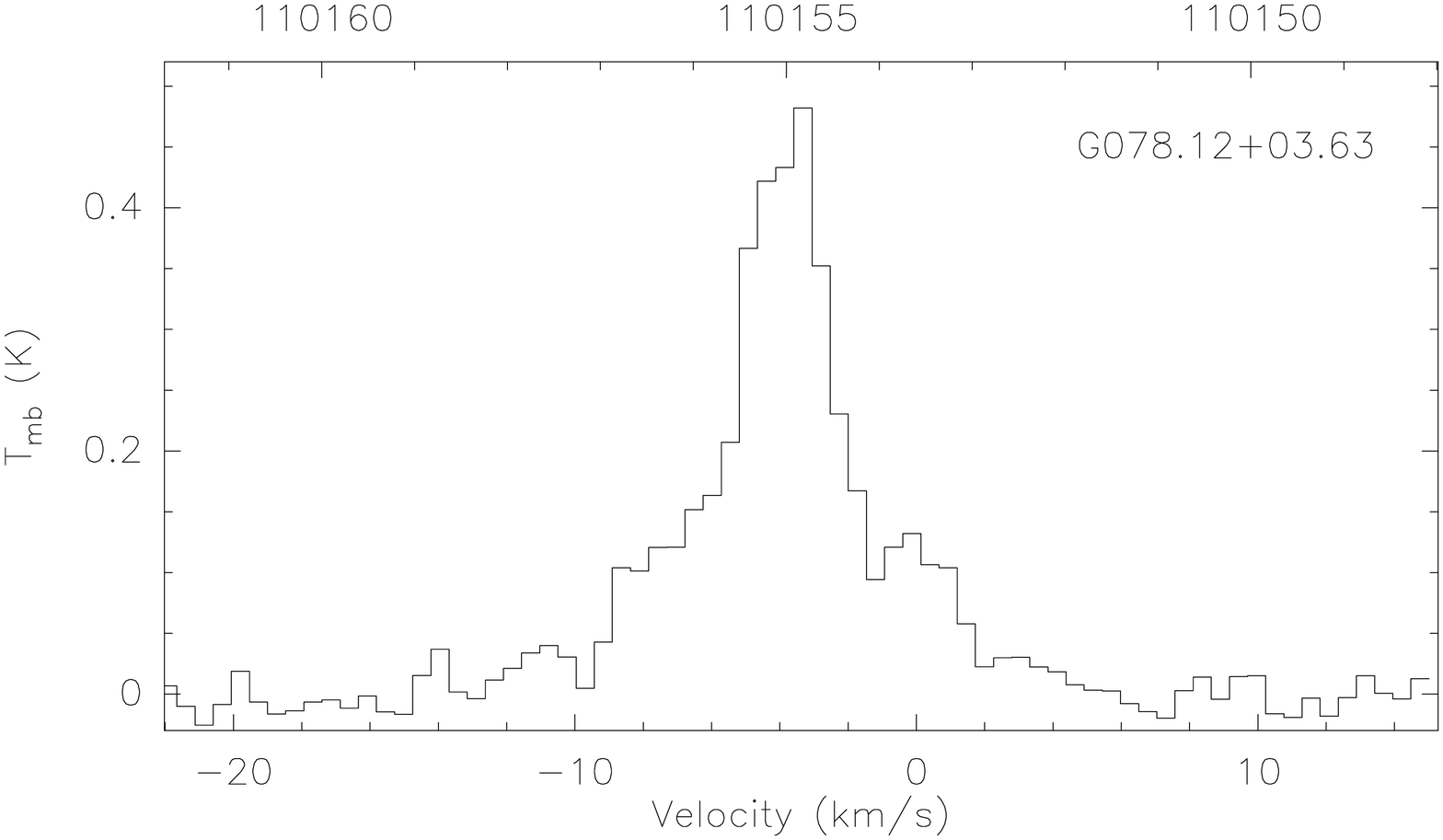}\\
	\caption{The spectra detected for NH$_2$D$1_{11}^a-1_{01}^s$.}
	\label{spectrum3}
\end{figure}

\begin{figure}
\centering
\includegraphics[width=0.5\textwidth]{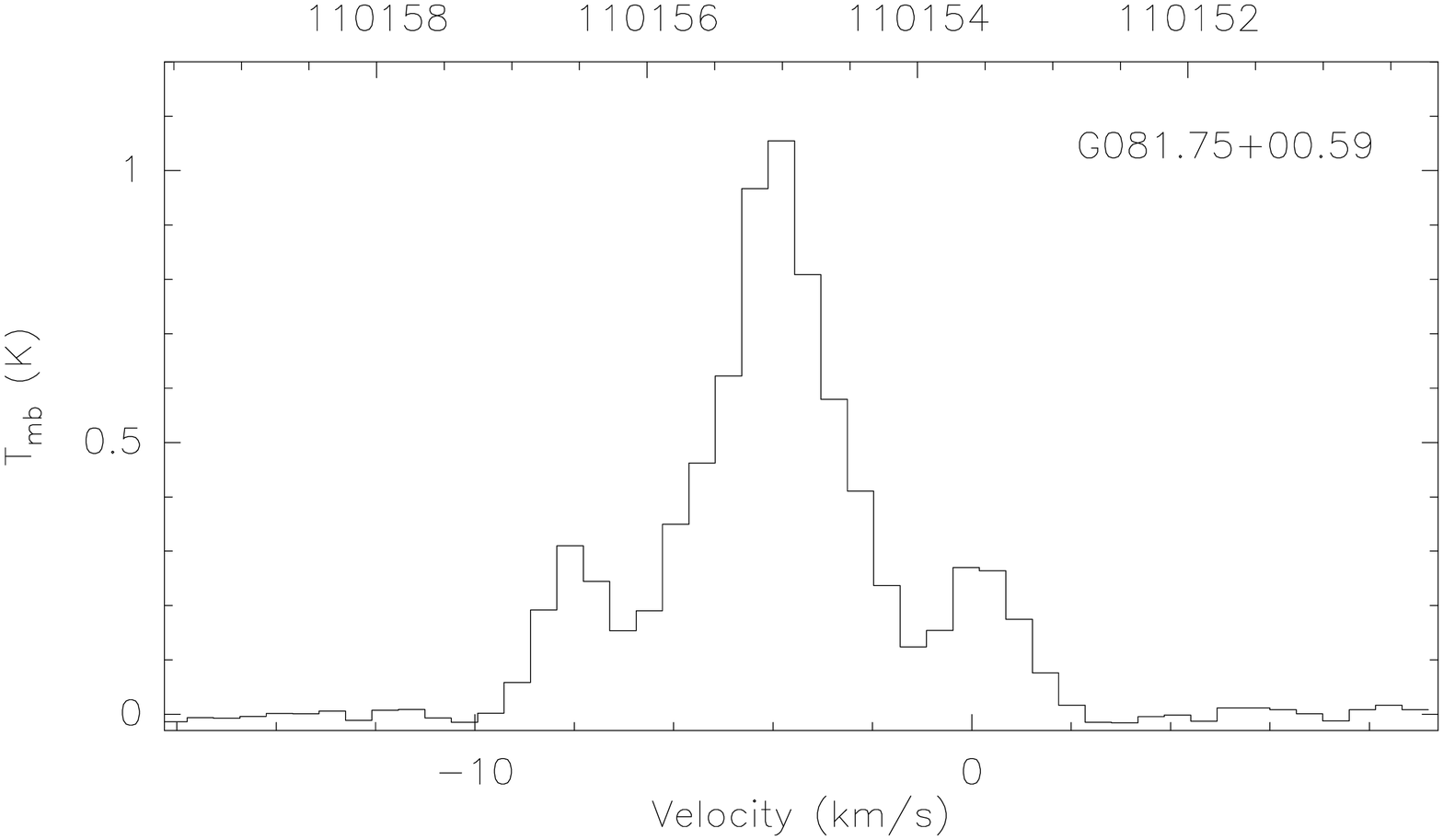}\includegraphics[width=0.5\textwidth]{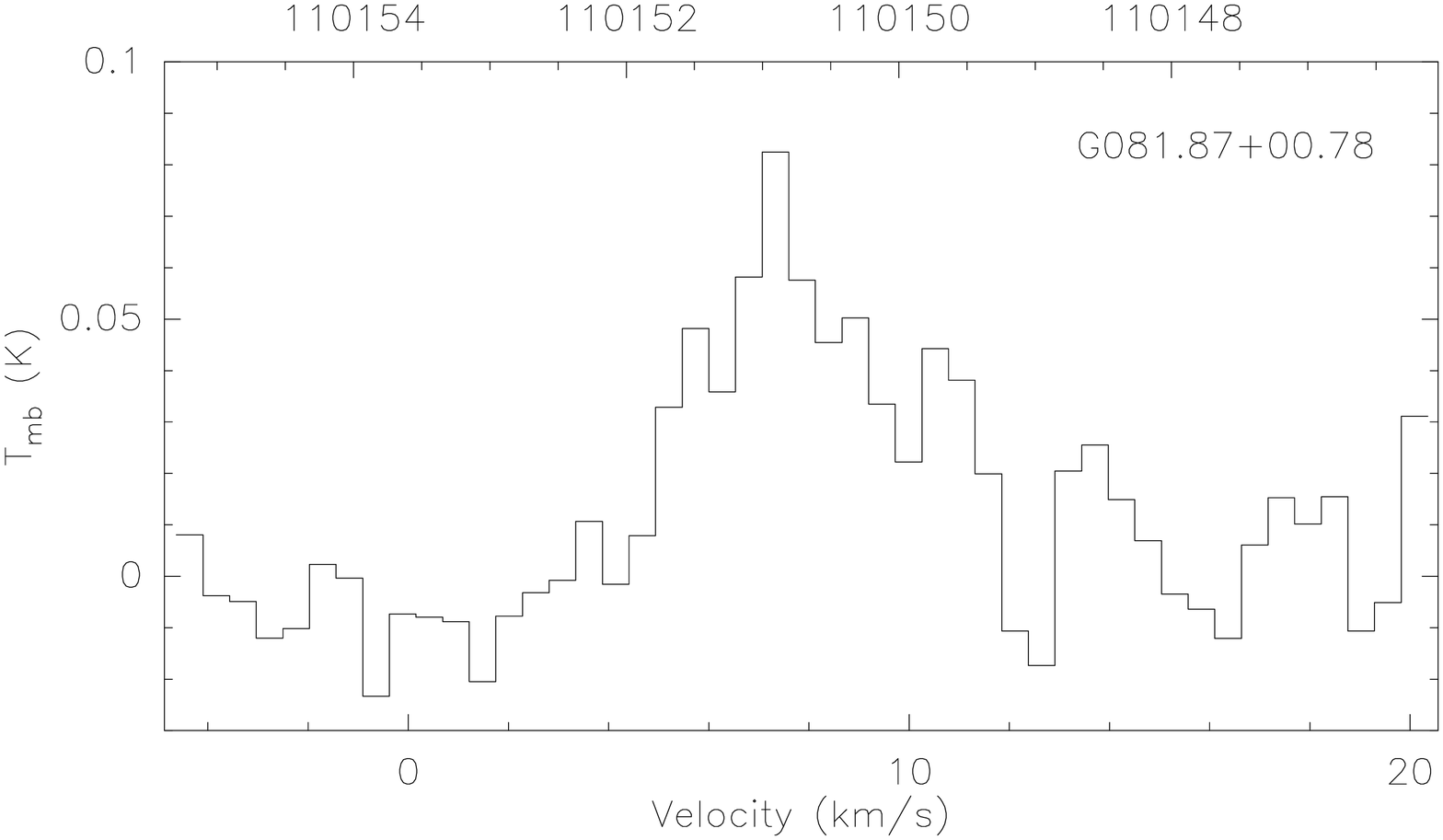}\\
\includegraphics[width=0.5\textwidth]{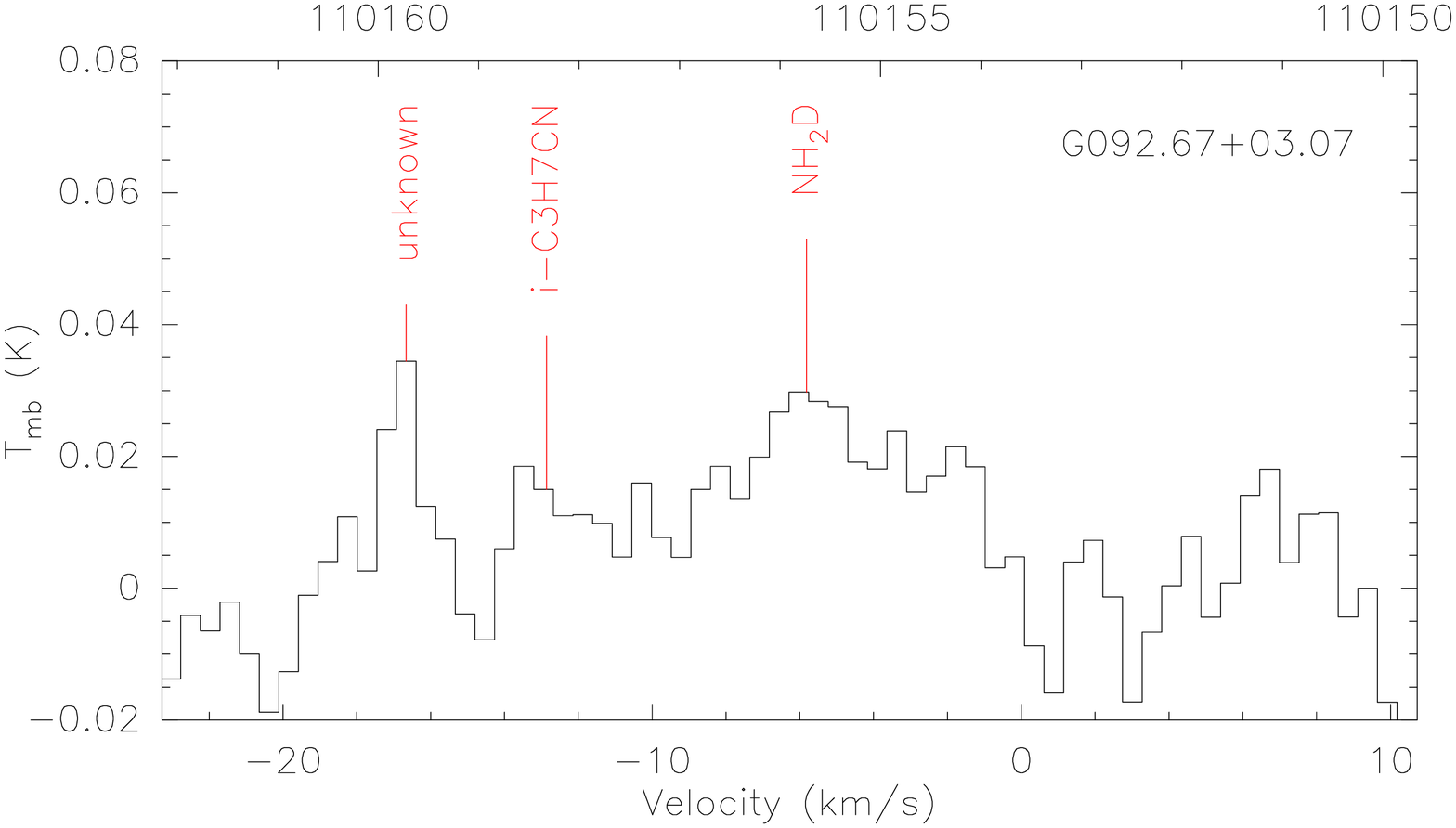}\includegraphics[width=0.5\textwidth]{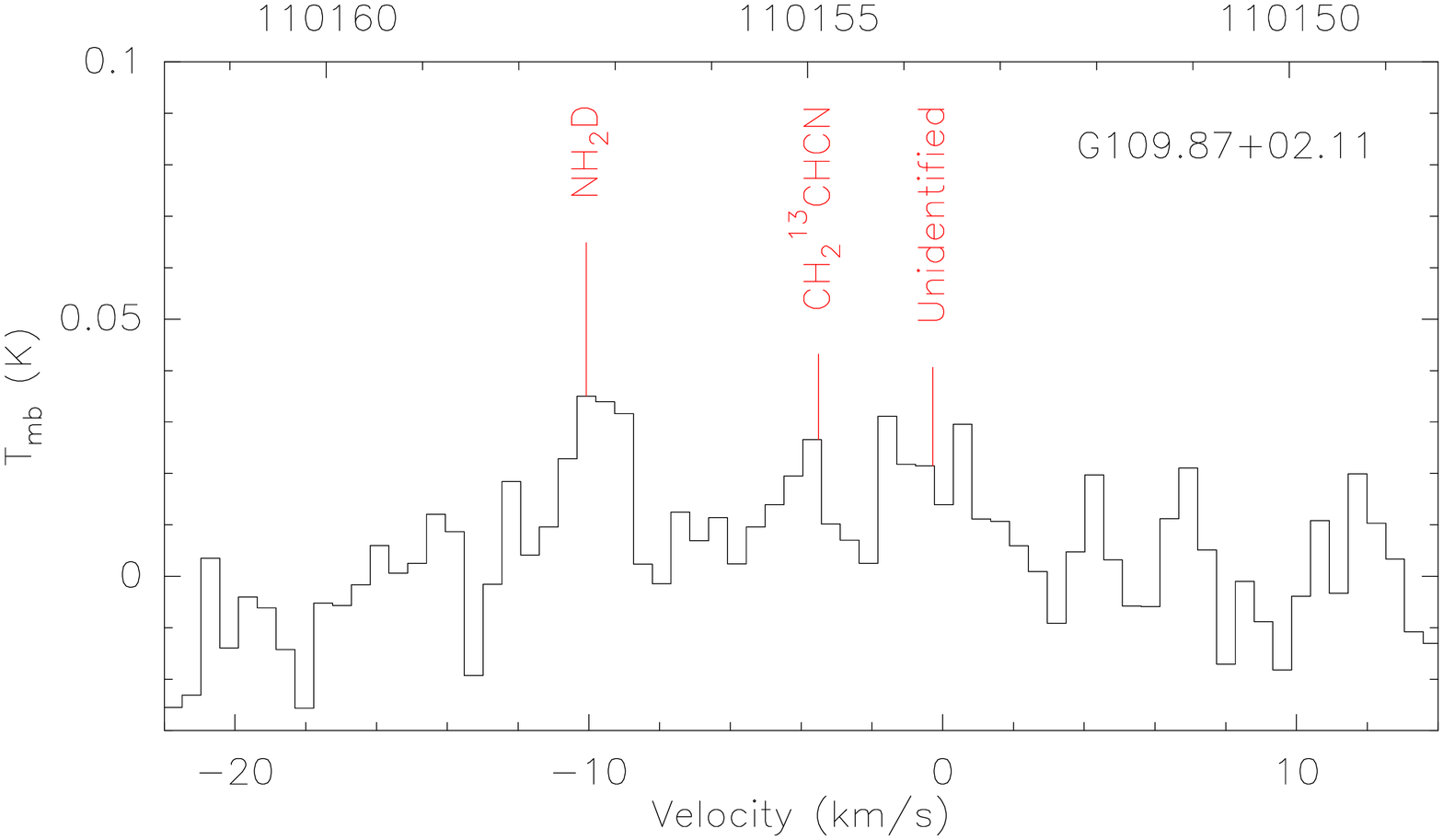}\\
\includegraphics[width=0.5\textwidth]{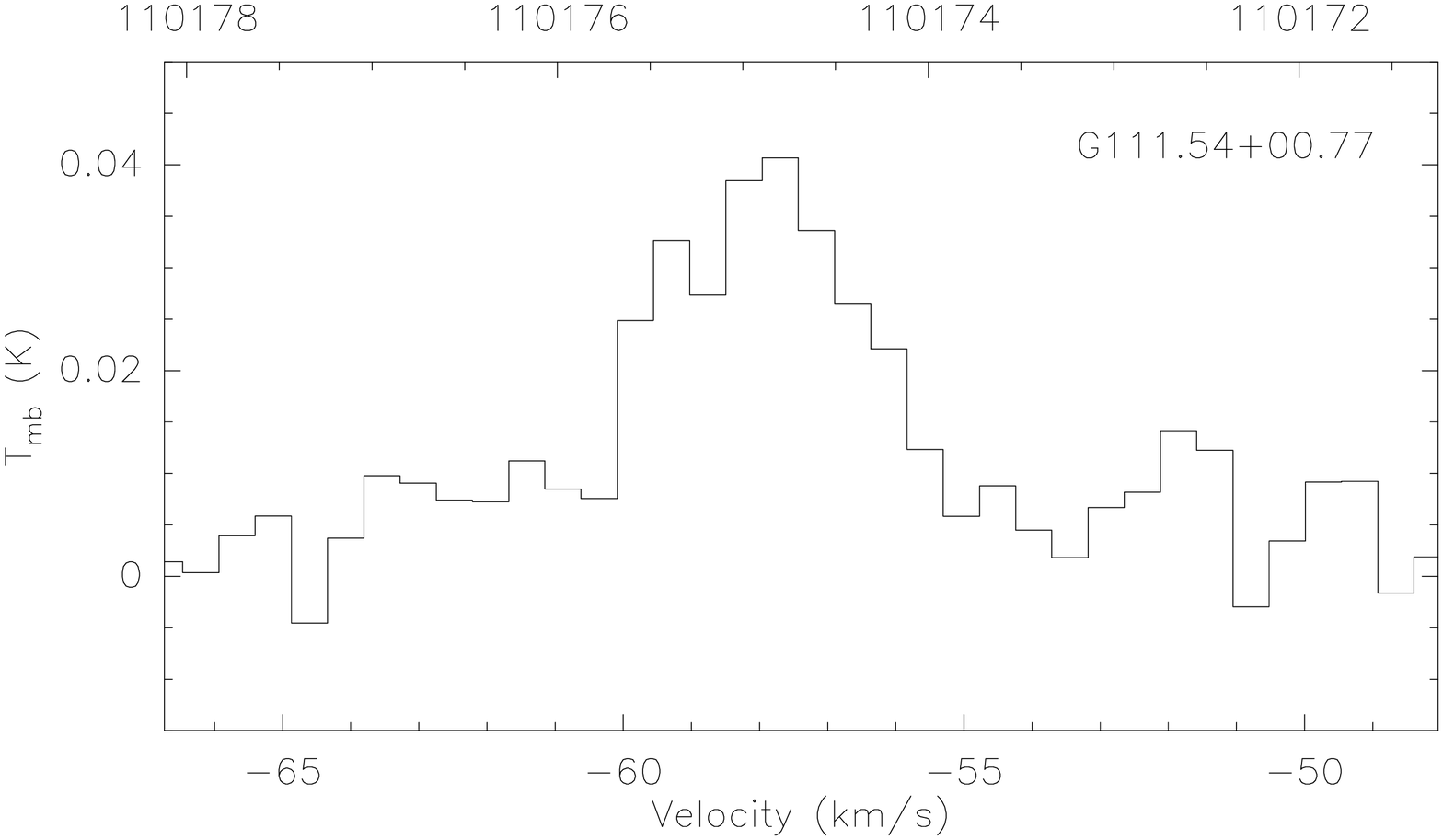}\includegraphics[width=0.5\textwidth]{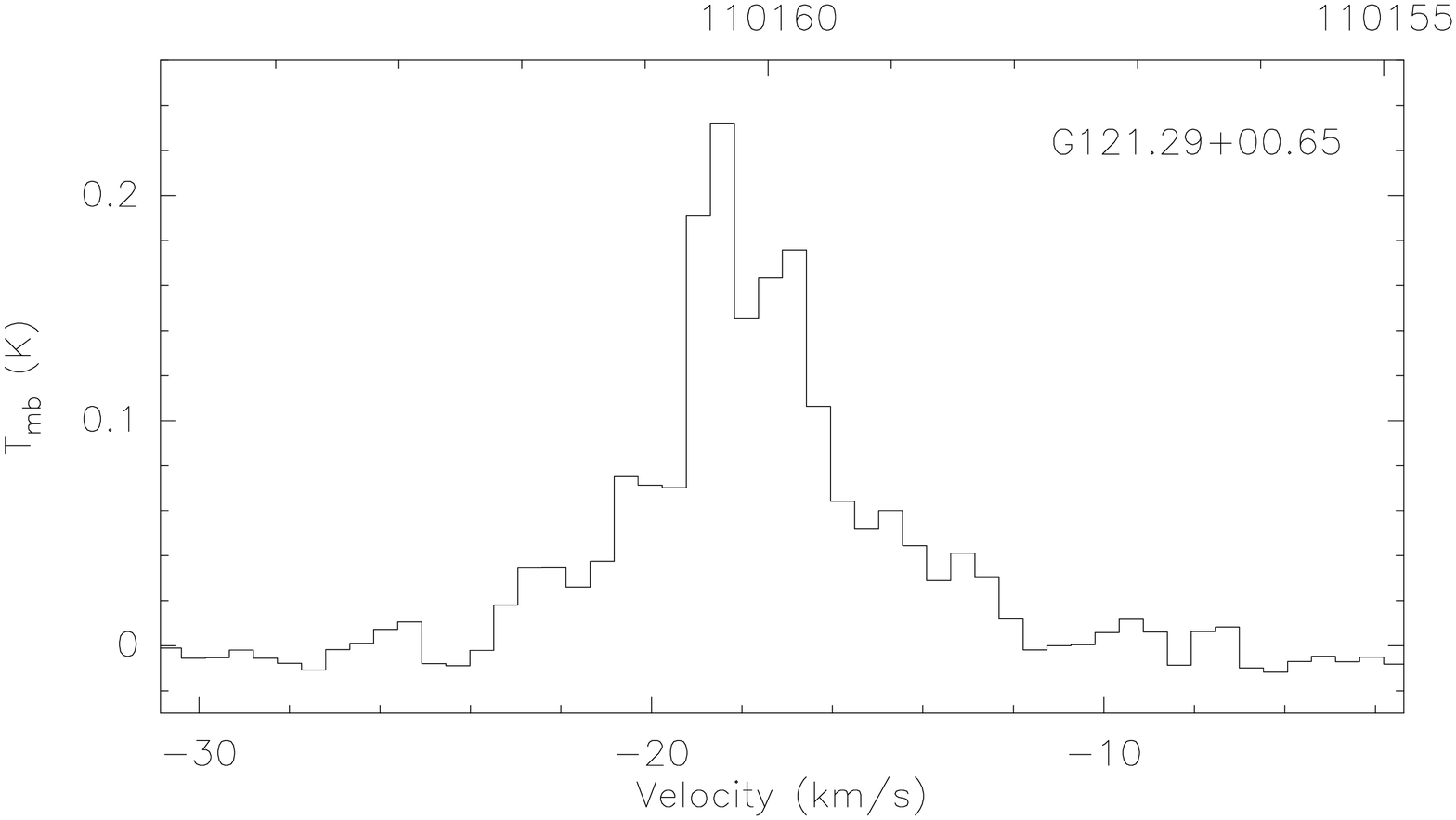}\\
\includegraphics[width=0.5\textwidth]{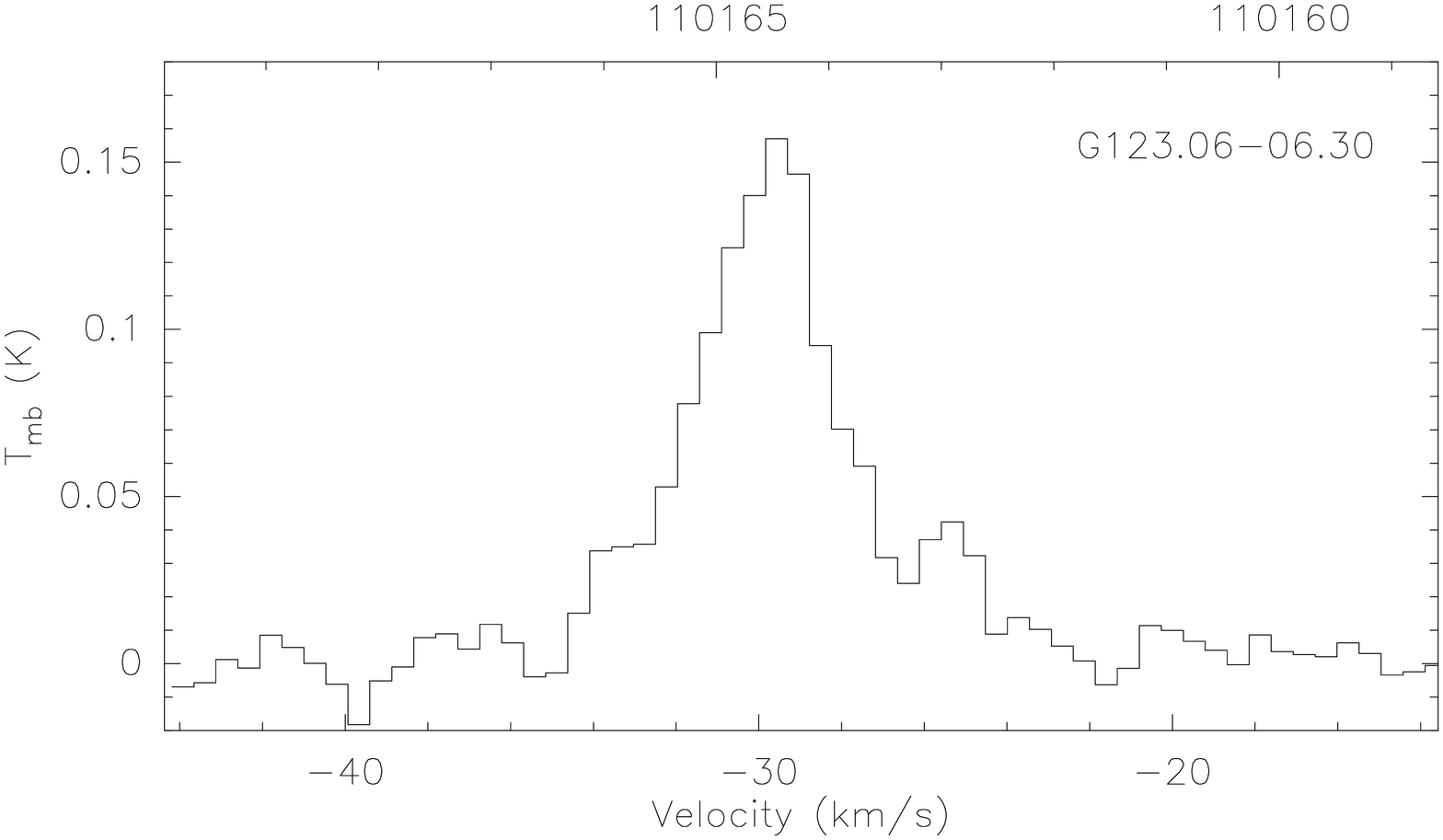}\includegraphics[width=0.5\textwidth]{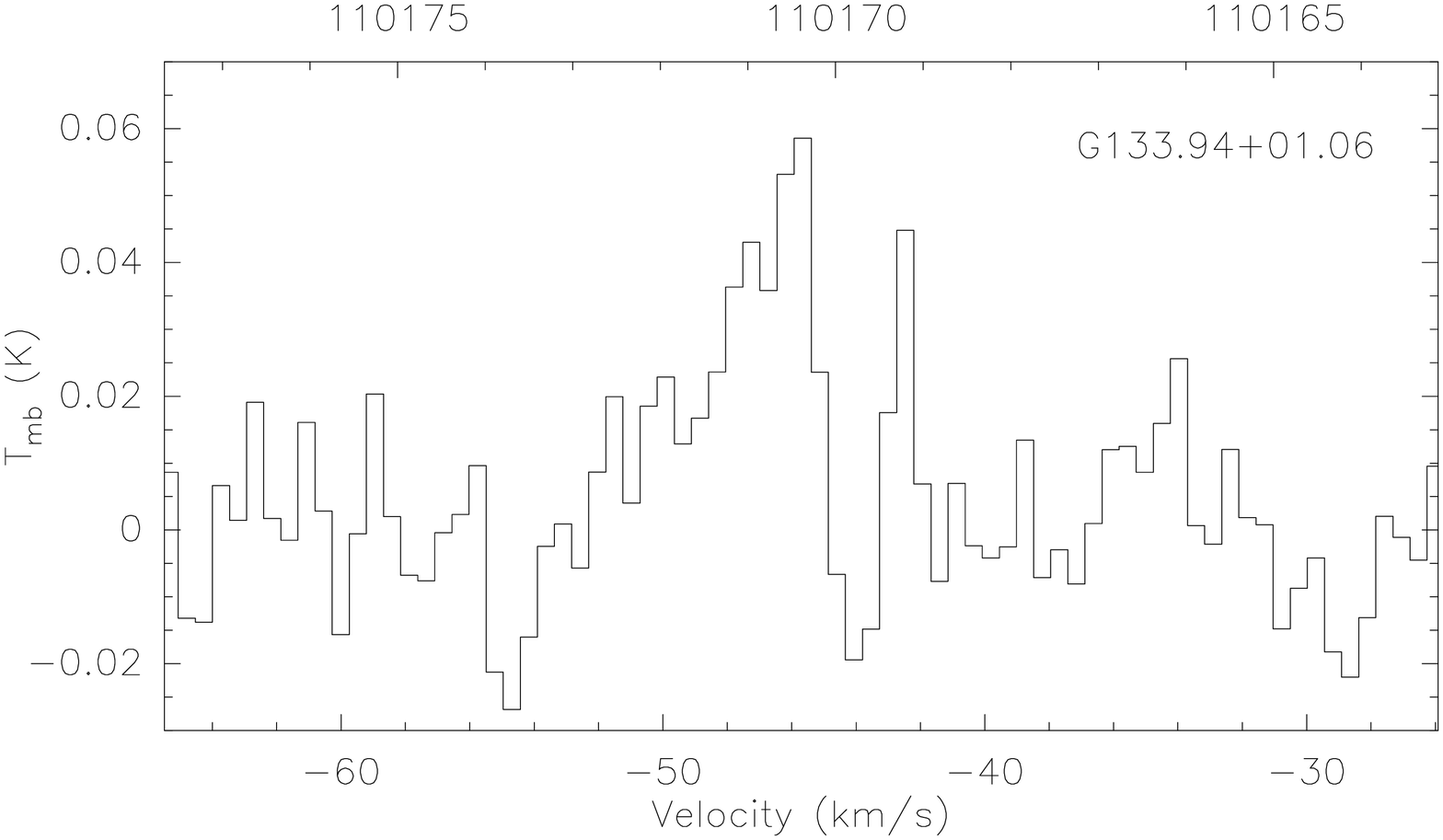}\\
	\caption{The spectra detected for NH$_2$D$1_{11}^a-1_{01}^s$.}
	\label{spectrum4}
\end{figure}

\begin{figure}
\centering
\includegraphics[width=0.5\textwidth]{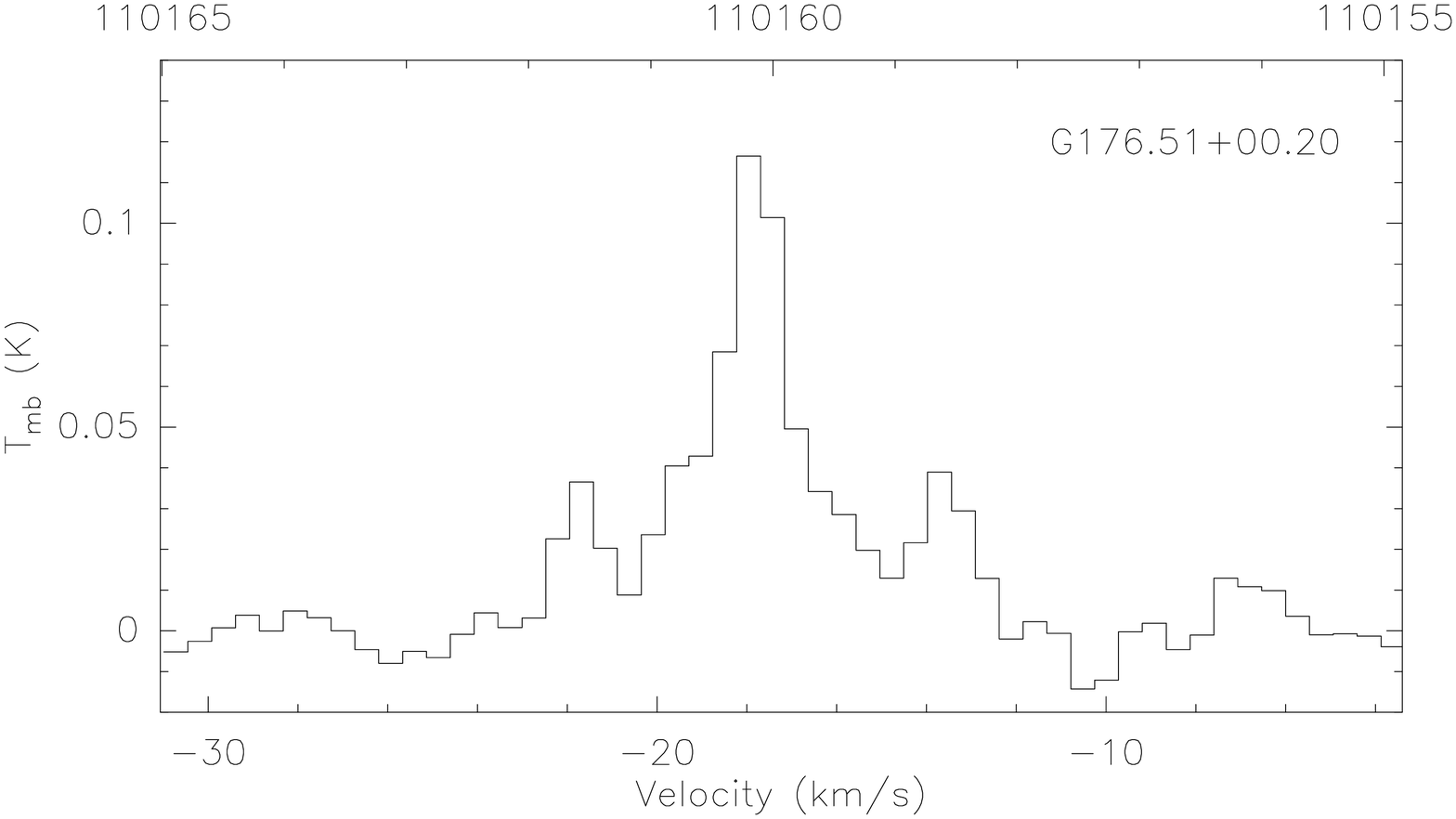}\includegraphics[width=0.5\textwidth]{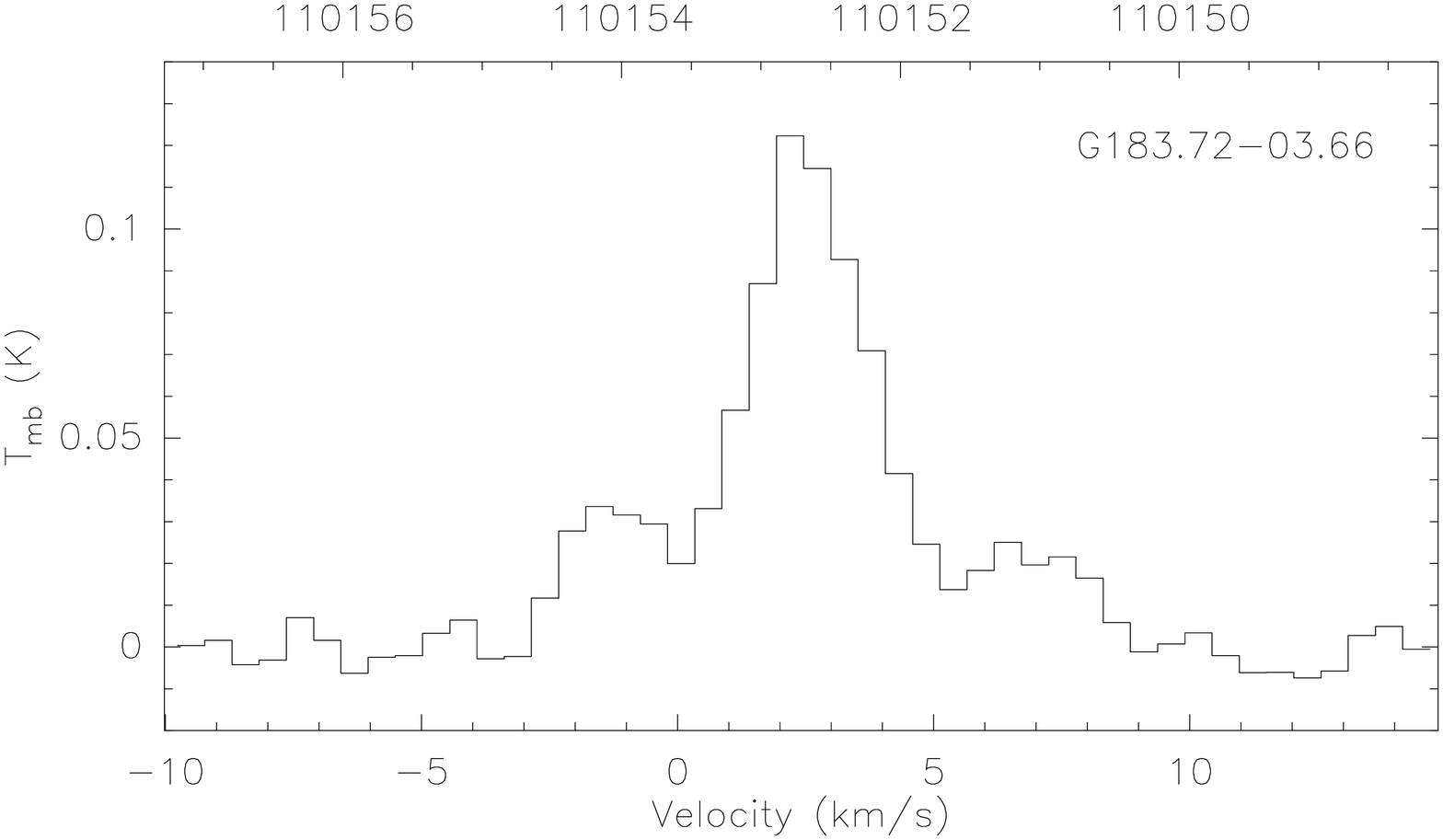}\\
\includegraphics[width=0.5\textwidth]{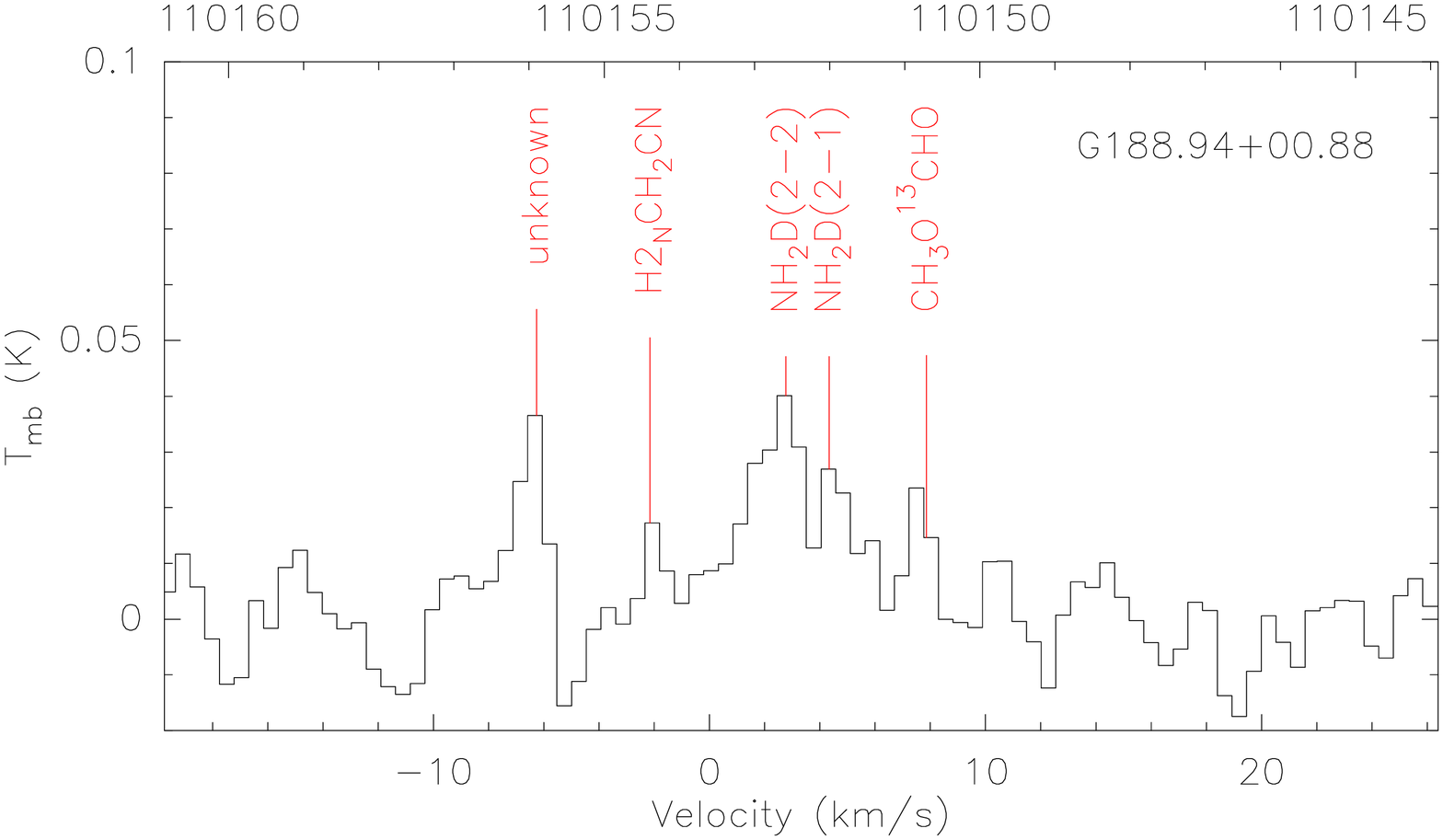}\includegraphics[width=0.5\textwidth]{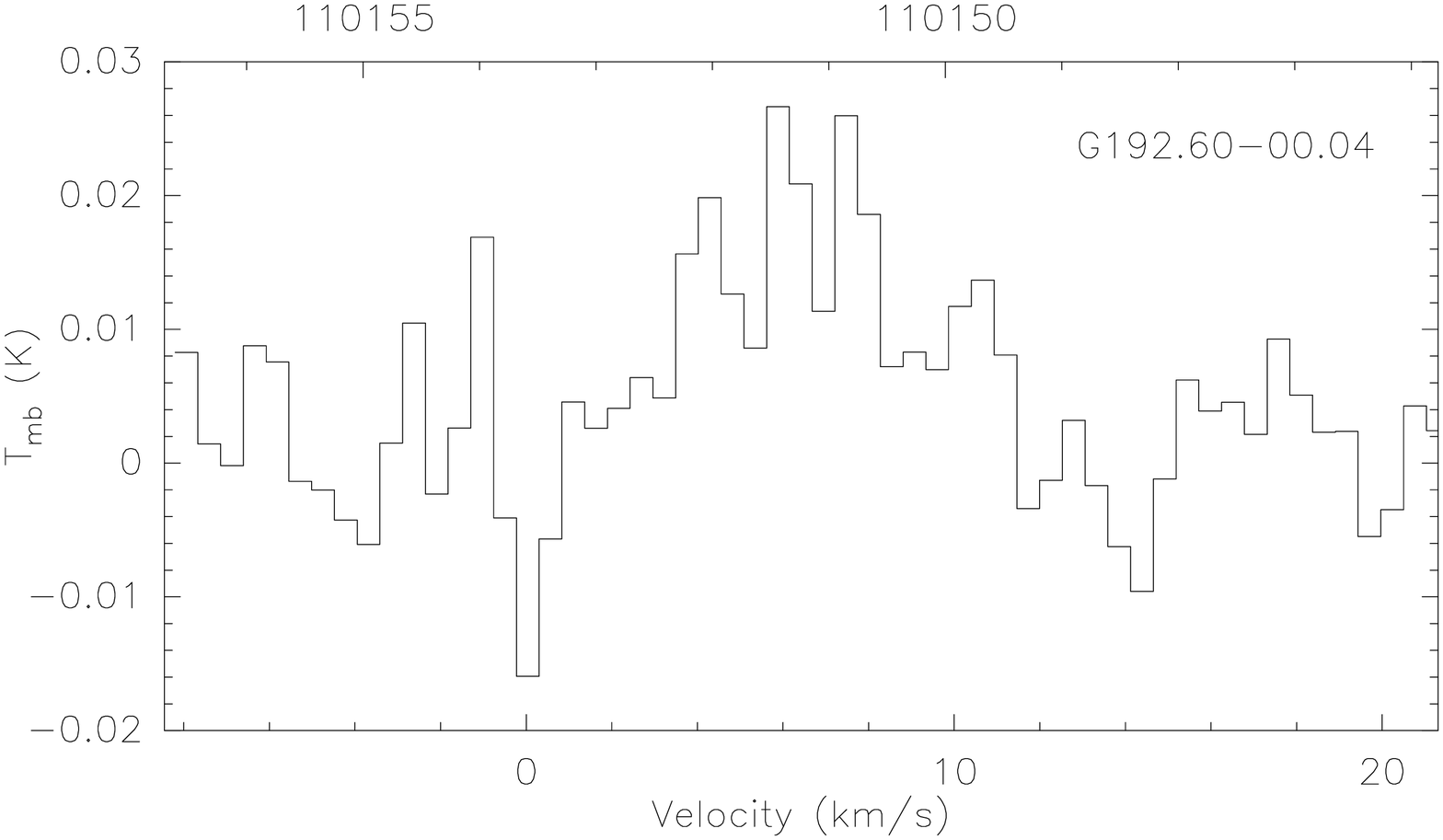}\\
	\caption{The spectra detected for NH$_2$D$1_{11}^a-1_{01}^s$.}
	\label{spectrum5}
\end{figure}

\bsp	

\label{lastpage}

\begin{thebibliography}{99}
\bibitem[\protect\citeauthoryear{Busquet et al.}{2010}]{2010A&A...517L...6B} Busquet G., Palau A., Estalella R., Girart J.~M., S{\'a}nchez-Monge {\'A}., Viti S., Ho P.~T.~P., et al., 2010, A\&A, 517, L6. doi:10.1051/0004-6361/201014866
\bibitem[\protect\citeauthoryear{Caselli et al.}{2002a}]{2002ApJ...565..344C} Caselli P., Walmsley C.~M., Zucconi A., Tafalla M., Dore L., Myers P.~C., 2002, ApJ, 565, 344. doi:10.1086/324302
\bibitem[\protect\citeauthoryear{Caselli et al.}{2002b}]{2002ApJ...565..331C} Caselli P., Walmsley C.~M., Zucconi A., Tafalla M., Dore L., Myers P.~C., 2002, ApJ, 565, 331. doi:10.1086/324301
\bibitem[\protect\citeauthoryear{Caselli et al.}{2008}]{2008A&A...492..703C} Caselli P., Vastel C., Ceccarelli C., van der Tak F.~F.~S., Crapsi A., Bacmann A., 2008, A\&A, 492, 703. doi:10.1051/0004-6361:20079009
\bibitem[\protect\citeauthoryear{Caselli \& Ceccarelli}{2012}]{2012A&ARv..20...56C} Caselli P., Ceccarelli C., 2012, A\&ARv, 20, 56. doi:10.1007/s00159-012-0056-x
\bibitem[\protect\citeauthoryear{Cazaux et al.}{2003}]{2003ApJ...593L..51C} Cazaux S., Tielens A.~G.~G.~M., Ceccarelli C., Castets A., Wakelam V., Caux E., Parise B., et al., 2003, ApJL, 593, L51. doi:10.1086/378038
\bibitem[\protect\citeauthoryear{Chen et al.}{2010}]{2010ApJ...713L..50C} Chen H.-R., Liu S.-Y., Su Y.-N., Zhang Q., 2010, ApJL, 713, L50. doi:10.1088/2041-8205/713/1/L50
\bibitem[\protect\citeauthoryear{Chen et al.}{2011}]{2011ApJ...743..196C} Chen H.-R., Liu S.-Y., Su Y.-N., Wang M.-Y., 2011, ApJ, 743, 196. doi:10.1088/0004-637X/743/2/196
\bibitem[\protect\citeauthoryear{Crapsi et al.}{2005}]{2005ApJ...619..379C} Crapsi A., Caselli P., Walmsley C.~M., Myers P.~C., Tafalla M., Lee C.~W., Bourke T.~L., 2005, ApJ, 619, 379. doi:10.1086/426472
\bibitem[\protect\citeauthoryear{Crapsi et al.}{2007}]{2007A&A...470..221C} Crapsi A., Caselli P., Walmsley M.~C., Tafalla M., 2007, A\&A, 470, 221. doi:10.1051/0004-6361:20077613
\bibitem[\protect\citeauthoryear{Emprechtinger et al.}{2009}]{2009A&A...493...89E} Emprechtinger M., Caselli P., Volgenau N.~H., Stutzki J., Wiedner M.~C., 2009, A\&A, 493, 89. doi:10.1051/0004-6361:200810324
\bibitem[\protect\citeauthoryear{Feng et al.}{2019}]{2019ApJ...883..202F} Feng S., Caselli P., Wang K., Lin Y., Beuther H., Sipil{\"a} O., 2019, ApJ, 883, 202. doi:10.3847/1538-4357/ab3a42
\bibitem[\protect\citeauthoryear{Fontani et al.}{2015}]{2015A&A...575A..87F} Fontani F., Busquet G., Palau A., Caselli P., S{\'a}nchez-Monge {\'A}., Tan J.~C., Audard M., 2015, A\&A, 575, A87. doi:10.1051/0004-6361/201424753
\bibitem[\protect\citeauthoryear{Frerking, Langer, \& Wilson}{1982}]{1982ApJ...262..590F} Frerking M.~A., Langer W.~D., Wilson R.~W., 1982, ApJ, 262, 590. doi:10.1086/160451
\bibitem[\protect\citeauthoryear{Gerner et al.}{2015}]{2015A&A...579A..80G} Gerner T., Shirley Y.~L., Beuther H., Semenov D., Linz H., Albertsson T., Henning T., 2015, A\&A, 579, A80. doi:10.1051/0004-6361/201423989
\bibitem[\protect\citeauthoryear{Gong et al.}{2015}]{2015A&A...581A..48G} Gong Y., Henkel C., Thorwirth S., Spezzano S., Menten K.~M., Walmsley C.~M., Wyrowski F., et al., 2015, A\&A, 581, A48. doi:10.1051/0004-6361/201526275
\bibitem[\protect\citeauthoryear{Harju et al.}{2017}]{2017A&A...600A..61H} Harju J., Daniel F., Sipil{\"a} O., Caselli P., Pineda J.~E., Friesen R.~K., Punanova A., et al., 2017, A\&A, 600, A61. doi:10.1051/0004-6361/201628463
\bibitem[\protect\citeauthoryear{Hatchell}{2003}]{2003A&A...403L..25H} Hatchell J., 2003, A\&A, 403, L25. doi:10.1051/0004-6361:20030297
\bibitem[\protect\citeauthoryear{Herbst}{1982}]{1982A&A...111...76H} Herbst E., 1982, A\&A, 111, 76
\bibitem[\protect\citeauthoryear{Jacq et al.}{1990}]{1990A&A...228..447J} Jacq T., Walmsley C.~M., Henkel C., Baudry A., Mauersberger R., Jewell P.~R., 1990, A\&A, 228, 447
\bibitem[\protect\citeauthoryear{Kirby}{2009}]{2009ApJ...694.1056K} Kirby L., 2009, ApJ, 694, 1056. doi:10.1088/0004-637X/694/2/1056
\bibitem[\protect\citeauthoryear{Li et al.}{2016}]{2016AJ....152...92L} Li F.~C., Xu Y., Wu Y.~W., Yang J., Lu D.~R., Menten K.~M., Henkel C., 2016, AJ, 152, 92. doi:10.3847/0004-6256/152/4/92
\bibitem[\protect\citeauthoryear{Linsky et al.}{1995}]{1995ApJ...451..335L} Linsky J.~L., Diplas A., Wood B.~E., Brown A., Ayres T.~R., Savage B.~D., 1995, ApJ, 451, 335. doi:10.1086/176223
\bibitem[\protect\citeauthoryear{Lis et al.}{2002}]{2002ApJ...569..322L} Lis D.~C., Gerin M., Phillips T.~G., Motte F., 2002, ApJ, 569, 322. doi:10.1086/339232
\bibitem[\protect\citeauthoryear{M{\"u}ller et al.}{2001}]{2001A&A...370L..49M} M{\"u}ller H.~S.~P., Thorwirth S., Roth D.~A., Winnewisser G., 2001, A\&A, 370, L49. doi:10.1051/0004-6361:20010367
\bibitem[\protect\citeauthoryear{M{\"u}ller et al.}{2005}]{2005JMoSt.742..215M} M{\"u}ller H.~S.~P., Schl{\"o}der F., Stutzki J., Winnewisser G., 2005, JMoSt, 742, 215. doi:10.1016/j.molstruc.2005.01.027
\bibitem[\protect\citeauthoryear{Oliveira et al.}{2003}]{2003ApJ...587..235O} Oliveira C.~M., H{\'e}brard G., Howk J.~C., Kruk J.~W., Chayer P., Moos H.~W., 2003, ApJ, 587, 235. doi:10.1086/368019
\bibitem[\protect\citeauthoryear{Pillai et al.}{2007}]{2007A&A...467..207P} Pillai T., Wyrowski F., Hatchell J., Gibb A.~G., Thompson M.~A., 2007, A\&A, 467, 207. doi:10.1051/0004-6361:20065682
\bibitem[\protect\citeauthoryear{Pillai et al.}{2011}]{2011A&A...530A.118P} Pillai T., Kauffmann J., Wyrowski F., Hatchell J., Gibb A.~G., Thompson M.~A., 2011, A\&A, 530, A118. doi:10.1051/0004-6361/201015899
\bibitem[\protect\citeauthoryear{Reid et al.}{2014}]{2014ApJ...783..130R} Reid M.~J., Menten K.~M., Brunthaler A., Zheng X.~W., Dame T.~M., Xu Y., Wu Y., et al., 2014, ApJ, 783, 130. doi:10.1088/0004-637X/783/2/130
\bibitem[\protect\citeauthoryear{Remijan, Markwick-Kemper, \& ALMA Working Group on Spectral Line Frequencies}{2007}]{2007AAS...21113211R} Remijan A.~J., Markwick-Kemper A., ALMA Working Group on Spectral Line Frequencies, 2007, AAS
\bibitem[\protect\citeauthoryear{Roberts \& Millar}{2000a}]{2000A&A...361..388R} Roberts H., Millar T.~J., 2000, A\&A, 361, 388
\bibitem[\protect\citeauthoryear{Roberts \& Millar}{2000b}]{2000A&A...364..780R} Roberts H., Millar T.~J., 2000, A\&A, 364, 780
\bibitem[\protect\citeauthoryear{Roueff et al.}{2005}]{2005A&A...438..585R} Roueff E., Lis D.~C., van der Tak F.~F.~S., Gerin M., Goldsmith P.~F., 2005, A\&A, 438, 585. doi:10.1051/0004-6361:20052724
\bibitem[\protect\citeauthoryear{Saito et al.}{2000}]{2000ApJ...535..227S} Saito S., Ozeki H., Ohishi M., Yamamoto S., 2000, ApJ, 535, 227. doi:10.1086/308818
\bibitem[\protect\citeauthoryear{Schneider et al.}{2010}]{2010A&A...520A..49S} Schneider N., Csengeri T., Bontemps S., Motte F., Simon R., Hennebelle P., Federrath C., et al., 2010, A\&A, 520, A49. doi:10.1051/0004-6361/201014481
\bibitem[\protect\citeauthoryear{Shah \& Wootten}{2001}]{2001ApJ...554..933S} Shah R.~Y., Wootten A., 2001, ApJ, 554, 933. doi:10.1086/321396
\bibitem[\protect\citeauthoryear{Turner}{1990}]{1990ApJ...362L..29T} Turner B.~E., 1990, ApJL, 362, L29. doi:10.1086/185840
\bibitem[\protect\citeauthoryear{Vanden Bout et al.}{1983}]{1983ApJ...271..161V} Vanden Bout P.~A., Loren R.~B., Snell R.~L., Wootten A., 1983, ApJ, 271, 161. doi:10.1086/161184
\bibitem[\protect\citeauthoryear{Wienen et al.}{2021}]{2021A&A...649A..21W} Wienen M., Wyrowski F., Walmsley C.~M., Csengeri T., Pillai T., Giannetti A., Menten K.~M., 2021, A\&A, 649, A21. doi:10.1051/0004-6361/201731208






\end{thebibliography}
\end{document}